\pdfoutput=1
\documentclass[12pt,titlepage]{article}

\usepackage{graphicx}
\usepackage{latexsym}
\usepackage{mathrsfs}
\font\grande=cmr9.5 scaled \magstep4
\font\medio=cmr9.5 scaled \magstep2
\outer\def\beginsection#1\par{\medbreak\bigskip
      \message{#1}\leftline{\bf#1}\nobreak\medskip
\vskip-\parskip
      \noindent}

\begin{document}
\bibliographystyle {unsrt}

\titlepage

\begin{flushright}
\end{flushright}

\vspace{1cm}
\begin{center}
{\grande Palatini approach and large-scale magnetogenesis}\\
\vspace{1cm}
 Massimo Giovannini 
 \footnote{Electronic address: massimo.giovannini@cern.ch} \\
\vspace{1cm}
{{\sl Department of Physics, CERN, 1211 Geneva 23, Switzerland }}\\
\vspace{0.5cm}
{{\sl INFN, Section of Milan-Bicocca, 20126 Milan, Italy}}
\vspace*{1cm}
\end{center}

\vskip 0.3cm
\centerline{\medio  Abstract}
\vskip 0.1cm
Large-scale magnetogenesis is analyzed within the Palatini approach when the gravitational action is supplemented by a contribution that is nonlinear in the Einstein-Hilbert term. While the addition of the nonlinear terms does not affect the scalar modes of the geometry during the inflationary phase, the tensor-to-scalar ratio is nonetheless suppressed. In this context it is plausible to have a stiff phase following the standard inflationary stage provided the potential has a quintessential form. The large-scale magnetic fields can even be a fraction of the nG over typical length scales of the order of the Mpc prior to the gravitational collapse of the protogalaxy.
\noindent

\vspace{5mm}
\vfill
\newpage

\renewcommand{\theequation}{1.\arabic{equation}}
\setcounter{equation}{0}
\section{Introduction}
\label{sec1}
Higher-order actions are a recurrent theme in various areas of physics and are 
often invoked as a short-scale modifications of the underlying theory. 
Their addition to the Einstein-Hilbert action is a key element of quantum
theories in curved background geometries \cite{Hone1,Hone2}. Quadratic gravitational actions 
also appear in the first string tension correction to the (tree-level) effective action  
\cite{Htwo1,Htwo2,Htwo3} and often involve the Euler-Gauss-Bonnet combination \cite{Htwo4,Htwo5}. In four space-time dimensions the generalized Gauss-Bonnet theorem states that the homonym combination of quadratic invariants is a topological term and coincides 
with the Euler number so that its metric variation vanishes exactly. In 
single-field inflationary models the conventional (tree level) scalar-tensor action 
 can be regarded as the first term of a generic effective field theory where the higher derivatives are suppressed by the negative powers of a large mass-scale associated with the fundamental theory 
 that underlies the effective description \cite{Htwo6}. In this case the leading 
 correction (containing all possible terms with four derivatives) 
 also includes higher-order curvature terms.

When gravitation is treated within the metric approach the higher-derivative terms are always associated, 
in spite of their physical origin, to a short-distance modification of the Einstein-Hilbert theory. In four-dimensions, unless the higher-order corrections do not contain the Euler-Gauss-Bonnet combination, the metric approach suggests that further degrees of freedom may appear \cite{Htwo8}. It has been noted long ago that  the Palatini and the metric formulations are inequivalent in the case of higher-order actions (see e.g. \cite{Htwo9a,Htwo9b} and references therein). If the inflationary models are treated within the Palatini approach (and in the presence of higher order gravitational contributions) the physical situation is quite intersting and has been recently explored in various papers
\cite{PAL1,PAL2,PAL3,PAL4,PAL4a,PAL5,PAL6,PAL7,PAL8}. In the Palatini 
 formulation the nonlinear gravity action does not introduce supplementary degrees of freedom but 
 it  modifies the relation among the existing ones \cite{PAL1,PAL2,PAL3}. In the context 
 of conventional inflationary models this means, in particular, that the relation between 
 the inflaton and the curvature scale may be modified as it happens when the 
 original higher-derivative Starobinsky model \cite{PAL9} is analyzed within the Palatini approach as 
 pointed out in Ref. \cite{PAL1}. If the inflaton is minimally coupled in the presence 
 of a higher-order gravity action various classes of inflationary models that are currently 
 under pressure because of an excessive tensor to scalar ratio \cite{RT1,RT2,RT2a} become viable again 
 within the Palatini approach \cite{PAL2,PAL3}. This idea has been developed in various frameworks 
 by changing both the matter contribution and the nonlinear gravitational action 
 \cite{PAL4,PAL5,PAL6,PAL7,PAL8}. 
 
Among the various potentials that could be rescued in the framework of the Palatini formulation some classes of quintessential inflationary models are particularly interesting \cite{PAL4} since, in their original formulation \cite{QUINT1}, they involve power law inflationary potentials (typically quartic \cite{QUINT2}) which are replaced by inverse power-laws \cite{QUINT3} during the conventional quintessential evolution \cite{QUINT4}. For some classes of inflationary potentials this scenario is at odds with the current observational evidence since the corresponding tensor-to-scalar ratio is too large. Indeed  the power-law potentials in the conventional metric analysis are severely constrained \cite{RT1,RT2,RT2a} (see also \cite{QUINT5})  and  this has been the case since the final data release of the WMAP collaboration  \cite{RT3,RT4}. The corresponding limits on the tensor to scalar ratio $r_{T}$ can be however evaded in the Palatini formulation \cite{PAL1,PAL2,PAL3,PAL4} exactly because the nonlinear gravity action changes the relation between the slow-roll parameters and ultimately suppresses $r_{T}$. 

If quintessential inflation is analyzed in the Palatini formalism and in the context of a nonlinear gravitational action the scalar modes of the geometry produced by the inflationary evolution and in the subsequent stiff epoch are not affected but the tensor power spectra and the related $r_{T}$ are suppressed. The high-frequency spike of the relic graviton spectrum is squeezed and the whole signal is suppressed \cite{QUINT6} at least when the higher-order contributions appearing in the action are explicitly decoupled from the inflaton \cite{PAL4}. The post-inflationary evolution involves a stiff epoch whose potential role has been suggested in various different scenarios. 
The first speculations suggesting that the early expansion rate could have been much slower than radiation go back to  the mid sixties \cite{QUINT7,QUINT8,QUINT9}. After the formulation of conventional inflationary models Ford \cite{QUINT10} noted that gravitational particle production at the end of inflation could account for the entropy of the present universe and observed that the backreaction effects of the created quanta constrain the length of a stiff post-inflationary phase by making the expansion dominated by radiation. It has been later argued by  Spokoiny \cite{QUINT11} that various classes of scalar field potentials exhibit a transition from inflation to a stiff phase dominated by the kinetic energy of the inflaton.

The purpose of this paper is to discuss the interplay between the evolution 
of the Abelian hypercharge fields possibly present during inflation and the Palatini formulation of quintessential inflation when the gravitational action contains contributions that are nonlinear in the Einstein-Hilbert term. Since the inflaton and the quintessence field are unified in a single quantity the physical description is more economical. Furthermore the presence of a long stiff phase increases the total number of inflationary $e$-folds presently accessible to large-scale observations \cite{QUINT12}. For the sake of concreteness we shall therefore analyze the following four-dimensional action:
\begin{eqnarray}
S &=& - \frac{1}{2 \ell_{P}^2} \int d^{4} x\,\,\sqrt{-G} \,f(\overline{R})  + \int d^{4} x\,\ \sqrt{-G}\,\, \biggl[ \frac{1}{2} G^{\alpha\beta} \partial_{\alpha} \varphi \partial_{\beta} \varphi - V(\varphi) \biggr]
\nonumber\\
&-&\frac{1}{16 \pi} \int d^{4} x \sqrt{- G} \biggl[\, \, \sigma(\varphi) \,\,Y_{\alpha\beta} \, Y^{\alpha\beta} + 
\overline{\sigma}(\varphi) Y_{\alpha\beta} \, \widetilde{\,Y\,}^{\alpha\beta}\biggr],
\label{AAA1}
\end{eqnarray}
where $\ell_{P} = \sqrt{8 \pi G}$ and $\overline{R}$ denotes, for convenience, the Ricci scalar 
defined in terms of the Palatini connections $\overline{\Gamma}_{\alpha\beta}^{\,\,\,\lambda}$ that 
do not share the same relation of the Christoffel symbols with the derivatives 
of the metric tensor $G_{\mu\nu}$.  
The first term of Eq. (\ref{AAA1}) denotes the gravitational action while the second and third 
contributions account for the dynamics of the inflaton $\varphi$ and of the gauge fields. 
In the case of the Einstein-Hilbert action (i.e. $f(\overline{R}) = \overline{R}$) 
the Palatini and the metric approach are equivalent. However, as already mentioned,  in the presence of higher derivatives 
this is not the case. In Eq. (\ref{AAA1}) $Y_{\alpha\beta}$ and $\widetilde{\, Y\, }^{\alpha\beta}$ are, respectively, the Abelian 
gauge field strength and its dual defined in terms of the Palatini metric:
\begin{equation}
\widetilde{\,Y\,}^{\alpha\beta} = \frac{1}{2} \, E^{\alpha\beta\rho\sigma} Y_{\rho\sigma}, \qquad E^{\alpha\beta\rho\sigma}= 
\frac{\epsilon^{\alpha\beta\rho\sigma}}{\sqrt{-G}},
\label{AAAdef}
\end{equation}
where $\epsilon^{\alpha\beta\rho\sigma}$ denotes the totally antisymmetric (Levi-Civita) symbol. 
If we define the four-dimensional gauge coupling as $g= \sqrt{4 \pi/\sigma}$ the action for the Abelian fields can always be expressed as:
\begin{equation}
S_{gauge} = - \int d^{4} x\, \frac{\sqrt{- G}}{4 g^2} \biggl[ Y_{\alpha\beta} \, Y^{\alpha\beta} + \biggl(\frac{\overline{\sigma}}{\sigma}\biggr)
Y_{\alpha\beta}\, \widetilde{\,Y\,}^{\alpha\beta} \biggr].
\label{AAA1a}
\end{equation}
While it will be assumed throughout that $\sigma= \sigma(\varphi)$ and $\overline{\sigma}= \overline{\sigma}(\varphi)$,  different situations can be imagined where the gauge coupling 
depend upon some other spectator as it happens in low-scale quintessential inflation 
\cite{QUINT13}. The presence of a direct coupling between the gauge fields and the inflaton-quintessence field breaks Weyl invariance
and may potentially produce large-scale magnetic fields whose origin is still under intense debate \cite{rev1,rev2,rev3}. 
 Various models of explicit breaking of the Weyl invariance have been 
investigated and it would be impossible to account here for all the proposals. The action (\ref{AAAdef}) actually generalises and unifies two different 
perspectives associated with the pseudoscalar \cite{PSC1,PSC2,PSC3} and scalar couplings \cite{PSC4,PSC5,PSC7} of the gauge kinetic term 
to the inflaton. The presence of the pseudo-scalar vertex in Eq. (\ref{AAAdef}) permits the production of hypermagnetic gyrotropy which may be ultimately responsible of the baryon asymmetry of the universe \cite{PSC9,PSC10,PSC11}. Unlike all the previous analyses of Eq. (\ref{AAAdef}) we shall consider here an explicit model based on quintessential inflation and complemented by a nonlinear action in the Palatini approach\footnote{We remark, in this respect, that the present framework cannot be obviously 
reduced to the case of a plateau-like potential supplemented by a kinetic phase (possibly added by hand).
Furthermore plateau-like potentials (even if deduced within a higher-derivative model like the one 
of Ref. \cite{PAL9}) are normally not analyzed within the Palatini approach and do not obviously 
lead to a quintessential phase when the inflaton goes to zero. }

The layout of this investigation is the following. In section \ref{sec2} the governing 
equations of the system are analyzed; particular attention is to  the overall consistency
between the Palatini and the Einstein frames. Section \ref{sec3} deals 
with the physical aspects of the inflationary and of the quitessential evolutions; the suppression of $r_{T}$ is revisited in the specific framework of this scenario. In the second part of this 
section we discuss the evolution of the gauge coupling and its parametrization. The quantum mechanical description of the hyperelectric and hypermagnetic fields appears in section \ref{sec4} 
where the power spectra are derived in the inflationary stage; in the second part of the section the power spectra are deduced in the stiff and in the radiation-dominated stages. The main 
phenomenological implications are analyzed in section \ref{sec5} where 
we separately consider the constraints on the gauge spectra during the inflationary stage and in the 
post-inflationary epoch. The magnetogenesis requirements are then examined by assuming that 
the evolution of the gauge coupling either freezes during the stiff epoch or in the radiation stage. 
The concluding remarks are contained in section \ref{sec6}.

\renewcommand{\theequation}{2.\arabic{equation}}
\setcounter{equation}{0}
\section{Palatini and Einstein frames}
\label{sec2}
To avoid confusions it is useful to appreciate that the Palatini frame introduced here is not related with the so-called Jordan frame customarily introduced when the scalar-tensor theories are analyzed in the metric approach. While in the Palatini frame ($P$-frame in what follows) the connections do not have the standard form in terms of the derivatives of the metric, in the Einstein frame ($E$-frame in what follows) the connections have the conventional Levi-Civita expression. For this reason one could refer to the  Levi-Civita fame but this terminology is uncommon. The differences and the analogies between  the Palatini and the Einstein frames will now be introduced.  Even if each of the two frames has its own advantages, the two descriptions can be used interchangeably provided their connection is clearly established and this is actually one of the purposes of this section.

\subsection{The $P$-frame}
The extremization of the action (\ref{AAA1}) with respect to the metric variation implies: 
\begin{equation}
F \, \overline{R}_{\mu\nu} - \frac{f}{2} \,G_{\mu\nu} = \ell_{P}^2 \, T_{\mu\nu}^{(P)},\qquad F(\overline{R}) = \frac{\partial f}{\partial \overline{R}},
\label{AAA2}
\end{equation} 
where $T_{\mu\nu}^{(P)}$ is the total energy momentum in the $P$-frame:
\begin{eqnarray}
 T_{\mu\nu}^{(P)} &=& \partial_{\mu} \varphi \partial_{\nu} \varphi - G_{\mu\nu} \biggl[\frac{1}{2} G^{\alpha\beta} \partial_{\alpha} \varphi \partial_{\beta} \varphi - V(\varphi) \biggr]
\nonumber\\
&+& \frac{1}{2 g^2} \biggl[ - Y_{\mu\alpha} \, Y_{\nu}^{\,\,\,\alpha} + \frac{1}{4} G_{\mu\nu} \, Y_{\alpha\beta} \, 
Y^{\alpha\beta} \biggr].
\label{AAA3}
\end{eqnarray}
The extremization of the action (\ref{AAA1}) with respect to the variation of the Palatini connections 
$\overline{\Gamma}_{\alpha\beta}^{\,\,\,\,\,\,\lambda}$ implies instead the following condition
\begin{equation}
\overline{\nabla}_{\lambda} \biggl( \sqrt{- G} \,\, G^{\alpha\beta} F \biggr) =0.
\label{AAA4}
\end{equation}
Equation (\ref{AAA4}) does not define a Levi-Civita connection however it can be brought in that 
form by defining a rescaled metric $\overline{G}_{\alpha\beta}$:
\begin{equation}
 \overline{G}_{\alpha\beta} = F \, G_{\alpha\beta}\qquad \Rightarrow \qquad \overline{\nabla}_{\lambda} \biggl(\sqrt{- \overline{G}} \,\, \overline{G}^{\alpha\beta}\biggr)=0.
\label{AAA4a}
\end{equation}
Equation (\ref{AAA4a}) now implies that the Palatini connection does have the Levi-Civita form in terms of the 
rescaled metric $\overline{G}_{\alpha\beta}$. The two different metrics $\overline{G}_{\alpha\beta}$ and $G_{\alpha\beta}$ define two complementary physical frames that will be generically referred to as the Einstein and the Palatini 
frames respectively. While the inflationary evolution is more conveniently studied  
in the $E$-frame, the stiff dynamics becomes simpler in the $P$-frame. The two 
frames will be shown to be equivalent both at the level of the background and for the 
corresponding inhomogeneities. If both sides of Eq. (\ref{AAA3}) are contracted with the 
Palatini metric the result is: 
\begin{equation}
\overline{R}(G) = - \ell_{P}^2 T^{(P)}= \ell_{P}^2 \biggl( G^{\alpha\beta} \partial_{\alpha} \varphi \partial_{\beta} \varphi - 4 V\biggr).
\label{AAA5}
\end{equation}
In Eq. (\ref{AAA5}) the notation $\overline{R}(G)$ reminds that the corresponding contraction is performed in terms of the Palatini metric, 
i.e. $\overline{R}(G) = G^{\alpha\beta} \, \overline{R}_{\alpha\beta}$. The same contraction can be performed with respect to the Einstein metric and this operation will then be denoted as $\overline{R}(\overline{G}) = \overline{G}^{\alpha\beta} \, \overline{R}_{\alpha\beta}$. It should be clear, from the above considerations, that $\overline{R}(G)$ and $\overline{R}(\overline{G})$ do not coincide.

\subsection{From the $P$-frame to the $E$-frame}
In the Palatini frame Eq. (\ref{AAA2}) is supplemented by the evolution of $\varphi$ and by the equations of the gauge 
fields. In particular by  extremizing the action (\ref{AAA1}) with respect to the variation of $\varphi$ we will have: 
\begin{equation}
\frac{1}{\sqrt{- G}} \partial_{\alpha} \biggl( \sqrt{- G} G^{\alpha\beta} \partial_{\beta} \varphi \biggr) + V_{\,,\varphi} +
\frac{1}{16 \pi} \biggl[ \sigma_{\,,\varphi} Y_{\alpha\beta} \, Y^{\alpha\beta} + \overline{\sigma}_{\,,\varphi} Y_{\alpha\beta} \, \widetilde{\,Y\,}^{\alpha\beta} \biggr] =0,
\label{AAA6}
\end{equation}
where $V_{\,,\varphi} = \partial_{\varphi} V$ and similarly for the susceptibilities $\sigma_{\,,\varphi} = \partial_{\varphi} \sigma$ and $\overline{\sigma}_{\,,\varphi} = \partial_{\varphi} \overline{\sigma}$. Always in the $P$-frame the evolution equation for the gauge fields follows 
from the corresponding variation of Eq. (\ref{AAA1})
\begin{eqnarray}
&& \partial_{\alpha} \biggl( \sqrt{-G} \, \sigma \, Y^{\alpha\beta} \biggr) + \partial_{\alpha}\biggl( \sqrt{- G}\, \,\overline{\sigma} \, 
\widetilde{\, Y\,}^{\alpha\beta} \biggr) =0, 
\label{AAA7}\\
&& \partial_{\alpha} \biggl(\sqrt{- G} \,\widetilde{\, Y\,}^{\alpha\beta} \biggr)=0.
\label{AAA8}
\end{eqnarray}
Equations (\ref{AAA6}) and (\ref{AAA7})--(\ref{AAA8}) demonstrate that the equations for the inflaton and for the gauge fields have a standard expression in the $P$-metric.  Conversely the evolution of the geometry becomes simpler in the $E$-frame metric $\overline{G}_{\mu\nu}$ where the connections $\overline{\Gamma}_{\alpha\beta}^{\,\,\,\,\mu}$ have 
the standard Levi-Civita form. In the $E$-frame the expressions of the gauge potentials and of the gauge fields are different; in particular for the transformation  
 $G_{\mu\nu} \to \overline{G}_{\mu\nu} = F \,G_{\mu\nu}$ suggested by Eq. (\ref{AAA4a}) the gauge potentials transform as 
\begin{equation}
Y_{\alpha} \to Z_{\alpha} = Y_{\alpha}, \qquad Y^{\alpha} \to Z^{\alpha} =  Y^{\alpha}/F.
\label{AAA9}
\end{equation}
The rationale for the result is quite simple but since it is potentially confusing we remind that the second expression in Eq. (\ref{AAA9}) follows from writing, in the new frame, $Z^{\alpha} = \overline{G}^{\alpha\beta} Z_{\beta}= \overline{G}^{\alpha\beta} Y_{\beta}$; since $\overline{G}^{\alpha\beta} = G^{\alpha\beta}/F$ we finally have, as expected,  $Z^{\alpha} = G^{\alpha\beta} Y_{\beta}/F = Y^{\alpha}/F$. In the same way we can deduce the transformations for the gauge field 
strengths and for their duals:
\begin{eqnarray}
&&Y_{\alpha\beta} \to Z_{\alpha\beta} = Y_{\alpha\beta}, \qquad Y^{\alpha\beta} \to Z^{\alpha\beta} =Y^{\alpha\beta}/F^2,
\label{AAA10}\\
&&\widetilde{\, Y\,}_{\alpha\beta} \to \widetilde{\, Z\,}_{\alpha\beta} = \widetilde{\, Y\,}_{\alpha\beta}, \qquad \widetilde{\, Y\,}^{\alpha\beta} \to \widetilde{\, Z\,}^{\alpha\beta} = \widetilde{\, Y\,}^{\alpha\beta}/F^2.
\label{AAA11}
\end{eqnarray}
Note that the dual field strength in the $P$-frame has been already introduced in 
Eq. (\ref{AAAdef}). In the $E$-frame $\widetilde{\, Z\,}^{\alpha\beta}$ will be defined 
in terms of the corresponding metric:
\begin{equation}
\widetilde{\, Z\,}^{\alpha\beta} = \frac{1}{2} \overline{E}^{\alpha\beta\rho\sigma} Z_{\rho\sigma}, \qquad\qquad \overline{E}^{\alpha\beta\rho\sigma}
= \frac{\epsilon^{\alpha\beta\rho\sigma}}{\sqrt{-\overline{G}}}.
\label{AAA13}
\end{equation}
But since $\sqrt{-\overline{G}} = F^2\, \sqrt{-G}$ we also have that Eqs. (\ref{AAAdef}) and (\ref{AAA13}) transform, as they must, according to Eq. (\ref{AAA11}). 

\subsection{The governing equations in the $E$-frame}
In the $E$-frame the analog of Eq. (\ref{AAA2}) follows from the corresponding 
transformation between the two metrics and the final result is:
\begin{equation}
F \overline{R}_{\mu\nu}(\overline{G}) = \frac{f}{2 F} \overline{G}_{\mu\nu} + \ell_{P}^2 T_{\mu\nu}^{(E)},
\label{AAA14}
\end{equation}
where $T_{\mu\nu}^{(E)}$ denotes the total energy-momentum tensor in the $E$-frame, i.e. 
\begin{eqnarray}
 T_{\mu\nu}^{(E)} &=& \partial_{\mu} \varphi \partial_{\nu} \varphi - \overline{G}_{\mu\nu} \biggl[\frac{1}{2} \overline{G}^{\alpha\beta} \partial_{\alpha} \varphi \partial_{\beta} \varphi - \frac{V}{F} \biggr]
\nonumber\\
&+& \frac{F}{2 g^2} \biggl[ - Z_{\mu\alpha} \, Z_{\nu}^{\,\,\,\alpha} + \frac{1}{4} \overline{G}_{\mu\nu} \, Z_{\alpha\beta} \, 
Z^{\alpha\beta} \biggr].
\label{AAA15}
\end{eqnarray}
The gauge part of the energy-momentum tensor is readily obtained by taking into account 
the transformations of Eqs. (\ref{AAA10})--(\ref{AAA11}).
As already mentioned $\overline{R}_{\mu\nu}(\overline{G})$ and $\overline{R}_{\mu\nu}(G)$ differ by a conformal 
transformation and, in particular,
\begin{equation}
\overline{R}_{\mu\nu}(G) = \overline{R}_{\mu\nu}(\overline{G}) -  2 \biggl[ \partial_{\mu} q \partial_{\nu} q - G_{\mu\nu} (\partial q)^2\biggr] + 2 \biggl[ \nabla_{\mu} \nabla_{\nu} q + \frac{G_{\mu\nu}}{2} \nabla^2 q \biggr],
\label{AAA16}
\end{equation}
where $q$ is the natural logarithm of the conformal factor (i.e.  $q = \ln{\sqrt{F}}$); moreover, for 
notational convenience, we adopted the shorthand notations $\nabla^2 q = G^{\alpha\beta} \nabla_{\alpha} \nabla_{\beta} q$ and $(\partial q)^2 = G^{\alpha\beta} \partial_{\alpha} q \partial_{\beta} q$. If we introduce $\overline{{\mathcal G}}_{\mu\nu}$ (i.e. 
the Einstein tensor in the $E$-frame) Eq.  (\ref{AAA14}) can also be expressed as:
\begin{equation}
\overline{{\mathcal G}}_{\mu\nu}  = - \frac{f}{2 F^2} \overline{G}_{\mu\nu} + \frac{\ell_{P}^2}{F} \biggl[ T_{\mu\nu}^{(E)} - 
\frac{T^{(E)}}{2} \overline{G}_{\mu\nu} \biggr],\qquad\overline{{\mathcal G}}_{\mu\nu} = \overline{R}_{\mu\nu}(\overline{G}) - \frac{1}{2} \overline{G}_{\mu\nu} \overline{R}(\overline{G}).
\label{AAA17}
\end{equation}
As already suggested after Eqs. (\ref{AAA6}) and (\ref{AAA7})--(\ref{AAA8}) the 
evolution of the inflaton will be slightly more cumbersome in the $E$-frame description and it is:
\begin{equation}
\overline{G}^{\alpha\beta} \overline{\nabla}_{\alpha} \overline{\nabla}_{\beta} \varphi + 
\frac{V_{,\,\varphi}}{F} = 
\frac{1}{F} \overline{G}^{\alpha\beta} \partial_{\alpha} F \partial_{\beta} \varphi - \frac{1}{16\pi} \biggl(\sigma_{\,, \varphi} Z_{\alpha\beta}\, Z^{\alpha\beta} + \overline{\sigma}_{\,, \varphi} Z_{\alpha\beta}\, \widetilde{\,Z\,}^{\alpha\beta}\biggr).
\label{AAA18}
\end{equation}
Note that while the covariant gradient of $\varphi$ coincides in both frames (i.e.  $\nabla_{\alpha} \varphi = \overline{\nabla}_{\alpha} \varphi$) we have instead 
$\overline{\nabla}_{\mu} \overline{\nabla}_{\nu} \varphi \neq \nabla_{\mu} \nabla_{\nu} \varphi$; furthermore 
in the $E$-frame $\overline{\nabla}_{\mu} \overline{\nabla}_{\nu} \varphi = (\partial_{\mu}\partial_{\nu} \varphi 
- \overline{\Gamma}_{\mu\nu}^{\,\,\,\,\alpha} \partial_{\alpha} \varphi)$ where now $\overline{\Gamma}_{\mu\nu}^{\,\,\,\,\alpha}$
has the standard Levi-Civita expression in terms of $\overline{G}_{\mu\nu}$. Finally, in the $E$-frame the 
evolution of the gauge fields reads:
\begin{equation}
\overline{\nabla}_{\alpha} \biggl( \sigma \, Z^{\alpha\beta} \biggr) + \overline{\nabla}_{\alpha} \biggl( \overline{\sigma} \, \widetilde{Z}^{\alpha\beta} \biggr) =0,
\label{AAA19}
\end{equation}
with $\overline{\nabla}_{\alpha} \widetilde{\,Z\,}^{\alpha\beta} =0$. By appropriately using Eqs. (\ref{AAA4a}) and (\ref{AAA10})--(\ref{AAA11}) 
it will always be possible to switch from one frame to the other.

\subsection{Consistency between the $P$- and the $E$-frame} 
Once a specific form of $f(\overline{R})$ is selected the two frames must be consistent not only in principle but also 
in practice. In what follows we shall focus on the 
 case where $f(\overline{R})$ contains a quadratic correction on top of the Einstein-Hilbert term:
\begin{equation}
f(\overline{R}) = \overline{R} - \overline{\alpha} \overline{R}^2, \qquad F(\overline{R}) = 1 - 2 \overline{\alpha} \, \overline{R},
\label{AAA20}
\end{equation}
with $\overline{\alpha}  = \alpha/\overline{M}^2$. In what follows on top of the Planck mass $\overline{M}_{P}$ there will be two further scales 
$\overline{M}\leq \overline{M}_{P}$ (associated with $\overline{\alpha}$) and $M\ll \overline{M} \leq \overline{M}_{P}$ (controlling the transition between the inflationary and the quintessential evolutions).
If Eq. (\ref{AAA2}) is contracted with the $P$-metric $G^{\mu\nu}$ we obtain: 
\begin{equation} 
\overline{R}(G) = - \ell_{P}^2 T^{(P)}, \qquad \Rightarrow \qquad F = 1 + 2 \overline{\alpha} \ell_{P}^2 \biggl[ 4 V - G^{\alpha\beta} \partial_{\alpha} \varphi\partial_{\beta} \varphi\biggr].
\label{AAA21}
\end{equation}
In the $E$-frame Eq. (\ref{AAA21}) translates into an relation determining 
$F$ in terms of the inflaton potential and of its kinetic term, i.e. 
\begin{equation}
F = 1 + 2 \overline{\alpha} \ell_{P}^2 \biggl[ 4 V - F \,\overline{G}^{\alpha\beta} \partial_{\alpha} \varphi\partial_{\beta} \varphi\biggr]
\qquad \Rightarrow\qquad F = \frac{1 + 2 Q}{F_{0}},
\label{AAA22}
\end{equation}
where the two auxiliary variables $Q$ and $F_{0}$ are defined as:
\begin{equation}
Q= \overline{\alpha} \, \ell_{P}^2 \, \overline{G}^{\alpha\beta} \partial_{\alpha} \varphi\partial_{\beta}\varphi, \varphi\qquad F_{0} = 1 + 8 \, \overline{\alpha}\, \ell_{P}^2 V.
\label{AAA23}
\end{equation}
We shall often need to evaluate $Q$ and $F_{0}$ on the 
background; to avoid potential ambiguities we now stress that 
$\overline{Q}$ and $\overline{F}_{0}$ denote the background values 
associated with the quantities appearing in Eq. (\ref{AAA23}). In section \ref{sec3}
we shall specifically consider the 
case where the metric is homogeneous in both frames, i.e. $\overline{G}_{\mu\nu} = a_{E}^2 \eta_{\mu\nu}$ and $ G_{\mu\nu} = a_{P}^2 \eta_{\mu\nu}$. As a consequence of the tansformation between the $E$-frame and the $P$-frame (see Eq. (\ref{AAA4a})),  all the homogeneous 
quantities will change accordingly; so for instance we will have that: 
\begin{equation}
a^2_{P} \to a^2_{E} = \biggl(\frac{1 + 2 \overline{Q}}{\overline{F}_{0}} \biggr)a^2_{P}, \qquad {\mathcal H}_{P} \to {\mathcal H}_{E} = {\mathcal H}_{P} + \overline{q}^{\prime},
\label{AAA23a}
\end{equation}
where $\overline{q}= (1/2) \ln{\overline{F}}$ indicates the background value of the quantity introduced in Eq. (\ref{AAA16}). We also note that in Eq. (\ref{AAA23a}), as usual,
${\mathcal H}_{E}= a_{E}^{\prime}/a_{E}$ and ${\mathcal H}_{P}= a_{P}^{\prime}/a_{P}$.
To avoid the proliferation of the indices we shall denote the scale factors in the two frames as $a$ and $b$, i.e. 
\begin{equation}
a\equiv a_{E}, \qquad b \equiv a_{P}, \qquad {\mathcal H}_{E} \equiv {\mathcal H}.
\label{AAA23b}
\end{equation}
Let us finally remind that the conformal time coordinate is the same in the two frames whereas the cosmic times will be different and, in particular, we will have that $d t_{E} = d\tau/a(\tau)$ and  $d t_{P} = d\tau/b(\tau)$. As soon as
$\overline{F} \to 1$  the cosmic times defined in the two frames coincide.

\subsection{Rescaled potential and rescaled action}
In the standard metric approach the higher derivatives introduce supplementary propagating degrees of freedom \cite{Htwo8} unless the 
nonlinear corrections either coincide with the Euler-Gauss-Bonnet 
term or they are proportional to it through an inflaton-dependent coupling. In the Palatini approach the presence of higher derivatives may instead change the relation between the matter action and the curvature scale. In this respect we consider Eq. (\ref{AAA17})  and note that, in the case of Eq. (\ref{AAA20}), the kinetic term of the scalar field and the corresponding potential can be rescaled through $F_{0}$ according to the following transformation (see e.g. \cite{PAL1,PAL2,PAL3,PAL4}):
\begin{equation}
\partial_{\mu} \varphi \to \partial_{\mu} \overline{\varphi} = \frac{\partial_{\mu} \varphi}{\sqrt{F_{0}}}, \qquad V \to W= \frac{V}{F_{0}}.
\label{AAA24}
\end{equation}
From Eq. (\ref{AAA22}) we can write $F = (1 + 2 Q)/F_{0}$ and if this expression is inserted into Eq. (\ref{AAA17}),  the rescalings of Eq. (\ref{AAA24}) lead to the following form of the governing equation:
\begin{eqnarray}
\overline{{\mathcal G}}_{\mu\nu} &=& \ell_{P}^2 \biggl\{ ( 1 + 2 Q ) \partial_{\mu} \overline{\varphi}\, \partial_{\nu} \overline{\varphi}  
+ \overline{G}_{\mu\nu} \biggl[ W - \frac{(\partial \overline{\varphi})^2}{2} ( 1 + Q)\biggr] \biggr\},
\nonumber\\
&+& \frac{\ell_{P}^2}{2 \, g^2} \biggl( - Z_{\mu\alpha} \, Z_{\nu}^{\,\,\,\alpha} + \frac{\overline{G}_{\mu\nu}}{4} Z_{\alpha\beta} \, Z^{\alpha\beta} \biggr).
\label{AAA25}
\end{eqnarray}
With the same strategy, starting directly from Eq. (\ref{AAA18}) we have:
\begin{eqnarray}
&& \frac{1}{\sqrt{-\overline{G}}} \partial_{\alpha} \biggl[ \sqrt{-\overline{G}}\,\, \overline{G}^{\alpha\beta}\, \, \partial_{\beta} \overline{\varphi} \,\,( 1 + 2 Q) \biggr] + (1 - 4 Q^2 ) W_{\,, \overline{\varphi} } 
\nonumber\\
&&+  \frac{1}{16\pi \, F_{0}} \biggl(\sigma_{\,, \overline{\varphi}} Z_{\alpha\beta}\, Z^{\alpha\beta} + \overline{\sigma}_{\,, \overline{\varphi}} Z_{\alpha\beta}\, \widetilde{\,Z\,}^{\alpha\beta}\biggr) = 0.
\label{AAA26}
\end{eqnarray}
It is a bit lengthy but not difficult to verify, as expected, that Eq. (\ref{AAA26}) follows from the Bianchi identity 
$\overline{\nabla}_{\mu} \overline{{\mathcal G}}^{\mu\nu} =0$. The equations 
of the gauge fields derived in Eq. (\ref{AAA19}) are not affected by the rescaling (\ref{AAA24}) 
which instead modifies the gauge energy-momentum tensor of Eq. (\ref{AAA25}).
The system written in the form given by Eqs. (\ref{AAA25})--(\ref{AAA26}) follows from the variation of the total action 
\begin{eqnarray}
S &=& - \frac{1}{2 \ell_{P}^2} \int d^{4} x \,\,\sqrt{- \overline{G}} \,\,\overline{R}(\overline{G}) + 
 \int d^{4} x \sqrt{- \overline{G}} \biggl[\frac{1}{2} \overline{G}^{\alpha\beta} \partial_{\alpha} \overline{\varphi} \partial_{\beta} \overline{\varphi} \biggl(1 + Q \biggr)- W]
 \nonumber\\
 &-&\int d^{4} x\,\, \frac{\sqrt{- \overline{G}}}{4 g^2} \biggl[  Z_{\alpha\beta} \, Z^{\alpha\beta} + \biggl(\frac{\overline{\sigma}}{\sigma} \biggr)Z_{\alpha\beta} \, \widetilde{\,Z\,}^{\alpha\beta} \biggr].
\label{AAA27}
\end{eqnarray}
The results of Eqs. (\ref{AAA25})--(\ref{AAA26}) and (\ref{AAA27}) 
also demonstrate that the addition of a quadratic term in the Palatini approach does not introduce supplementary degrees of freedom in the $E$-frame. The same conclusion is reached by working in the $P$-frame. 

\renewcommand{\theequation}{3.\arabic{equation}}
\setcounter{equation}{0}
\section{Inflationary and quintessential evolutions}
\label{sec3}
The background evolution can be studied either in the $E$- or in the $P$-frame but during inflation
the former is more convenient than the latter. We shall bound our attention to the case 
where $V(\varphi)$ dominates at early times and leads to stage of accelerated expansion while it 
becomes quickly subleading after the end of inflation. This is what happens in quintessential inflationary scenarios
where the inflaton and the quintessence field coincide. The possible suppression of the late-time potential has been implicitly suggested in Refs. \cite{QUINT10,QUINT11}.
When the inflaton and the quintessence field are identified
this class of (dual) potentials arise quite naturally since they imply and a stage of accelerated expansion at early and at 
recent times \cite{QUINT1,QUINT2,QUINT3,QUINT10,QUINT11}.

\subsection{Evolution equations of the system} 
We consider now the general system of equations in Friedmann-Robertson-Walker metric 
written in the conformal time coordinate $\tau$. In particular we follow the conventions 
established in Eqs. (\ref{AAA23a})--(\ref{AAA23b}) for the homogeneous backgrounds: 
$a$ denotes the scale factor in the $E$-frame and ${\mathcal H} = a^{\prime}/a$.
From the $(00)$ component of Eq. (\ref{AAA25}) we have:
\begin{equation}
{\mathcal H}^2 + \kappa = \frac{\ell_{P}^2}{3} \biggl[ \frac{\overline{\varphi}^{\,\prime\,2}}{2} \biggl( 1 + 3 \overline{Q}\biggr) + W\, a^2\biggr]
+ \frac{\ell_{P}^2\, a^2}{3} (\rho_{B} + \rho_{E}),
\label{BBB1}
\end{equation}
where $\kappa$ denotes the spatial curvature and, by definition, $\overline{Q} = \overline{\alpha} \, \ell_{P}^2 \, (\overline{\varphi}^{\prime\,\,2}/a^2) \, \overline{F}_{0}$ is the value of $Q$ evaluated on the homogeneous background (see Eq. (\ref{AAA23}) and discussion thereafter). In Eq. (\ref{BBB1})  
 $\rho_{B}$ and $\rho_{E}$ are the hypermagnetic and hyperelectric energy densities, respectively.
Even if $\rho_{E}$ and $\rho_{B}$ will turn out to be 
fully inhomogeneous it is useful to keep track of all these terms 
to evaluate explicitly the backreaction of the produced gauge fields on the homogeneous
background. As we shall discuss in sections \ref{sec4} and \ref{sec5}, in 
average $\rho_{B}$ and $\rho_{E}$ must not exceed the homogeneous background of Eq. (\ref{BBB1}).

The explicit form of $\rho_{E}$ and $\rho_{B}$ follows from the expressions 
of the gauge field strengths in the $E$-frame where 
the components of $Z^{\alpha\beta}$ read $Z^{i\,0} = e^{i}/a^2$ and $Z^{i \,j} = - \epsilon^{i\, j\, k} b_{k}/a^2$ so that and corresponding energy densities are expressed as:
\begin{equation}
\rho_{B} = \frac{B^2}{8 \pi a^4}, \qquad \rho_{E} = \frac{E^2}{8 \pi a^4}, \qquad \vec{B} = \sqrt{\sigma} \, a^2\vec{b}, \qquad 
\vec{E} = \sqrt{\sigma} a^2 \vec{e},
\label{BBB2}
\end{equation}
where $\vec{B}$ and $\vec{E}$ are the canonical fields whose evolution is derived 
from Eq. (\ref{AAA19}) and it is given by:
\begin{eqnarray}
&& \vec{\,\nabla\,} \times \vec{\,B\,} = \frac{1}{\sqrt{\sigma}} \partial_{\tau} \bigl(\sqrt{\sigma} \, \vec{\,E\,} \bigr) + \frac{\overline{\sigma}^{\,\prime}}{\sigma} \vec{\,B\,},
\label{BBB3}\\
&& \vec{\,\nabla\,} \times \vec{\,E\,} + \sqrt{\sigma}  \partial_{\tau} \biggl(\frac{\vec{\,B\,}}{\sqrt{\sigma}}\biggr) =0,
\label{BBB4}\\
&& \vec{\,\nabla\,} \cdot\vec{\,E\,} =0, \qquad\qquad \vec{\,\nabla\,} \cdot\vec{\,B\,} =0,
\label{BBB5}
\end{eqnarray}
where $\vec{\,\nabla\,}$ now denotes the ordinary three-dimensional gradient. 
Equations (\ref{BBB3})--(\ref{BBB4}) have been written under the assumption that $\sigma$ and $\overline{\sigma}$ are homogeneous and they describe the amplification of the 
quantum fluctuations of the gauge fields during the inflationary stage.  The backreaction terms 
associated with $\rho_{E}$, $\rho_{B}$ and $\vec{E}\cdot\vec{B}$ also 
appear in Eq. (\ref{AAA26}) whose explicit expression becomes: 
\begin{eqnarray}
\overline{\varphi}^{\,\prime\prime} + 2 {\mathcal H} \overline{\varphi}^{\prime} + \frac{ 2 \overline{Q}^{\prime} \overline{\varphi}^{\prime}}{2 \overline{Q} +1}  + ( 1 - 2 \overline{Q})  W_{\,,\overline{\varphi}}  
+  \frac{1}{ 1 + 2 \overline{Q}} \biggl[ \frac{a^2 \sigma_{,\,\varphi}}{\sigma} \biggl(\rho_{B} - \rho_{E}\biggr) + \frac{\overline{\sigma}_{\,,\varphi}}{ 4 \pi a^2 \sigma} \vec{E}\cdot\vec{B} \biggr] =0.
\label{BBB6}
\end{eqnarray}
In Eq. (\ref{BBB6}) $\rho_{B}$ and $\rho_{E}$ scale as in Eq. (\ref{BBB2}) but both terms are multiplied 
by $a^2$; this is only a consequence of the conformal time parametrization. 

The evolution of the background geometry can also be phrased in the cosmic time parametrization. In this respect the only caveat is that, while the conformal time coordinate is the same in both frames, the cosmic time does change between the $E$-frame and the $P$-frame (see also Eq. (\ref{AAA23a}) and discussion thereafter). 
In the cosmic time coordinate (related to $\tau$ as $a(\tau) d\tau= d t$) Eq. (\ref{BBB1}) reads:
\begin{equation}
H^2 + \frac{\kappa}{a^2} = \frac{\ell_{P}^2}{3} \biggl[ \frac{\dot{\overline{\varphi}}^{2}}{2} \biggl( 1 + 3 \overline{Q}\biggr) + W\biggr]
+ \frac{\ell_{P}^2}{3} (\rho_{B} + \rho_{E}),
\label{BBB7}
\end{equation}
where now $\overline{Q} =  \overline{\alpha} \, \ell_{P}^2 \, \dot{\overline{\varphi}}^{\,2} \overline{F}_{0}$  and the overdot 
denotes a derivation with respect to $t$. In the cosmic time coordinate Eq. (\ref{BBB6}) becomes: 
\begin{eqnarray}
 \ddot{\overline{\varphi}} + 3 H \dot{\overline{\varphi}} + \frac{2 \dot{\overline{\varphi}} \dot{\overline{Q}}}{ 1 + 2 \overline{Q}}
+ ( 1 - 2 \overline{Q})  W_{\,,\overline{\varphi}}  
+  \frac{1}{ 1 + 2 \overline{Q}} \biggl[ \frac{ \sigma_{,\,\varphi}}{\sigma} \biggl(\rho_{B} - \rho_{E}\biggr) + \frac{\overline{\sigma}_{\,,\varphi}}{ 4 \pi a^4 \sigma} \vec{E}\cdot\vec{B} \biggr] =0.
\label{BBB8}
\end{eqnarray}
Finally the $(ij)$ component of Eq. (\ref{AAA25}) in the cosmic time parametrization reads 
\begin{equation}
2 \dot{H} + 3 H^2 = \frac{\kappa}{a^2} + \ell_{P}^2 \biggl[ W - \frac{(1 + \overline{Q})}{2} \dot{\overline{\varphi}}^2\biggr]
+ \ell_{P}^2 (p_{B} + p_{E}),
\label{BBB9}
\end{equation}
where $p_{E} =\rho_{E}/3$ and $p_{B}= \rho_{B}/3$. For the sake of conciseness we did not report the analog 
of Eq. (\ref{BBB9}) in the conformal time but it is immediate to obtain by transforming back from the 
cosmic time to the $\tau$ parametrization.

\subsection{The early inflationary stage}
The form of Eq. (\ref{AAA24}) suggests that in the limit $\alpha \gg 1$ the rescaled potential is suppressed in spite of the shape of $V(\varphi)$; more precisely we can write:
\begin{equation}
\lim_{\alpha \gg 1} W = \frac{V}{1 + 8 \overline{\alpha} \,\ell_{P}^2 V} \to \frac{\overline{M}_{P}^2 \, \overline{M}^2}{8 \alpha},
\label{PPP1}
\end{equation}
where, as discussed above, $\overline{\alpha}= \alpha/\overline{M}^2$. 
According to Eq. (\ref{PPP1}), if $V$ inflates $W$ will describe an inflationary evolution with a suppressed slow-roll parameter and a lower tensor to scalar ratio. For the same reason we can 
write $\overline{Q}$ as: 
\begin{equation}
\overline{Q} =  \frac{ 2\, \epsilon_{0}\,\overline{\alpha} \, \ell_{P}^2\, V}{ 3(1 + 8 \overline{\alpha} \ell_{P}^2 V)}\ll 1,\qquad\qquad \epsilon_{0} = \frac{\overline{M}_{P}^2}{2} \biggl(\frac{V_{\,,\varphi\varphi}}{V}\biggr)^2,
\label{PPP2}
\end{equation}
where $\epsilon_{0}<1$ is the slow-roll parameter associated with the potential $V$; since we are 
here considering the inflationary branch of the potential, Eq. (\ref{PPP2}) also implies $\overline{\,Q\,} \ll 1$. It is relevant to stress that 
$\epsilon_{0}$ is now the slow-roll parameter in the limit $\alpha \to 0$, i.e. when the 
quadratic corrections disappear. When $\alpha \neq 0$ the slow-roll parameter 
will be simply denoted by $\epsilon$. The relation between $\epsilon$ and $\epsilon_{0}$ directly 
follows from the definition of $W$ in terms of $V$ given in Eq. (\ref{AAA24}) (see also Eq. (\ref{PPP1})); the result is:
\begin{equation}
\epsilon= \frac{\overline{M}_{P}^2}{2} \biggl(\frac{W_{,\,\,\overline{\varphi}}}{W}\biggr)^2 = 
\frac{\epsilon_{0}}{\overline{F}_{0}}.
\label{PPP3}
\end{equation}
It follows from Eqs. (\ref{PPP2})--(\ref{PPP3}) that for $\alpha \gg 1$ we have that $\epsilon \ll \epsilon_{0} < 1$;
this means that  Eqs. (\ref{BBB7})--(\ref{BBB8}) and (\ref{BBB9}) take the form:
\begin{equation}
3 \,H^2 \, \overline{M}_{P}^2 = W, \qquad \qquad 3\, H\, \dot{\overline{\varphi}} + W_{,\,\overline{\varphi}}=0, \qquad \qquad 2\,\dot{H} \overline{M}_{P}^2 = - \dot{\overline{\varphi}}^2,
\label{PPP4}
\end{equation}
during the inflationary stage of expansion. 
Equations (\ref{PPP4}) are nothing but the standard slow-roll equations written in the case of a rescaled potential $W(\overline{\varphi})$. 
According to Eq. (\ref{PPP4}), when $\alpha\neq 0$  we can also introduce the second slow-roll
parameter conventionally denoted by $\eta$:
\begin{equation}
\eta = \overline{M}_{P}^2 \biggl(\frac{W_{\,,\overline{\varphi}\,\overline{\varphi}}}{W}\biggr) = \eta_{0} - \frac{24 \epsilon_{0} \overline{\alpha} \, \ell_{P}^2\, V}{1 + 8\, \overline{\alpha} \, \ell_{P}^2 \, V} \to \eta_{0} - 3 \epsilon_{0},
\label{PPP5}
\end{equation}
where the last (physical) limit in Eq. (\ref{PPP5}) refers to  $\alpha \gg 1$. From the analysis of scalar modes
during the inflationary stage \cite{PAL4} it follows that
the scalar spectral index and the corresponding amplitude of the scalar power spectrum will be given by:
\begin{equation}
n -1 \simeq - 6 \epsilon + 2 \eta + {\mathcal O}(\epsilon^2), \qquad \qquad 
{\mathcal A}^{(s)} = \frac{ W}{ 24 \pi^2 \epsilon \overline{M}_{P}^4},
\label{PPP6}
\end{equation}
where $n$ denotes the scalar spectral index in the case $\alpha\neq 0$.
In the limit $\alpha \to 0$ the analog of Eq. (\ref{PPP6}) will be instead given by:
\begin{equation}
n_{0} -1 \simeq - 6 \epsilon_{0} + 2 \eta_{0} + {\mathcal O}(\epsilon_{0}^2), \qquad \qquad 
{\mathcal A}^{(s)}_{0} = \frac{ V}{ 24 \pi^2 \epsilon_{0} \overline{M}_{P}^4},
\label{PPP7}
\end{equation}
where $n_{0}$ denotes the scalar spectral index for $\alpha \to 0$.
Thanks to Eq. (\ref{PPP5}), Eqs. (\ref{PPP6})--(\ref{PPP7}) imply that  the scalar spectral indices coincide $n = n_{0}$; furthermore
since $\epsilon = \epsilon_{0}/\overline{F}_{0}$ and $W = V/\overline{F}_{0}$, Eqs. (\ref{PPP6})--(\ref{PPP7}) 
demand that ${\mathcal A}^{(s)}_{0}= {\mathcal A}^{(s)}$. The same analysis 
can be repeated for the tensor modes \cite{PAL4}. In particular, as expected, when $\alpha \neq 0$ 
\begin{equation}
n^{(t)}  \simeq - 2 \epsilon  + {\mathcal O}(\epsilon^2)
\, \qquad {\mathcal A}^{(t)} =\frac{2 \, W}{3 \,\pi^2\,\overline{M}_{P}^4},
\label{PPP8}
\end{equation}
where the exact definition of the tensor spectral index has been included 
together with its slow-roll limit. If  $\alpha \to 0$ the tensor spectral index and the corresponding
 amplitude are instead:
\begin{equation}
n_{0}^{(t)} = - 2 \epsilon_{0}, \qquad {\mathcal A}_{0}^{(t)} = \frac{2\, V}{3\,\pi^2 \, \overline{M}_{P}^4}.
\label{PPP9}
\end{equation}
The mutual relation of Eq. (\ref{PPP8}) and Eq. (\ref{PPP9}), to lowest order in the slow-roll approximation reads
\begin{equation}
n^{(t)} = n_{0}^{(t)}/\overline{F}_{0}, \qquad  {\mathcal A}^{(t)}= {\mathcal A}_{0}^{(t)}/\overline{F}_{0}. 
\label{PPP10}
\end{equation}
If we now divide the tensor amplitude ${\mathcal A}^{(t)}$ given in Eq. (\ref{PPP8}) by the scalar amplitude 
of Eq. (\ref{PPP6})  the explicit expression of the tensor to scalar ratio $r_{T}$ in the case $\alpha\neq 0$ becomes:
\begin{equation} 
r_{T} = \frac{{\mathcal A}^{(t)}}{{\mathcal A}^{(s)}} = 16 \, \epsilon= \frac{16 \epsilon_{0}}{\overline{F}_{0}},
\label{PPP11}
\end{equation}
showing the tensor-to-scalar ratio in the case 
$\alpha \neq 0$ get reduced in comparison with  $\alpha \to 0$. 

\subsection{Dual potentials and quintessential inflation}
The $\alpha$-suppression of Eq. (\ref{PPP1}) is particularly relevant for those potentials that are at variance 
with the actual determinations of the tensor-to-scalar ratio \cite{RT1,RT2,RT2a} like the monomials(i.e. $V \propto \varphi^{n}$) whose associated 
tensor-to-scalar ratio is too large to be compatible with the values inferred from 
the temperature and polarization anisotropies of the microwave background. The monomials are 
also interesting since they can be used to build minimal models of quintessential inflation where 
the inflaton and the quintessence field are unified by reducing, in this way, the number 
of parameters. For these two reasons we shall focus, for the sake of concreteness,
on the situation where $V(\varphi)$ has a dual form: a power-law shape for $\varphi < 0$ 
while it is suppressed as an inverse power of $\varphi$ for $\varphi \geq 0$. A particular example 
along this direction is \cite{QUINT1}:
\begin{eqnarray}
V(\varphi) &=&  \lambda (\varphi^4 + M^4), \qquad \qquad \varphi \leq 0,
\nonumber\\
V(\varphi) &=& \lambda M^8/(\varphi^4 + M^4), \qquad\qquad \varphi >0,
\label{QQQ1}
\end{eqnarray}
where $M$ is a further scale related with the dominance of the quintessence field and already mentioned after Eq. (\ref{AAA20}). The initial data for the inflationary evolution are assigned in the regime $\varphi \ll - \overline{M}_{P}$ with 
$\dot{\varphi}^2 \ll V$. In the case $\alpha\to 0$ we have that the total number of $e$-folds and 
the initial value of the field are determined, as usual, by: 
\begin{equation}
N = \frac{\varphi_{i}^2 - \varphi_{f}^2}{8 \, \overline{M}_{P}^2}, \qquad \qquad \biggl(\frac{\varphi_{i}}{\overline{M}_{P}}\biggr)^2 \simeq 8 (N +1),
\label{QQQ2}
\end{equation}
where $\varphi_{f}$ follows  from the condition $\epsilon_{0}(\varphi_{f})  \simeq 1$. If the total number of $e$-folds 
is between $70$ and $100$ the initial value of $\varphi$ in units of $\overline{M}_{P}$ will be between $-20$ and $-30$ ad this is why we assumed $\varphi\ll - \overline{M}_{P}$. The specific value of $\lambda$ is fixed by considering 
the scalar modes that exited the Hubble radius about $60$ $e$-folds before the end of inflation implying that 
$\lambda \simeq 3.9 \times 10^{-14}$ where the amplitude of the scalar mode has been estimated as\footnote{For the actual numerical value of the scalar and tensor amplitudes 
we shall always refer to the pivot scale $k_{p} = 0.002 \mathrm{Mpc}^{-1}$.} ${\mathcal A}^{(s)} = 2.41\times 10^{-9}$. 
The unsuppressed value of the slow-roll parameter  is therefore given by 
$\overline{\epsilon}_{0}\simeq (N+1)^{-1}$ where the field has been evaluated 
 approximately $N$-$e$folds before the end of inflation; the corresponding value for the tensor to scalar ratio is $16/(N+1)$. In the case 
$N \sim {\mathcal O}(60)$ we would have that $\overline{\epsilon}_{0} \simeq 0.016$ and a tensor to scalar ratio $ {\mathcal O}(0.26)$ 
which is not compatible with the current observational determinations \cite{RT1,RT2,RT2a}. 
If $\alpha\neq 0$ 
all the properties of the inflationary branch of the potential (\ref{QQQ1}) are preserved 
with the difference that $\epsilon$ and $r_{T}$ are now suppressed. For typical values 
of $\alpha = {\mathcal O}(100)$ and $\overline{M} = {\mathcal O}(10^{-3}) \overline{M}_{P}$ 
the value of $\epsilon$ diminishes from $0.016$ to $\epsilon \simeq 0.001$;
similarly the tensor-to-scalar ratio passes from $0.26$ to $0.03$. Other specific cases of the parameters 
could be considered; however the general idea illustrated here
is largely independent on the specific parameters and, to some extent, even on the shape of the inflationary potential.  In this sense Eq. (\ref{QQQ1}) might have a different power during inflation or even a different shape. We finally note that the rescaled potential of Eq. (\ref{PPP1}) could be also 
expressed as:
\begin{equation}
(W/V) = \frac{1}{1 + V/W_{1}}, \qquad \qquad W_{1} = \frac{\overline{M}^2 \, \overline{M}_{P}^2}{8 \alpha},
\label{QQQ3}
\end{equation}
so that a suppression of the tensor-to-scalar ratio may occur provided $V(\varphi_{i})/W_{1} \gg 1$.
This condition becomes more explicit by using Eq. (\ref{QQQ1}) for $\varphi<0$: 
\begin{equation}
8 \alpha \lambda \biggl( \frac{\varphi_{i}}{\overline{M}_{P}^4}\biggr)^4 \, \biggl(\frac{\overline{M}_{P}}{\overline{M}}\biggr)^2 \gg 1.
\label{QQQ4}
\end{equation}
But if we assume a quartic form of the potential $(\varphi_{i}/\overline{M}_{P})^2 = 8 (N+1)$ and, 
with this specification the condition (\ref{QQQ4}) is equivalent to:
\begin{equation}
\biggl(\frac{\overline{M}}{\overline{M}_{P}}\biggr) \ll \sqrt{ 512 \, \alpha \, \lambda} \, (N+1).
\label{QQQ5}
\end{equation}
Equation (\ref{QQQ5}) can be made even more explicit by estimating the value of $\lambda$ from the 
observed value of the scalar power spectrum. In other words, to first-order in the slow-roll approximation 
we will have that $\lambda= 3\pi^2 {\mathcal A}^{(s)}/[8 (N+1)^3]$ where ${\mathcal A}^{(s)} = 2.41 \times 10^{-9}$;
for $N= 60$ the previous expression evaluates to $\lambda = 3.9 \times 10^{-14}$ and Eq. 
(\ref{QQQ5}) becomes:
\begin{equation}
\biggl(\frac{\overline{M}}{\overline{M}_{P}}\biggr) < \sqrt{\frac{192 \, \pi^2 \, {\mathcal A}^{(s)}}{ 8 (N+1)} } \, \sqrt{\alpha} = 2.1 \times 10^{-3} \,\,\biggl(\frac{{\mathcal A}^{(s)}}{2.41\times 10^{-9}}\biggr)^{1/2} \,\,\sqrt{\frac{\alpha}{N +1}}.
\label{QQQ6}
\end{equation}

\subsection{Post-inflationary evolution}
For the dual potentials illustrated in Eq. (\ref{QQQ1}) the post-inflationary evolution 
(i.e. $\varphi \geq 0$) will be initially dominated by the kinetic energy of the inflaton-quintessence field. From Eq. (\ref{QQQ3}) we see that, as soon as inflation ends, $V \ll W_{1}$; thus, after an irrelevant transition regime, the background will be dominated by the kinetic energy 
of the inflaton/quintessence field. As already mentioned above the typical scale $M= {\mathcal O}(10^{6})$ GeV is related to the dark energy dominace \cite{QUINT1} so that $\varphi_{f} \gg M$ and the ratio 
\begin{equation}
\frac{V(\varphi_{f})}{W_{1}} = \frac{8 \alpha \lambda\, \bigl(M/\overline{M}_{P}\bigr)^{2}\,\bigl(M/\overline{M}\bigr)^2}{[ (\varphi_{f}/M)^4 +1]} \ll 1,
\label{QQQ7}
\end{equation}
is negligible for all practical purposes. Indeed, for typical values of $\lambda$ and $\overline{M}$ [i.e. $\lambda= {\mathcal O}(10^{-14})$, $\overline{M} ={\mathcal O}(10^{-3}) \overline{M}_{P}$] we would have that while the denominator will bring a further suppression, the numerator at the right hand side of Eq. (\ref{QQQ7}) 
is already of the order of $\alpha 10^{-60}$; this means that, in spite of the values of $\alpha$ and $M$ 
$V\ll W_{1}$ will always be negligible at the end of inflation and right after it. Thus, after the end of inflation the potential can be safely neglected; furthermore if $V \ll W_{1}$ in Eq. (\ref{QQQ3}) we also 
have that: 
\begin{equation}
W(\overline{\varphi}) \simeq V(\varphi), \qquad\qquad \overline{F}_{0}\to 1, \qquad \qquad \overline{\varphi} \simeq \varphi.
\label{RRR1}
\end{equation}
Equation (\ref{RRR1})  means that we can directly estimate $\overline{F}$ in the $P$-frame from Eq. (\ref{AAA21}) where 
\begin{equation}
F = 1 + 2 \overline{\alpha} \ell_{P}^2 \biggl[ 4 V - G^{\alpha\beta} \partial_{\alpha} \varphi\partial_{\beta} \varphi\biggr] \to 1 - 2 \overline{\alpha} \frac{\varphi^{\prime\, \, 2}}{\overline{M}_{P}^2 \, b^2}.
\label{RRR1a}
\end{equation}
In the absence of potential and assuming that the amplified gauge fields are always small (as we shall see it is the case in the following section) Eq. (\ref{AAA6}) 
can be immediately solved
\begin{equation} 
\varphi^{\prime} = \varphi_{1}^{\prime} \biggl(\frac{b_{1}}{b}\biggr)^2 \qquad \Rightarrow \qquad 
\overline{F} = 1 - 12 \alpha \biggl(\frac{H_{1}}{\overline{M}}\biggr)^2 \biggl(\frac{b_{1}}{b}\biggr)^{6} \to 1.
\label{RRR1b}
\end{equation}
Since after inflation the two frames coincide Eqs. (\ref{RRR1a})--(\ref{RRR1b})
also imply 
\begin{equation}
a(\tau) = b(\tau) = \sqrt{\frac{\tau}{\tau_{1}}}, \qquad \varphi(\tau) = \sqrt{2} \, \overline{M}_{P} \, \ln{(a/a_{1})}.
\label{RRR1c}
\end{equation}
In the regime defined by Eqs. (\ref{RRR1b})--(\ref{RRR1c}) we have that
 $\overline{Q} = \alpha (H_{1}/M)^2 (a_{1}/a)^6 \to 0$ and this is consistent with Eq. (\ref{AAA21})  because $\overline{F}_{0} \to 1$ and $F\to 1$.
 From now on the main physical steps 
coincide with the ones of the standard quintessential evolution (see e.g. \cite{QUINT1,QUINT2,QUINT3,QUINT13}). 
In the cosmic time coordinate the evolution (\ref{RRR1c}) corresponds to $a(t) \simeq (t/t_{1})^{\beta}$ with $\beta =1/3$.
It will be practical, for phenomenological purposes (see next section), 
to parametrize the stiff evolution in terms of the following two parameters:
\begin{equation}
\beta = \frac{d \ln{a}}{d\ln{t}}, \qquad\qquad 1/3\leq\beta  < 1/2, \qquad \qquad \xi_{r} = \frac{H_{r}}{H_{1}}< 1.
\label{PARAM1}
\end{equation}
The parameter $\beta$ controls the expansion 
rate during the stiff phase and it will be taken to be $1/3 \leq 1/3 < 1/2$.
Of course Eqs. (\ref{RRR1b})--(\ref{RRR1c}) imply that $\beta = 1/3$; 
we shall however consider different values of $\beta$ just to account for the 
more general situation where, after inflation, a stiff phase is present. 
In Eq. (\ref{PARAM1}) $H_{1}$ and $H_{r}$ denote the expansion rates at the beginning and at the end of the stiff phase, respectively. 
While the upper bound on $\xi_{r}$ is obvious by definition (since $H_{r} < H_{1}$) 
there is also a lower bound that is fixed by the backreaction considerations as originally discussed by Ford \cite{QUINT10}. If we suppose that 
during the inflationary phase there are ${\mathcal N}$ non-conformally coupled species 
that are amplified with a quasi-flat spectrum their energy density will scale as ${\mathcal N}\, H_{1}^4 (a_{1}/a)^4$ and 
this value will dominate the background at a putative scale $\xi_{r}$ defined by:
\begin{equation}
\overline{\xi}_{r} = \biggl(\frac{\pi^2\, {\mathcal N}}{6} \, r_{T} \, {\mathcal A}^{(s)} \biggr)^{1/(2 - 4\beta)}.
\label{YYY6}
\end{equation}
Since ${\mathcal N}$ is  much larger than $1$ [e.g. ${\mathcal N} = {\mathcal O}(100)$] we have that $\overline{\xi}_{r}\leq \xi_{r} < 1$.  As the curvature scale decreases $\varphi$ will evolve in a background dominated by radiation. It can be shown in general terms that, in both cases, $\varphi$ will increase as $\varphi \propto (t/t_{*})^{1/3}$ 
where $t$ denotes the cosmic time coordinate \cite{QUINT1,QUINT3}; the precise rate 
of increase actually depends on the inverse power of the potential and the power $1/3$ 
refers to the case $V = \lambda M^8/(\varphi^4 + M^4)$ of Eq. (\ref{QQQ1}).
The quintessence background will therefore dominate, by definition, at a 
typical curvature scale comparable with the present value of the Hubble rate. In the 
quintessential stage of the model $\overline{F}_{0} = \overline{F}=  1 + 24 \alpha (H_{0}/\overline{M})^2$ where $ H_{0}/\overline{M} = {\mathcal O}(10^{-57})$. This means, incidentally, that in spite of the value of $\alpha$ the $P$-frame and the $E$-frame 
coincide.

\subsection{The evolution of the gauge coupling}

There have been various proposals for the evolution 
of the gauge coupling during a stage of inflationary expansion. The simplest possibility compatible 
with the present scenario
is that  $g = \sqrt{4\pi/\sigma}$ increases all along the inflationary stage and then freezes later on. In this case $\sigma(\tau)$ will decrease as a function of $\tau$ and the evolution the susceptibility can be assigned either as a function of the inflaton field (i.e. $\sigma = \sigma(\varphi)$) or in terms of the scale factor (e.g. $\sigma = \sigma(a)$). Taking into account Eqs. (\ref{PPP1})--(\ref{PPP4}) the evolution during inflation is simply described by: 
\begin{equation}
3 H^2 \overline{M}_{P}^2 = W \simeq \frac{\overline{M}^2 \overline{M}_{P}^2}{8 \alpha}, 
\qquad\dot{\overline{\varphi}} = - \sqrt{2\epsilon} \, H\, \overline{M}_{P} \simeq 
-\frac{\sqrt{\epsilon}}{ 2 \sqrt{3\alpha}}  \overline{M} \, \overline{M}_{P}.
\label{QQQ8}
\end{equation}
For $\alpha\gg1$ the expansion rate is $H\simeq \overline{M}/\sqrt{8 \alpha}$ and $a(\tau) = (- \tau H)^{-1}$. If $g(\tau)$ increases for $\tau \to - 1/H$
we could always phrase the evolution of the susceptibility in terms of the (expanding) scale factor, i.e. $\sigma_{inf} \propto a^{- c_{inf}}$ where $c_{inf}> 0$ is a parameter that depends on the explicit form the of the coupling. The objective here is to constrain the rate of 
evolution of $\sigma$ by imposing all the relevant limits together with the magnetogenesis 
requirements. In this approach the obtained gauge fields will be solely expressed in terms of the properties of the background supplemented 
by the rates of evolution of $\sigma$; according to this logic in the stiff epoch 
a similar expression will hold, e.g. $\sigma_{stiff} \propto a^{- c_{stiff}}$. Finally, during the radiation-dominated stage $\sigma_{rad} \propto a^{- c_{rad}}$.  
Since the evolution of the scale factor is continuous across the various 
stages of the model the three parameters $c_{inf}$, $c_{stiff}$ and $c_{rad}$ are always positive semidefinite and will be ultimately related to the evolution in the conformal time coordinate 
$\tau$. For instance if $\sigma_{inf}(a) \simeq a^{-c_{inf}}$ we will have that, in terms 
of the conformal time coordinate $\sigma_{inf} \propto (-\tau/\tau_{1})^{2 \gamma}$ 
with $\gamma = c_{inf}/[ 2 (1 - \epsilon)]$. We are therefore led to consider the following evolution for the gauge susceptibilities:
\begin{eqnarray}
\sigma_{in}(\tau) &=& \sigma_{1} \biggl(- \frac{\tau}{\tau_{1}}\biggr)^{2 \gamma}, \qquad \qquad \tau\leq - \tau_{1},
\label{TTT1}\\
\sigma_{st}(\tau) &=& \sigma_{1} \biggl[1+ \frac{\gamma}{\delta} \biggl(\frac{\tau}{\tau_{1}} +1\biggr)\biggr]^{-2\delta} , \qquad \qquad -\tau_{1} < \tau\leq \tau_{r},
\label{TTT2}\\
\sigma_{rad}(\tau) &=& \sigma_{st}(\tau_{r}) \biggl[ 1 + \frac{(\gamma/\zeta)\, (\tau_{r}/\tau_{1})\, (\tau/\tau_{r} -1)}{(\gamma/\delta)(\tau_{r}/\tau_{1}+1) 
+1}\biggr]^{- 2\zeta},\qquad\qquad \tau \geq \tau_{r},
\label{TTT3}
\end{eqnarray}
where $\gamma>0$ while $\delta \geq 0$ and $\zeta \geq 0$; in Eqs. (\ref{TTT1})--(\ref{TTT2}) and (\ref{TTT3}) the subscripts parametrize the evolution during the inflationary stage and during the stiff phase. Equation (\ref{TTT3}) corresponds to the regime where the gauge coupling flattens during the radiation epoch and $\zeta \ll 1$.  During inflation (i.e. for $\tau\leq - \tau_{1}$) the rate of variation 
${\mathcal F}_{inf} = \sqrt{\sigma}^{\,\prime}/\sqrt{\sigma}$ is simply given by $\gamma/\tau$. In the subsequent 
stiff and radiation epochs ${\mathcal F}$ can be instead expressed as:
\begin{eqnarray}
{\mathcal F}_{st} &=& - \frac{\delta}{y(\tau)}, \qquad \tau_{r} < \tau - \tau_{1},\qquad \mathrm{where} \qquad y(\tau) = \tau + (q_{1} + 1) \tau_{1},
\nonumber\\
{\mathcal F}_{rad} &=& - \frac{\delta}{z(\tau)}, \qquad \tau \geq \tau_{r},\qquad \mathrm{where} \qquad z(\tau) = \tau + (q_{2} - 1) \tau_{r},
\label{TTT4}
\end{eqnarray}
where $q_{1}$ and $q_{2}$ are defined in terms of $\gamma$, $\delta$ and $\zeta$:
\begin{equation}
q_{1} = \delta/\gamma, \qquad \qquad q_{2} = \frac{(\gamma/\delta)(\tau_{r}/\tau_{1} +1) +1}{(\gamma/\zeta)(\tau_{r}/\tau_{1})}.
\label{TTT6}
\end{equation}
It is relevant to mention that since $\sigma$ is continuous and differentiable also ${\mathcal F}$ will be 
continuous. Note that, at the and of inflation, $y(-\tau_{1}) = q_{1} \tau_{1}$ 
while at the onset of the radiation stage $z(\tau_{r})= q_{2} \, \tau_{r}$. 

The evolution of the gauge coupling is indirectly constrained by the thermalization process. 
The produced particles during the stiff phase interact via the exchange of gauge bosons 
and their concentration will be roughly given by $n = {\mathcal N} T^3$ where $T \simeq H_{1}(a_{1}/a)$ 
is the kinetic temperature which eventually coincides with the thermodynamic temperature after thermalization \cite{QUINT1,QUINT10,QUINT11}. If the interactions take place via the exchange of gauge bosons the 
cross section will be proportional to $\alpha_{g}/T^2$ where $\alpha_{g} = g^2/(4\pi) = 1/\sigma$. When the cross section multiplied by the concentration 
becomes comparable with the expansion rate PAPERrmalization takes place and this occurs, approximately, when $\alpha_{g} \,{\mathcal N} T \simeq H$.
In the case of a generic stiff phase we have that $H \simeq H_{1} (a_{1}/a)^{1/\beta}$ so that we have that the thermalization 
will take place at a typical curvature scale $H_{th}$:
\begin{equation}
\xi_{th} = \biggl(\frac{H_{th}}{H_{1}}\biggr) = \biggl({\mathcal N} \alpha_{g,\,th}\biggr)^{1/(1 -\beta)}.
\label{TTT7}
\end{equation}
where, by definition, $\alpha_{g,\,th}= \alpha_{g}(\tau_{th})$ is the value of the gauge coupling at thermalization. The estimate (\ref{TTT7}) together with the timeline 
of the scenario suggest that either the gauge coupling freezes during the 
stiff phase or it will get to a constant value in the radiation epoch. In both cases we will have to 
demand that thermalization takes place before radiation dominates. This condition will then implies that 
$\xi_{th}  > \xi_{r} \geq \overline{\xi}_{r}$. If we now recall the explicit expression of (\ref{YYY6}) we have that 
\begin{equation}
{\mathcal N}^{(1- 3\beta)/[(1 -\beta) (1- 2 \beta)]}\, \alpha_{g, \, th}^{2/(1-\beta)}> \biggl(\frac{\pi \, r_{T} \, {\mathcal A}^{(s)}}{6}\biggr)^{1/(1 - 2\beta)}.
\label{TTT8}
\end{equation}
Equation (\ref{TTT8}) is always verified in the case the evolution of the 
gauge coupling is given by Eqs. (\ref{TTT1})--(\ref{TTT2}) and (\ref{TTT3}).
For instance in the case $\beta =1/3$ (which is probably the most realistic 
situation) we have that  $\alpha_{g, \, th} > ( \pi \, r_{T}\, {\mathcal A}^{(s)}/6)$.
If the gauge coupling freezes when $\alpha_{g} = {\mathcal O}(0.01)$, Eq. (\ref{TTT8})
will be safely satisfied. 

According to Eq. (\ref{TTT8}) the gauge coupling should not be too small around 
thermalisation and this requirement excludes, in practice, 
the possibility that $g(\tau)$ decreases by always remaining perturbative throughout all the stages of the evolution.  Let us consider, in this respect, the dual evolution of Eqs. 
(\ref{TTT1})--(\ref{TTT2}) and (\ref{TTT3}); the dual case 
can be simply obtained by transforming 
\begin{equation}
\gamma \to - \widetilde{\gamma}, \qquad \delta \to - \widetilde{\delta}, \qquad 
\zeta\to -\widetilde{\zeta},
\label{TTT9}
\end{equation}
with $ \widetilde{\gamma}>0$, $\widetilde{\delta} \geq 0$ and $\widetilde{\zeta} \geq 0$.
If the transformation (\ref{TTT9}) is applied to Eqs. 
(\ref{TTT1})--(\ref{TTT2}) and (\ref{TTT3}) we have that $\sigma \to 1/\sigma$ 
(and $g \to 1/g$).  Incidentally the transformation (\ref{TTT9}) exchanges also 
the magnetic and electric power spectra as we shall briefly see later on \cite{NINEaa,TENaa}.
From Eq. (\ref{TTT9}) the gauge coupling always 
decreases and we therefore have two possibilities: either $g_{i} = {\mathcal O}(1)$ 
at the onset of inflation or $g_{i} \gg 1 $. If $g_{i} ={\mathcal O}(1)$ 
the constraint (\ref{TTT8}) cannot be satisfied; conversely if  $g_{i} \gg 1$ will 
be strongly coupled at the beginning. Recalling that $g_{i}= \sqrt{4\pi/\sigma_{i}}$, 
from the dual version of Eq. (\ref{TTT1}) we will have 
\begin{equation}
\sqrt{\sigma_{i}} = \sqrt{\sigma_{1}} \biggl(\frac{a_{i}}{a_{f}}\biggr)^{\widetilde{\,\gamma\,}} \ll 1 \Rightarrow
g_{i} = \frac{\sqrt{4\pi}}{\sqrt{\lambda_{i}}} \gg 1.
\label{TTT10}
\end{equation}
Since  $N= {\mathcal O}(60)$ is the total number of inflationary $e$-folds,
$(a_{i}/a_{f})^{\widetilde{\,\gamma\,}} = e^{ - N \,\widetilde{\,\gamma\,}}\ll 1$. 
Equation (\ref{TTT10}) implies that the evolution of the gauge coupling starts from a non-perturbative regime unless $\sqrt{\lambda_{1}}$ is 
extremely large: only in this way we would have $\sqrt{\lambda_{i}} = {\mathcal O}(1)$. Whenever 
$\sqrt{\lambda_{1}} \gg 1$ the gauge coupling will be extremely minute at the end of inflation and this is at odds with the fact that during the decelerated stage of expansion 
we would like to have $g^2 = {\mathcal O}(10^{-2})$ but not much smaller.
 In what follows we shall focus on the case where the gauge coupling increases 
and then freezes (either during the stiff phase or in the radiation epoch) by always remaining perturbative. More contrived cases could be imagined 
but the purpose here is just to illustrate the simplest situation 
compatible with quintessential inflation in the Palatini formulation.

\renewcommand{\theequation}{4.\arabic{equation}}
\setcounter{equation}{0}
\section{The spectra of the gauge fields}
\label{sec4}
Assuming, for the sake of concreteness, that the electroweak temperature $T_{ew}$  is ${\mathcal O}(100)$ GeV,  when $T < T_{ew}$ the $SU_{L}(2)\otimes U_{Y}(1)$ symmetry is broken down to $U_{\mathrm{em}}(1)$ and the produced gauge fields will survive ordinary magnetic fields evolving in an electrically neutral plasma. 
For  $T> T_{ew}$ the electroweak symmetry is restored while around $T\simeq T_{ew}$ the  ordinary magnetic fields are proportional 
to the hypermagnetic fields through the cosine of the Weinberg's angle $\theta_{W}$, i.e. $\cos{\theta_{W}} \, \vec{B}$.  The modes reentering for $T > T_{ew}$ may affect 
the baryon asymmetry \cite{PSC7,PSC9,PSC10,PSC11}. In what 
follows, for reasons of opportunity, we shall mainly be concerned with the modes reentering just prior to decoupling and affecting the magnetogenesis requirements. 

\subsection{Quantum description of the gauge fields}
To compute the hypermagnetic and hyperelectric power spectra we start from the gauge part of Eq. (\ref{AAA27}) and deduce the explicit form of the Hamiltonian under the assumption that  both $\sigma$ and $\overline{\sigma}$ are homogeneous: 
\begin{equation}
\widehat{H}_{Z}(\tau) = \frac{1}{2} \int d^3 x \biggl[ \widehat{\pi}_{i}^{2}  + 
{\mathcal F}\biggl( \widehat{\pi}_{i}\,\widehat{\mathcal Z}_{i} + \widehat{\mathcal Z}_{i}\,\widehat{\pi}_{i}\biggr)+
\partial_{i}\widehat{{\mathcal Z}}_{k} \,\,\partial^{i} \widehat{{\mathcal Z}}_{k} - \biggl(\frac{\overline{\sigma}^{\, \prime}}{ \sigma} \biggr)\, \widehat{{\mathcal Z}}_{i} \,\,\partial_{j} \widehat{{\mathcal Z}}_{k} \,\, \epsilon^{i \, j\, k}\biggr],
\label{UUU2}
\end{equation}
where $ \widehat{{\mathcal Z}}_{i}$ is the quantum field operator corresponding to the (rescaled) 
vector potential ${\mathcal Z}_{i} = \sqrt{\sigma/(4\pi)} \,\, Z_{i}$ defined in the Coulomb gauge \cite{fordcoul} which is the most convenient since it is invariant 
under Weyl rescaling. In Eq. (\ref{UUU2}) $\widehat{\pi}_{i} = \widehat{{\mathcal Z}}_{i}^{\,\,\prime} - {\mathcal F} \, \widehat{{\mathcal Z}}_{i}$ denotes the canonical momentum operator; to make the notation more concise, the rate of variation of the gauge coupling  ${\mathcal F} = \sqrt{\sigma}^{\,\prime}/\sqrt{\sigma}$ has been introduced throughout. The evolution equations of the field operators following form the Hamiltonian (\ref{UUU2}) are (units $\hbar = c =1$ will be adopted):
\begin{eqnarray}
\widehat{\pi}_{i}^{\,\,\prime} &=& i\, \biggl[ \widehat{H}_{Z}, \widehat{\pi}_{i} \biggr] = - {\mathcal F} \, \widehat{\pi}_{i} + \nabla^2 \widehat{{\mathcal Z}}_{i} + \frac{\overline{\sigma}^{\,\prime}}{\sigma} \, \epsilon_{i\, j\, k} \partial^{j} \, \widehat{{\mathcal Z}}^{k},
\nonumber\\
\widehat{{\mathcal Z}}_{i}^{\,\,\prime} &=&   i\, \biggl[ \widehat{H}_{Z}, \widehat{{\mathcal Z}}_{i} \biggr] = \widehat{\pi}_{i} + {\mathcal F} \, \widehat{{\mathcal Z}}_{i}.
\label{UUU3}
\end{eqnarray}
The field operators of Eq. (\ref{UUU3}) obey the canonical commutation relations at equal times 
\begin{equation}
\biggl[\widehat{{\mathcal Z}}_{i}(\vec{x}_{1}, \tau),\widehat{\pi}_{j}(\vec{x}_{2}, \tau)\biggr] = i \Delta_{ij}(\vec{x}_{1} - \vec{x}_{2}),
\label{UUU4}
\end{equation}
where, as usual in the Coulomb gauge, $\Delta_{ij}(\vec{x}_{1} - \vec{x}_{2}) = \int d^{3}k e^{i \vec{k} \cdot (\vec{x}_{1} - \vec{x}_2)} p_{ij}(\hat{k})/(2\pi)^3$ [with $p_{ij}(\hat{k}) = (\delta_{ij} - \hat{k}_{i} \hat{k}_{j})$] is the transverse generalization of the Dirac delta function ensuring that  both the field operators and the canonical momenta are divergenceless. The mode expansion for the hyperelectric and hypermagnetic fields in the circular basis is:
\begin{eqnarray}
\widehat{E}_{i}(\vec{x},\tau) &=&  -   \sum_{\alpha= \pm} \, \,\int\frac{d^{3} k}{(2\pi)^{3/2}}\,\,
\biggl[ g_{k,\,\alpha}(\tau) \, \widehat{a}_{k,\alpha} \,\,  \varepsilon^{(\alpha)}_{i}(\hat{k})\,\,e^{- i \vec{k} \cdot\vec{x}} + \mathrm{h.c.}\biggr],
\label{UUU5}\\
\widehat{B}_{k}(\vec{x}, \tau)  &=&  - i \, \,\epsilon_{i\,j\,k} \,   \sum_{\alpha= \pm}\,\,\int\, \frac{d^{3} k}{(2\pi)^{3/2}}\,\, k_{j} \,\,
\biggl[ f_{k,\, \alpha}(\tau) \, \widehat{a}_{k,\,\alpha}\, \,\,  \varepsilon^{(\alpha)}_{i}(\hat{k})\, e^{- i \vec{k} \cdot\vec{x}} - \mathrm{h.c.} \biggr].
\label{UUU6}
\end{eqnarray}
In Eq. (\ref{UUU6}) the right (i.e. $R$) and left (i.e. $L$) polarizations are defined, respectively, by:
\begin{equation}
\hat{\varepsilon}^{(\pm)}(\hat{k}) = \frac{\hat{e}^{\oplus}(\hat{k}) \pm i \, \hat{e}^{\otimes}(\hat{k})}{\sqrt{2}}, 
\qquad  \hat{\varepsilon}^{(+)}(\hat{k}) \equiv \hat{\varepsilon}_{R}(\hat{k}), \qquad 
\hat{\varepsilon}^{(-)}(\hat{k}) \equiv \hat{\varepsilon}_{L}(\hat{k}),
 \label{POL1}
 \end{equation}
and $\hat{k}$, $\hat{e}_{\oplus}$ and $\hat{e}_{\otimes}$ denote a triplet of mutually orthogonal 
unit vectors defining, respectively, the direction of propagation and  the two linear (vector) polarizations.  From Eq. (\ref{POL1}) the vector product of $\hat{k}$ with the circular polarizations will be given by $\hat{k} \times \hat{\varepsilon}^{(\pm)} = \mp \, i\,  \hat{\varepsilon}^{(\pm)}$. 
The hyperelectric field operator coincides (up to a sign) with the canonical momentum 
[i.e. $\widehat{E}_{i} =  - \widehat{\pi}_{i} = -  \sqrt{\sigma} (\widehat{{\mathcal Z}}_{i}/\sqrt{\sigma})^{\, \prime}$] while the hypermagnetic field operator is simply $\widehat{B}_{k}=\epsilon_{i\,j \, k} \,\partial_{i}\, \widehat{{\mathcal Z}}_{j}$. The  hypermagnetic and hyperelectric mode functions [i.e. $f_{k,\, \alpha}(\tau)$ and $g_{k,\, \alpha}(\tau)$ respectively] must preserve the commutation relations (\ref{UUU4}) 
and this is why their Wronskian $W_{\alpha} = f_{k,\, \alpha} \, g^{\ast}_{k,\, \alpha} - f_{k,\, \alpha}^{\ast} \, g_{k,\, \alpha}\to i$  for each of the two circular polarizations. 
The actual evolution of the mode functions follows by inserting the expansions (\ref{UUU5})--(\ref{UUU6}) into Eq. (\ref{UUU3}) and the final result is: 
\begin{eqnarray}
f_{k,\, \pm}^{\,\prime} &=& g_{k,\,\pm} + {\mathcal F} f_{k,\, \pm},
\label{UUU7}\\
g_{k,\,\pm}^{\,\prime} &=& - k^2 \, f_{k,\, \pm} - {\mathcal F} \,g_{k,\,\pm}  \mp \,  \biggl(\frac{\overline{\sigma}^{\,\prime}}{\sigma}\biggr)\, k \, f_{k,\, \pm}.
\label{UUU8}
\end{eqnarray}
Inserting the explicit expressions of the field operators of Eqs. (\ref{UUU5})--(\ref{UUU6})  
into Eqs. (\ref{BBB3})--(\ref{BBB4}) we can reobtain, after some simple algebra, the 
equations for the mode function already derived in Eqs. (\ref{UUU7})--(\ref{UUU8}) 
from the evolution of the operators in the Heisenberg description.

The quantum mechanical initial conditions imposed on the mode functions 
are fully justified since we shall consider the situation where the total number of inflationary $e$-folds is always larger than ${\mathcal O}(60)$. This mens that potentially relevant 
classical inhomogeneities have been already dissipated. In the case of minimal duration 
of inflation the logic will be necessarily different (see e.g. \cite{mginc1,mginc2}). We also note that in the case of a stiff post-inflationary 
stage the maximal number of $e$-folds accessible to large-scale measurements 
may be even much larger than ${\mathcal O}(60)$ \cite{QUINT12,QUINT13}; in fact we have 
\begin{eqnarray}
 N_{max} &=& 60 + \frac{1}{4} \ln{\biggl(\frac{h_{0}^2 \Omega_{R 0}}{4.15 \times 10^{-5}} \biggr)} - \ln{\biggl(\frac{h_{0}}{0.7}\biggr)}
 \nonumber\\
 &+& \frac{1}{4} \ln{\biggl(\frac{{\mathcal A}^{(s)}}{2.4 \times 10^{-9}}\biggr)} + \frac{1}{4} \ln{\biggl(\frac{r_{T}}{0.01}\biggr)} - (1/2 - \beta) \ln{\xi_{r}}.
 \label{UUU9N}
\end{eqnarray} 
As in Eq. (\ref{PARAM1}) $\beta$ denotes the rate of expansion during the stiff phase and $\xi_{r} = H_{r}/H_{1}$. Note that, on a purely phenomenological ground, $H_{r}$ cannot be smaller than the one of nucleosynthesis (i.e. approximately
 $H_{r}>10^{-44} M_{\mathrm{P}}$); thus, in principle, $\xi_{r}$ can  be smaller than $\overline{\xi}_{r}$ provided the reheating occurs just prior to the formation of the light nuclei\footnote{When we say that, in principle, $\xi_{r}$ can be smaller than $\overline{\xi}_{r} $ we simply mean 
    that this situation is possible but not plausible. If this should happen the dominance of radiation 
     must anyway take place before the synthesis of the light nuclei. This must be true in spite of the (possibly contrived) dynamics leading 
     to such a situation.}. Thus, if $\beta - 1/2 <0$ (as it happens in $\beta = 1/3$ 
when the post-inflationary background is dominated by stiff sources), $N_{max}$ may increase even by $15$ $e$-folds.

\subsection{General forms of the gauge power spectra}
 From the Fourier transform of the field operators 
(\ref{UUU5})--(\ref{UUU6}) 
\begin{eqnarray}
\widehat{E}_{i}(\vec{q},\tau) &=& - \,\sum_{\alpha= \pm}\biggl[ \varepsilon_{i}^{(\alpha)}(\hat{q}) \, g_{q,\,\alpha} \,\widehat{a}_{\vec{q},\,\alpha} +\varepsilon_{i}^{(\alpha)\ast}(-\hat{q}) \, g_{q,\,\alpha}^{\ast} \,\widehat{a}_{-\vec{q},\,\alpha}^{\dagger}\biggr],
\label{UUU9a}\\
\widehat{B}_{k}(\vec{p},\tau) &=& - \, i\, \epsilon_{i\,j\,k} \,\sum_{\alpha= \pm}\biggl[ p_{i} \, \varepsilon_{j}^{(\alpha)}(\hat{p}) \, f_{p,\,\alpha}\, \widehat{a}_{\vec{p},\,\alpha} + p_{i} \, \varepsilon_{j}^{(\alpha)\ast}(-\hat{p})\, f_{p,\,\alpha}^{\ast}\, \widehat{a}_{-\vec{p},\,\alpha}^{\dagger} \biggr],
\label{UUU9b}
\end{eqnarray}
where $\epsilon_{i j k}$ is the Levi-Civita symbol in three-dimensions.
 As a consequence the  two-point functions constructed from Eqs. (\ref{UUU9a}) and (\ref{UUU9b}) will consists of the symmetric contribution and of the corresponding antisymmetric part\footnote{The expectation values are  computed from Eqs. (\ref{UUU9a})--(\ref{UUU9b}), by recalling that $2 \varepsilon_{i}^{(+)}(\hat{k}) \varepsilon_{j}^{(-)}(\hat{k}) = [ p_{ij}(\hat{k}) - i \, \epsilon_{i j \ell} \, \hat{k}^{\ell}]$. }
\begin{eqnarray}
&& \langle \widehat{E}_{i}(\vec{k},\tau)\, \widehat{E}_{j}(\vec{p},\tau) \rangle = \frac{ 2 \pi^2 }{k^3} \biggl[\, P_{E}(k,\tau) \, p_{ij}(\hat{k}) 
+ i\,P_{E}^{(G)}(k,\tau)\,\epsilon_{i\, j\, \ell} \, \hat{k}^{\ell}\biggr] \, \delta^{(3)}(\vec{p} + \vec{k}),
\label{UUU10}\\
&& \langle \widehat{B}_{i}(\vec{k},\tau)\, \widehat{B}_{j}(\vec{p},\tau) \rangle = \frac{ 2 \pi^2 }{k^3} \biggl[\, P_{B}(k,\tau) \, p_{ij}(\hat{k}) 
+ i\,P_{B}^{(G)}(k,\tau)\,\epsilon_{i\, j\, \ell} \, \hat{k}^{\ell}\biggr] \, \delta^{(3)}(\vec{p} + \vec{k}),
\label{UUU11}
\end{eqnarray}
where $p_{ij}(\hat{k}) = \delta_{i j} - \hat{k}_{i} \hat{k}_{j}$ is the standard traceless projector.
The superscript $(G)$ reminds that power spectra of Eq. (\ref{UUU13}) 
 determine the corresponding gyrotropies defined, respectively, by the 
 expectation values of the two pseudoscalar quantities $\langle \vec{B} \, 
 \cdot \vec{\nabla}\times \vec{B} \rangle$ and $\langle \vec{E} \, \cdot \vec{\nabla}\times \vec{E} \rangle$.  In Eqs. (\ref{UUU10})--(\ref{UUU11}) $P_{E}(k,\tau)$ and $P_{B}(k,\tau)$ denote the hyperelectric and the hypermagnetic power spectra
whose explicit expression is given by:
 \begin{equation}
P_{E}(k,\tau) = \frac{k^{3}}{4 \pi^2} \biggl[ \bigl| g_{k,\,-}\bigr|^2 + \bigl| g_{k,\,+}\bigr|^2 \biggr], \qquad
P_{B}(k,\tau) = \frac{k^{5}}{4 \pi^2} \biggl[ \bigl| f_{k,\,-}\bigr|^2 + \bigl| f_{k,\,+}\bigr|^2 \biggr].
\label{UUU12}
\end{equation}
In the limit $\overline{\sigma} \to 0$ the gyrotropic contributions of Eqs. (\ref{UUU10})--(\ref{UUU11})
\begin{equation}
P_{E}^{(G)}(k,\tau) =  \frac{k^{3}}{4 \pi^2} \biggl[  \bigl| g_{k,\,-}\bigr|^2 - \bigl| g_{k,\,+}\bigr|^2 \biggr],\qquad
P_{B}^{(G)}(k,\tau) =  \frac{k^{5}}{4 \pi^2} \biggl[ \bigl| f_{k,\,-}\bigr|^2  -  \bigl| f_{k,\,+}\bigr|^2\biggr]
\label{UUU13}
\end{equation}
vanish. The reason is that, in this limit, the evolution of the two circularly 
 polarized mode functions coincide (see Eqs. (\ref{UUU7})--(\ref{UUU8})) so that
\begin{equation}
\overline{\sigma} \to 0, \qquad P_{E}^{(G)}(k,\tau)\to 0, \qquad P_{B}^{(G)}(k,\tau)\to  0.
\label{UUU14}
\end{equation}
If $\overline{\sigma}\neq 0$ the mechanism discussed in this investigation generates  
the spectrum of the hypermagnetic gyrotropy.  The modes reentering prior to the electroweak phase transition must be released into fermions later on \cite{PSC9} so that the presence of $\overline{\sigma}$ provides a mechanism for the baryon asymmetry generation that has been investigated in various frameworks. Only for reasons of space and opportunity we leave aside these important applications that complements the phenomenological considerations of the following section. It should be however remarked that the slopes of the superhorizon hypermagnetic spectra are practically unaffected by the relative strength of the parity-breaking terms. Therefore we do not expect that the addition of $\overline{\sigma}$ will change 
the phenomenological considerations of the following section.
We finally recall that for the forthcoming applications what matters are not 
the comoving spectra of Eqs. (\ref{UUU2})--(\ref{UUU13}) but rather their physical counterpart. From the relations between the physical and the comoving fields introduced prior to Eqs.  the physical power spectra are given by:
\begin{equation}
P_{X}^{(phys)}(k,\tau) = \frac{P_{X}(k,\tau)}{\sigma(\tau) \, a^4(\tau)},
\label{PXphys}
\end{equation}
where $P_{X}(k,\tau)$ generically denotes either the hypermagnetic or the 
hyperelectric field. Let us finally remind that the energy density of the gauge fields 
follows from the corresponding energy-momentum tensor 
derived from the action. Using Eqs. (\ref{UUU10})--(\ref{UUU11}) we can obtain 
the averaged energy density of the parametrically amplified gauge fields; to
compare this estimate 
with the energy density of the background geometry we introduce the so-called spectral energy density in critical units:
\begin{eqnarray}
\Omega_{Z}(k,\tau) &=& \frac{1}{\rho_{crit}} \frac{ d \langle \hat{\rho} \rangle}{d \ln{k}} = \frac{2}{3 H^2 \, M_{P}^2 a^4}  \biggl[ P_{E}(k,\tau) + P_{B}(k,\tau)\biggr]
\nonumber\\
&=& \frac{2 \sigma}{3 H^2 \, M_{P}^2} \,\biggl[ P^{(phys)}_{E}(k,\tau) + P_{B}^{(phys)}(k,\tau)\biggr],
\label{OMZ}
\end{eqnarray}
where we expressed $\Omega_{Y}(k,\tau)$ both in terms of the comoving and of the physical power spectra.
To guarantee the absence of dangerous backreaction terms 
in  Eqs. (\ref{BBB1})--(\ref{BBB6}) and (\ref{BBB7})--(\ref{BBB9}) the value of  $\Omega_{Y}(k,\tau)$ must always be subcritical throughout the various stages of the evolution and for all relevant scales; this requirement must be separately verified both during and after inflation.

\subsection{The power spectra during inflation}
In the case $\overline{\sigma} \to 0$  the solution of Eqs. (\ref{UUU7})--(\ref{UUU8})  during the inflationary stage the expression of the mode functions 
follows from Eq. (\ref{TTT1}) and it is given by:
\begin{eqnarray}
f_{k}(\tau) &=& \frac{N_{f}}{\sqrt{2 k}} \, \sqrt{- k\tau} \, H_{|\gamma -1/2|}^{(1)}(-k\tau),\qquad 
N_{f} = \sqrt{\frac{\pi}{2}} e^{i \pi ( 1 + |2 \gamma -1|)/4},
\nonumber\\
g_{k}(\tau) &=& N_{g} \,\sqrt{\frac{k}{2}} \,  \sqrt{- k\tau} \, H_{\gamma+1/2}^{(1)}(-k\tau),\qquad N_{g} =\sqrt{\frac{\pi}{2}} e^{i \pi \gamma/2},
\label{VVV1}
\end{eqnarray}
where $g_{k,\,+}= g_{k,\,-} = g_{k}$ and $f_{k,\,+}= f_{k,\,-} = f_{k}$. In Eq. (\ref{VVV1}) $H_{\alpha}^{(1)}(z)$ 
are the Hankel functions of first  kind \cite{abr1,abr2}; as a consequence the mode 
functions satisfy the Wronskian normalization condition $f_{k}(\tau) g_{k}^{\ast}(\tau) - f_{k}^{\ast}(\tau) g_{k}(\tau) = i$ (recall, in this respect, 
that $H_{\alpha}^{(1)\ast}(z)= H_{\alpha}^{(2)}(z)$). 
From Eq. (\ref{VVV1}) the comoving 
power spectra during inflation are given by:
\begin{eqnarray}
P_{B}(k,\tau) &=& a^4 \,H^4 \, D(|\gamma-1/2|) \, \biggl(\frac{k}{a\, H}\biggr)^{ 5 - | 2 \gamma -1|},
\label{VVV2a}\\
P_{E}(k,\tau) &=& a^4 \, H^4 \, D(\gamma +1/2) \, \biggl(\frac{k}{a\, H}\biggr)^{ 4 - 2\gamma},
\label{VVV3a}
\end{eqnarray}
where the function $D(x)$ is defined as $D(x) = 2^{2 x -3} \Gamma^2(x)/\pi^3$. 
The spectral energy density of Eq. (\ref{VVV4}) must be always 
subcritical (i.e. $\Omega_{Z}(k,\tau) \ll 1$)  for $\tau \leq -\tau_{1}$ and $|k \tau| \leq 1$; this requirement is not always 
satisfied even if, during the inflationary phase, $H \ll M_{P}$. 
The first term inside the square bracket at the right hand side of Eq. (\ref{VVV4}) denotes the magnetic contribution while the second term is the electric result. We finally mention that from Eqs. (\ref{VVV2a})--(\ref{VVV3a}) it is immediate to obtain the gauge spectra in the case 
of decreasing gauge coupling. Following the logic of Eq. (\ref{TTT9}) we actually
have that for $\gamma \to - \widetilde{\gamma}$ the power spectra 
are exchanged i.e. $P_{B}(k,\tau) \to \widetilde{P}_{E}(k,\tau)$ 
and $P_{E}(k,\tau) \to \widetilde{P}_{B}(k,\tau)$; therefore the explicit 
expressions of $ \widetilde{P}_{B}(k,\tau)$ and $ \widetilde{P}_{E}(k,\tau)$ are:
\begin{eqnarray}
\widetilde{\,P\,}_{B}(k,\tau) &=& a^4 \,H^4 \, D(\widetilde{\gamma}+1/2) \, \biggl(\frac{k}{a\, H}\biggr)^{ 4 -  2 \widetilde{\gamma}|},
\label{VVV2adu}\\
\widetilde{\,P\,}_{E}(k,\tau) &=& a^4 \, H^4 \, D(|\widetilde{\gamma} -1/2|) \, \biggl(\frac{k}{a\, H}\biggr)^{ 5 -  |2 \widetilde{\gamma} -1|}.
\label{VVV3adu}
\end{eqnarray}
The results of Eqs. (\ref{VVV2adu})--(\ref{VVV3adu}) are a manifestation 
of the duality symmetry \cite{NINEaa,TENaa} and follow directly from the 
transformation properties of Eqs. (\ref{UUU7})--(\ref{UUU8}) under\footnote{The transformations 
transforming Eq. (\ref{UUU7}) into Eq. (\ref{UUU8}) (and vice versa) are $f_{k,\,\alpha} \to g_{k,\,\alpha}/k$ and $g_{k,\,\alpha} \to - k f_{k,\,\alpha}$.}
$\sigma \to 1/\sigma$ and in the limit $\overline{\sigma} \to 0$. 
The same symmetry can be used to obtain 
the explicit form of the power spectra in the other stages of the model.
As already discussed after Eq. (\ref{TTT9}) we shall preferentially 
discuss the case of increasing gauge coupling since this is 
comparatively lass constrained.

\subsection{The power spectra during the stiff phase}
During the stiff phase (i.e. for $\tau_{r}> \tau \geq  |\tau_{1}|$)
the continuous parametrization of $\sqrt{\sigma}$ given in Eq. (\ref{TTT2}) 
implies that the solutions of Eqs. (\ref{UUU7})--(\ref{UUU8}) with the correct asymptotic conditions are:
\begin{equation}
\left(\matrix{ f_{k}(\tau) &\cr
g_{k}(\tau)/k&\cr}\right) = \left(\matrix{ A_{f\, f}(k, \tau, \tau_{1})
& A_{f\,g}(k,\tau, \tau_{1})&\cr
A_{g\,f}(k,\tau, \tau_{1}) &A_{g\,g}(k,\tau, \tau_{1})&\cr}\right) \left(\matrix{ \overline{f}_{k} &\cr
\overline{g}_{k}/k&\cr}\right),
\label{VVV3}
\end{equation}
where 
\begin{equation}
\overline{f}_{k}= f_{k}(-\tau_{1})\qquad \mathrm{and}\qquad \overline{g}_{k} = g_{k}(-\tau_1)
\end{equation}
denote the values of the mode functions at end of the inflationary phase and the matrix elements at the right hand side of Eq. (\ref{VVV3}) can be found in appendix \ref{APPA}. 
The entries of the matrix appearing in Eq. (\ref{VVV3}) are not all of the same 
order; indeed from the explicit expressions of Eqs. (\ref{VVV4})--(\ref{VVV5}) we can prove that, in the limit $x_{1} = |k \tau_{1} | \ll 1$, the following hierarchy is always verified:  
\begin{equation}
\frac{A_{f\,f}(k, \tau, \tau_{1})\, k \, \overline{f}_{k}}{A_{f\,g}(k, \tau, \tau_{1})\, \overline{g}_{k}}  \simeq 
\frac{A_{g\, f}(k,\tau, \tau_{1})\, k\,  \overline{f}_{k}}{A_{g\,g}(k,\tau, \tau_{1})\, \overline{g}_{k}} = {\mathcal O}( x_{1}^{\alpha}),
\label{VVV9}
\end{equation}
where we introduced the combination\footnote{This combination has obviously nothing to do 
with the coefficient appearing in the nonlinear gravitational action; we think that this remark will 
prevent any possible confusion.}$\alpha = \gamma +3/2- |\gamma-1/2|$. In practice the condition $x_{1} \ll 1$ is always 
verified  since it only amounts to requiring that the various $k$-modes are 
smaller than the maximal frequency of the spectrum. The result of Eq. (\ref{VVV9}) implies that, in spite of the value of $\gamma$,
 the correction ${\mathcal O}(x_{1}^{\alpha})$ in Eq. (\ref{VVV8}) is always negligible for $x_{1} \ll 1$ so that the explicit form of the mode functions from Eq. (\ref{VVV3}) are:
\begin{eqnarray}
f_{k}(\tau) &=& \frac{A_{f\,g}(k,\tau,\tau_{1}) \,\overline{g}_{k}}{k}\biggl[ 1 + {\mathcal O}(x_{1}^{\alpha})\biggr],
\nonumber\\
g_{k}(\tau) &=& A_{g\,g}(k, \tau, \tau_{1}) \, \overline{g}_{k}\biggl[ 1 + {\mathcal O}(x_{1}^{\alpha})\biggr].
\label{VVV8}
\end{eqnarray}
Since $f_{k,\,+}= f_{k,\,-} = f_{k}$ and $g_{k,\,+}= g_{k,\,-} = g_{k}$ the explicit 
form of the comoving power spectra for $\tau_{r}< \tau < - \tau_{1}$  is obtained after inserting Eq. (\ref{VVV8}) into Eq. (\ref{UUU12}):
\begin{eqnarray}
P_{B}(k,\tau) &=& a_{1}^4\, H_{1}^4 D(\gamma +1/2) \bigl| A_{f\, g}(k, \tau, \tau_{1}) \bigr|^2 \, \biggl(\frac{k}{a_{1} \, H_{1}}\biggr)^{4 - 2 \gamma},
\label{VVV10}\\
P_{E}(k,\tau) &=& a_{1}^4\, H_{1}^4 D(\gamma +1/2) \bigl| A_{g\, g}(k,  \tau, \tau_{1}) \bigr|^2 \, \biggl(\frac{k}{a_{1} \, H_{1}}\biggr)^{4 - 2 \gamma}.
\label{VVV11}
\end{eqnarray}
These spectra can be computed for any range of the parameters. However since the relevant physical scales for magnetogenesis will reenter 
prior to equality (and generally after BBN) the  physical range 
of the parameters corresponds to the scales that are still larger than the effective horizon (i.e. $|k \tau| \ll 1$) for $|\tau_{1}|  \tau < \tau_{r}$; in this case 
 $A_{f\, g}(k, \tau,\tau_{1})$ and $A_{g\, g}(k, \tau, \tau_{1}) $ can be 
 approximated as:
\begin{eqnarray}
A_{f\, g}(k,  \tau, \tau_{1}) &=&  \frac{x}{(2 \delta +1)} \, \biggl(\frac{x}{q_{1}\, x_{1}}\biggr)^{\delta} \biggl[ 1 + {\mathcal O}( x^2)\biggr],
\nonumber\\
A_{g\, g}(k,\tau, \tau_{1}) &=&  \biggl(\frac{x}{q_{1}\, x_{1}}\biggr)^{\delta} \biggl[ 1 + {\mathcal O}( x^2)\biggr].
\label{VVV12}
\end{eqnarray}
Equation (\ref{VVV12}) is valid in the limit $x_{1} \ll 1$ and $x/x_{1} \simeq |\tau/\tau_{1}| \gg 1$. Therefore the explicit form of the 
power spectra to leading order is: 
\begin{eqnarray}
P_{B}(k,\tau) &=& a_{1}^4 \, H_{1}^4 {\mathcal C}_{B}(\gamma,\delta) \biggl(\frac{k}{a_{1} H_{1}}\biggr)^{4 - 2 \gamma - 2 \delta} \,\, \biggl( \frac{k}{a H}\biggr)^{ 2 \delta + 2},
\label{VVV13}\\
P_{E}(k,\tau) &=& a_{1}^4 \, H_{1}^4 {\mathcal C}_{E}(\gamma,\delta) \biggl(\frac{k}{a_{1} H_{1}}\biggr)^{4 - 2 \gamma - 2 \delta} \,\, \biggl( \frac{k}{a H}\biggr)^{ 2 \delta }.
\label{VVV14}
\end{eqnarray}
The numerical coefficients ${\mathcal C}_{E}(\gamma, \delta)$ and ${\mathcal C}_{B}(\gamma,\delta)$ are relevant for the final result but only depend on $\gamma$ and $\delta$; in the first approximation we can consider they give a contribution 
ranging between $10^{-2}$ and $1$:
\begin{equation}
{\mathcal C}_{B}(\gamma,\delta) = \frac{2^{2\gamma -2}}{\pi^3 \, (2 \delta + 1)^2}\biggl(\frac{\delta}{\gamma}\biggr)^{- 2\, \delta}\, \Gamma^2(\gamma +1/2), \qquad {\mathcal C}_{E}(\gamma,\delta) = (2 \delta+1)^2 \, {\mathcal C}_{B}(\gamma,\delta).
\label{VVV15}
\end{equation}
In practice the interesting physical range of Eqs. (\ref{VVV13})--(\ref{VVV14}) and (\ref{VVV15}) is realized for $\delta > 1/2$ and $k\tau\ll 1$. When $ 0 < \delta \ll \gamma$ the explicit expressions of the hypermagnetic 
and hyperelectric power spectra are:
\begin{eqnarray}
P_{B}(k,\tau) &=& a_{1}^{4} \, H_{1}^4 \, D(\gamma + 1/2) \, \biggl(\frac{k}{a_{1} \, H_{1}}\biggr)^{4 - 2 \gamma - 2 \delta} \, F^2_{B}(k \tau, \delta),
\nonumber\\
P_{E}(k,\tau) &=& a_{1}^{4} \, H_{1}^4 \, D(\gamma + 1/2) \, \biggl(\frac{k}{a_{1} \, H_{1}}\biggr)^{4 - 2 \gamma - 2 \delta} \, F^2_{E}(k \tau, \delta),
\label{VVV16}
\end{eqnarray}
where $F_{B}(x,\delta)$ and  $F_{E}(x,\delta)$ are defined as:
\begin{eqnarray}
F_{B}(x,\delta) &=& \biggl(\frac{q_{1}}{2}\biggr)^{-\, \delta} \, \sqrt{\frac{x}{2}} \, \Gamma(\delta+1/2) \, J_{\delta+1/2}(x), 
\nonumber\\
F_{E}(x,\delta) &=& \biggl(\frac{q_{1}}{2}\biggr)^{-\, \delta} \, \sqrt{\frac{x}{2}} \, \Gamma(\delta+1/2) \, J_{\delta-1/2}(x). 
\label{VVV17}
\end{eqnarray}
The results of Eqs. (\ref{VVV16})--(\ref{VVV17}) only assume $x_{1} <1 $ and $0\leq \delta \ll \gamma$ and can be evaluated either for $k\tau \ll 1$ or for $k\tau \gg 1$. 

\subsection{The power spectra during in the radiation stage}
For $\tau > \tau_{r}$ we have that $\sigma(\tau)$ is given by Eq. (\ref{TTT3}) the correctly normalized solutions of Eqs. (\ref{UUU7})--(\ref{UUU8}) are:
\begin{equation}
\left(\matrix{ f_{k}(\tau) &\cr
g_{k}(\tau)/k&\cr}\right) = \left(\matrix{ B_{f\, f}(k, \tau, \tau_{1}, \tau_{r})
& B_{f\,g}(k, \tau, \tau_{1}, \tau_{r})&\cr
B_{g\,f}(k, \tau, \tau_{1}, \tau_{r}) &B_{g\,g}(k, \tau, \tau_{1}, \tau_{r})&\cr}\right) \left(\matrix{ \widetilde{\,f\,}_{k} &\cr
\widetilde{\,g\,}_{k}/k&\cr}\right),
\label{UUU1}
\end{equation}
where this time 
\begin{equation}
\widetilde{\,f\,}_{k}= f_{k}(\tau_{r})\qquad \mathrm{and} \qquad \widetilde{\,g\,}_{k} = g_{k}(\tau_r)
\end{equation} 
denote the values of the mode functions at end of the stiff phase. The  matrix elements appearing in Eq. (\ref{UUU1}) are reported in appendix \ref{APPA}.
This time the hierarchy between the different terms of the solution is more complicated. 
To simplify the various terms appearing in Eq. (\ref{UUU2a})--(\ref{UUU3a}) we then consider the physical limit $x_{r} = |k \tau_{r}| \ll 1$:
\begin{eqnarray}
&&B_{f\, f}(k,\tau, \tau_{1}, \tau_{r})= \biggl( \frac{q_{2} x_{r}}{2}\biggr)^{\zeta} \,\,\biggl[ \sqrt{\frac{x}{2}} \Gamma(1/2 -\zeta) J_{- \zeta -1/2}(x) + 
{\mathcal O}(x_{r}) \biggr] + {\mathcal O}\biggl[(q_{2}\,x_{r})^{1 - \zeta}\biggr],
\nonumber\\
&&B_{f\, g}(k,\tau,\tau_{1},\tau_{r}) = \biggl( \frac{q_{2} x_{r}}{2}\biggr)^{-\zeta} \,\,\biggl[ \sqrt{\frac{x}{2}} \Gamma(1/2 +\zeta) J_{\zeta +1/2}(x) + {\mathcal O}(x_{r}) \biggr] + {\mathcal O}\biggl[ (q_{2}\,x_{r})^{1 + \zeta}\biggr],
\nonumber\\
&&B_{g\, f}(k,\tau,\tau_{1},\tau_{r}) = \biggl( \frac{q_{2} x_{r}}{2}\biggr)^{\zeta} \,\,\biggl[- \sqrt{\frac{x}{2}} \Gamma(1/2 -\zeta) J_{1/2- \zeta}(x) + 
{\mathcal O}(x_{r}) \biggr] + {\mathcal O}\biggl[ (q_{2}\,x_{r})^{1 - \zeta}\biggr],
\nonumber\\
&&B_{g\, g}(k,\tau, \tau_{1},\tau_{r})= \biggl( \frac{q_{2} x_{r}}{2}\biggr)^{-\zeta} \,\,\biggl[\sqrt{\frac{x}{2}} \Gamma(1/2 +\zeta) J_{\zeta -1/2}(x) + 
{\mathcal O}(x_{r}) \biggr] + {\mathcal O}\biggl[(q_{2}\,x_{r})^{1 + \zeta}\biggr].
\label{UUU4a}
\end{eqnarray}
If the gauge coupling freezes during the stiff phase then $\zeta =0$; conversely 
if it freezes during the radiation epoch we will have that $\zeta\ll \delta$. This is why Eq. (\ref{UUU4a}) has been obtained in the case $\zeta<1/2$ which encompasses both situations,  as we shall also see more precisely from section \ref{sec5}.
If we now recall the 
hierarchies  of Eq. (\ref{VVV9}) we can write $\widetilde{\,f\,}_{k}$ and $\widetilde{\,g\,}_{k}$ as:
\begin{equation}
\widetilde{\,f\,}_{k} = A_{f\, g}(k,\, \tau_{r},\, \tau_{1}) \frac{\overline{g}_{k}}{k}, \qquad 
\widetilde{\,g\,}_{k} = A_{g\, g}(k,\, \tau_{r},\, \tau_{1})\overline{g}_{k}.
\label{UUU6a}
\end{equation}
From Eqs. (\ref{UUU4a})--(\ref{UUU6a}) 
the mode functions for $\tau> \tau_{r}$ become:
\begin{eqnarray}
f_{k}(\tau) &=& \frac{\overline{g}_{k}}{k} \biggl[ B_{f\, f}(k, \tau, \tau_{1}, \tau_{r}) A_{f\, g}(k, \tau_{r}, \tau_{1}) + 
B_{f\, g}(k, \tau, \tau_{1}, \tau_{r}) A_{g\, g}(k, \tau_{r}, \tau_{1})\biggr],
\nonumber\\
g_{k}(\tau) &=& \overline{g}_{k}\biggl[ B_{g\, f}(k, \tau, \tau_{1}, \tau_{r}) A_{f\, g}(k, \tau_{r}, \tau_{1}) + 
B_{g\, g}(k, \tau, \tau_{1}, \tau_{r}) A_{g\, g}(k,  \tau_{r}, \tau_{1})\biggr].
\label{UUU7a}
\end{eqnarray}
Recalling that $f_{k,\,+}= f_{k,\,-} = f_{k}$ and $g_{k,\,+}= g_{k,\,-} = g_{k}$ the explicit 
form of the comoving power spectra for $\tau> \tau_{r}$  is obtained after inserting Eq. (\ref{UUU7a}) into Eq. (\ref{UUU12}):
\begin{eqnarray}
P_{B}(k, \tau) &=& \frac{k^3 \, \bigl|\overline{g}_{k}\bigr|^2}{2 \pi^2 q_{1}^{2 \delta}} \biggl| \frac{k \tau_{r}}{k \tau_{1}} \biggr|^{2 \delta} 
\biggl[ \frac{B_{f\, f}(k,\tau, \tau_{1},\tau_{r})}{1 + 2 \delta} |k \tau_{r}| + B_{f\, g}(k, \tau, \tau_{1}, \tau_{r})\biggr]^2,
\label{UUU8aa}\\
P_{E}(k, \tau) &=& \frac{k^3 \, \bigl|\overline{g}_{k}\bigr|^2}{2 \pi^2 \,q_{1}^{2 \delta}} \biggl| \frac{k \tau_{r}}{k \tau_{1}} \biggr|^{2 \delta} 
\biggl[ \frac{B_{g\, f}(k, \tau, \tau_{1},\tau_{r})}{1 + 2 \delta} |k \tau_{r}| + B_{g\, g}(k,\tau, \tau_{1}, \tau_{r})\biggr]^2.
\label{UUU9aa}
\end{eqnarray}
If we take the limits $k \tau_{1} \ll 1$ and $k \tau_{r} \ll 1$ by keeping $ k \tau$ free to vary we have 
that Eqs. (\ref{UUU8aa})--(\ref{UUU9aa}):
\begin{eqnarray}
P_{B}(k,\tau) &=& a_{1}^4 \, H_{1}^4 \overline{C}(\tau_{1}, \tau_{r},\gamma,\delta,\zeta) \biggl(\frac{k}{a_{r} H_{r}}\biggr)^{4 - 2 \gamma - 2 \zeta} 
\biggl( \frac{a_{r} H_{r}}{ a_{1} \, H_{1}}\biggr)^{ 4 - 2 \gamma - 2 \delta} \, Q_{B}^2(k\tau, k \tau_{r},\zeta),
\label{UUU10a}\\
P_{E}(k,\tau) &=& a_{1}^4 \, H_{1}^4 \overline{C}(\tau_{1}, \tau_{r},\gamma,\delta,\zeta)\biggl(\frac{k}{a_{r} H_{r}}\biggr)^{4 - 2 \gamma - 2 \zeta} 
\biggl( \frac{a_{r} H_{r}}{ a_{1} \, H_{1}}\biggr)^{ 4 - 2 \gamma - 2 \delta}  \, Q_{E}^2(k\tau, k \tau_{r},\zeta).
\label{UUU11a}
\end{eqnarray}
The two functions 
$Q_{B}(k\tau, k\tau_{r},\zeta)$ and $Q_{B}(k\tau, k\tau_{r},\zeta)$ appearing in Eqs. (\ref{UUU10})--(\ref{UUU11}) can be defined for generic values of their arguments and they are:
\begin{eqnarray}
&&Q_{B}(x,y,\zeta)=\sqrt{\frac{x}{2}}\biggl[\Gamma(\zeta+1/2)  J_{\zeta+1/2}(x) + \frac{ (q_{2}/2)^{ 2\zeta}}{(1+ 2 \delta)} \Gamma(1/2- \zeta) y^{2 \zeta +1}J_{-\zeta-1/2}(x)\biggr],
\label{UUU13a}\\
&&Q_{E}(x,y,\zeta)=\sqrt{\frac{x}{2}}\biggl[ \Gamma(\zeta+1/2)  J_{\zeta-1/2}(x) - \frac{ (q_{2}/2)^{2 \zeta}}{(1+ 2 \delta)} \Gamma(1/2- \zeta) y^{2 \zeta +1}J_{1/2 -\zeta}(x)\biggr].
\label{UUU14a}
\end{eqnarray}
Note, furthermore, that $\overline{C}(\tau_{1}, \tau_{r},\gamma,\delta, \zeta)$ does not depend on $\tau$ and it is given by:
\begin{equation}
\overline{C}(\tau_{1}, \tau_{r},\gamma,\delta, \zeta) = \frac{2^{2(\gamma + \zeta) -3}}{\pi^3} \, \Gamma^2(\gamma+1/2) (\gamma/\delta)^{-2\delta} \biggl[ \frac{(\gamma/\delta)(\tau_{r}/\tau_{1} +1)+1}{(\gamma/\zeta) (\tau_{r}/\tau_{1})}\biggr]^{- 2 \zeta} q_{2}^{2 \zeta}.
\label{UUU12a}
\end{equation}
The expression for $\overline{C}(\tau_{1}, \tau_{r},\gamma,\delta, \zeta)$ looks quite complicated but it simplifies considerably in the case $\tau_{r} \gg \tau_{1}$:
\begin{equation}
\lim_{\tau_{r} \gg \tau_{1}} \overline{C}(\tau_{1}, \tau_{r},\gamma,\delta, \zeta) = \frac{2^{2(\gamma + \zeta) -3}}{\pi^3} \, \Gamma^2(\gamma+1/2) (\gamma/\delta)^{-2\delta},
\label{UUU12b}
\end{equation}
where it has been used that, according to Eq. (\ref{UUU3a}), $q_{2} \simeq (\zeta/\delta)$ when $\tau_{r} \gg \tau_{1}$. From the phenomenological viewpoint we have that this is indeed the relevant regime since  
\begin{equation}
\frac{\tau_{r}}{\tau_{1}} = \frac{a_{1} \, H_{1}}{a_{r}\, H_{r}} \equiv \xi_{r}^{\beta - 1} \gg 1
\end{equation}
where the last inequality follows when $\xi_{r} \gg 1$ and $\beta < 1/2$ (see also Eq. 
(\ref{PARAM1})).

\renewcommand{\theequation}{5.\arabic{equation}}
\setcounter{equation}{0}
\section{Some phenomenological aspects}
\label{sec5}
The quintessential inflationary 
scenarios analyzed within the Palatini approach are  compatible with the current 
observational data and it is therefore interesting to discuss in some detail 
the possible constraints stemming from the magnetogenesis considerations. 
The two complementary phenomenological possibilities that 
will be scrutinized correspond to the results already derived on section \ref{sec4}. The first option is that the gauge coupling 
freezes during the stiff phase; in practice this 
case corresponds to $\delta \ll \gamma$ while $\zeta =0$.
The second possibility is that the gauge coupling is still evolving during 
the stiff phase but gets to a constant value in the radiation stage; this second situation 
is realized when $\zeta \ll \delta$. From the numerical viewpoint, as we shall see, values of $\delta < {\mathcal O}(0.1)$ will be effectively indistinguishable from the case $\delta\to 0$ and the same comment also holds for the variation of $\zeta$. 

\subsection{Constraints on the parameter space}
We shall now consider the bounds on the variation of the gauge coupling by separately examining the constraints during inflation and in the 
subsequent stages of expansion.  For $\tau> \tau_{r}$ the tenets  
of the concordance paradigm \cite{cosmow} will be adopted together with the conventional notations\footnote{In particular $\Omega_{M0}$ and $\Omega_{R0}$ denote the (present)
critical fractions of the matter and radiation; $h_{0}$ is the Hubble rate in units 
of $\mathrm{Hz}\,\mathrm{Mpc}/\mathrm{km}$; $r_{T}$ denotes the tensor to scalar ratio. These quantities have been already introduced in the previous sections.}. 

\subsubsection{Inflationary constraints}
During inflation the spectral energy density associated with the gauge fields can be 
immediately deduced by inserting Eqs. (\ref{VVV1}) into Eq. (\ref{OMZ}):
\begin{equation}
\Omega^{(inf)}_{Z}(k,\tau) = \frac{ 1}{12\pi} \biggl(\frac{H}{M_{P}}\biggr)^2 (- k \tau)^{5} \biggl[ \bigl| H^{(1)}_{|\gamma-1/2|}(-k\tau)\bigr|^2 + \bigl| H^{(1)}_{\gamma +1/2}(-k\tau)\bigr|^2 \biggr],
 \label{YYY0}
 \end{equation}
 where the superscript reminds that the spectral energy density is computed in the inflationary stage. 
 The relevant physical limit of Eq. (\ref{YYY0})  is when the 
 wavelengths are larger than the expansion rate or, in terms of the wavenumbers, 
 $k< a\, H$; in this case Eq. (\ref{YYY0}) becomes:
\begin{equation}
\Omega^{(inf)}_{Z}(k,\tau) = \frac{2}{3} \biggl(\frac{H}{M_{P}}\biggr)^2 \biggl[ D(|\gamma -1/2|) \biggl| \frac{k}{a\, H}\biggr|^{5 - |2 \gamma-1|} + D(\gamma +1/2) \biggl|\frac{k}{a \, H} \biggr|^{4 - 2 \gamma} \biggr].
\label{YYY1}
\end{equation}
From Eqs. (\ref{VVV2a})--(\ref{VVV3a}) 
and (\ref{YYY1}) the hyperelectric spectra are  scale-invariant for $\gamma\to 2$ while the hypermagnetic component steeply increases [i.e. $P_{B}(k,\tau) \propto |k \tau|^2$ for $k \tau \ll 1$]
so that  the condition $\Omega^{(inf)}_{Z}(k,\tau) \ll 1$ is safely satisfied.
If  $1/2 < \gamma \leq 2$ the requirement $\Omega^{(inf)}_{Z}(k,\tau) \ll 1$ always holds for $| k \tau \ll 1$.
Finally, if $\gamma > 2$ the hypermagnetic spectrum becomes even steeper while the hyperelectric spectrum diverges in the limit $k\tau \ll 1$: in this case the bound $\Omega^{(inf)}_{Z} \ll 1$ is not satisfied so that the whole class of models $\gamma > 2$ must be excluded. 
Equation (\ref{YYY1}) has been deduced in the most constraining regime, namely when the relevant wavelengths are larger than the Hubble radius. The steepness of the hypermagnetic spectrum during the 
inflationary stage in the limit $\gamma\to 2$ does not imply that the magnetic fields will also be minute at the galactic scale after the gauge coupling flattens out. This swift conclusion would only be true provided the hypermagnetic magnetic power spectrum at the end of inflation is not modified at late times 
(i.e. for $\tau > - \tau_{1}$ or even for $\tau > \tau_{r}$). This is, however, not the case as we saw 
from the results of the previous section and of appendix \ref{APPA}. We shall nonetheless require that $0 < \gamma < 2$ since, for $\gamma > 2$, the spectral energy density is overcritical already during 
the inflationary stage.

\subsubsection{Constraints from the stiff phase}
If the gauge coupling freezes during the stiff phase the spectral energy density follows by inserting Eqs. (\ref{VVV3}) into Eq. (\ref{OMZ}). As discussed in Eqs. (\ref{VVV8})--(\ref{VVV12})
the general expressions could be simplified in the relevant 
physical limits where $k \tau_{1} \ll 1$ and $|\tau_{1}| <\tau < \tau_{r}$. 
Therefore from Eqs. (\ref{VVV16})--(\ref{VVV17}) the explicit form of
$\Omega^{(st)}_{Z}(k,\tau)$ is given by:
\begin{equation}
\Omega^{(st)}_{Z}(k,\tau) = \frac{\pi r_{T}\, {\mathcal A}^{(s)}}{24}\, \biggl(\frac{a_{1}}{a}\biggr)^{4 -2/\beta} \, D(\gamma+1/2) 
\biggl(\frac{k}{a_{1}\, H_{1}} \biggr)^{4 - 2\gamma - 2 \delta} {\mathcal U}(k\tau, \delta), 
\label{YYY2}
\end{equation}
where we introduced ${\mathcal U}(k\tau, \delta) = F_{E}^2(k\tau, \delta) + F_{B}^2(k\tau, \delta)$. We stress that the superscript in $\Omega^{(st)}_{Z}(k,\tau)$ now reminds that the magnetic fields are computed during the stiff phase.  As already mentioned, in Eq. (\ref{YYY2}) $\beta$ measures the  expansion rate between $\tau_{1}$ and $\tau_{r}$ (see 
 Eq. (\ref{PARAM1}) and discussion therein). Equation (\ref{YYY2}) does not apply if the gauge coupling 
is still evolving during radiation and this will be the last class of constraints discussed in the following subsection. It is relevant to remark, in this respect, that the freezing of the gauge coupling does not 
imply the reentry of a given scale. On the contrary the gauge coupling may freeze at different 
epochs while the typical scales relevant to the magnetogenesis considerations 
are bound to reenter right before matter-radiation equality, as we shall more specifically
discuss hereunder. Because the 
gauge coupling reaches its flat limit for $\delta \to 0$,
Eq. (\ref{YYY2}) holds, in practice, for $\delta \ll \gamma$ and for $\delta \ll \beta$; in this limit, in spite of the value of $k\tau$ we have that the contribution of ${\mathcal U}(k\tau, \delta)$ can be neglected since:
\begin{equation}
\lim_{\delta \to 0} \biggl[F_{E}^2(k\tau, \delta) + F_{B}^2(k\tau, \delta)\biggr] \to 1.
\label{YYY3}
\end{equation}
Furthermore, thanks to the inflationary bound we must also require that $\gamma < 2$ which implies that $\Omega^{(st)}_{Z}(k,\tau)$ is an increasing or quasi-flat function of the wavenumber.
If we evaluate Eq. (\ref{YYY2}) at the end of inflation and for the maximal amplified wavenumber of the spectrum, the critical energy bound reads:
\begin{equation}
\Omega^{(st)}_{Z}(a_{1} H_{1}, \tau_{1}) = \frac{\pi  \, r_{T}}{24}\, {\mathcal A}^{(s)}  D(\gamma+1/2) \ll 1.
\label{YYY4a}
\end{equation}
The condition of Eq. (\ref{YYY4a}) depends on $\gamma$ very mildly since $D(\gamma+1/2) < {\mathcal O}(1)$ in the range $0<\gamma <2$  where the spectral energy density is subcritical during inflation. Since $r_{T}< 0.01$
Eq. (\ref{YYY4a}) implies $\Omega^{(st)}_{Z}(a_{1} H_{1}, \tau_{1}) < {\mathcal O}(10^{-10})$.
The bound on the spectral energy density is more constraining at $\tau_{r}$ because, during the stiff phase, the expansion rate is slower than in the 
radiation stage; from Eq. (\ref{YYY2}) the explicit expression of $\Omega^{(st)}_{Z}(k,\tau_{r})$ is
\begin{equation}
\Omega^{(st)}_{Z}(k,\tau_{r}) = \frac{\pi}{24}\,D(\gamma+1/2) \, r_{T}\,  \,{\mathcal A}^{(s)} \biggl(\frac{a_{1}}{a_{r}}\biggr)^{4 - 2/\beta} 
\biggl(\frac{k}{a_{1}\, H_{1}} \biggr)^{4 - 2\gamma - 2 \delta} \, {\mathcal U}(k\tau_{r}, \delta).
\label{YYY4}
\end{equation}
The spectral energy density can be evaluated for the maximal frequency (which is also the most constraining one).
Again the contribution of  ${\mathcal U}(k\tau_{r}, \delta)$ just gives a constant contribution which goes to $1$ in the case $\delta \ll 1$. 
Thus, for  $k = k_{1} = a_{1} H_{1}$ Eq. (\ref{YYY4}) demands:
\begin{equation}
\Omega^{(st)}_{Z}(k_{1},\tau_{r}) = \frac{\pi}{24}\,D(\gamma+1/2) \, r_{T}\,  \,{\mathcal A}^{(s)} \, \xi_{r}^{4 \beta -2} <1.
\label{YYY5}
\end{equation}
Since  $ \xi_{r} = H_{r}/H_{1} <1$ and $ \beta < 1/2$ (see Eq. (\ref{PARAM1})) the condition $\Omega^{(st)}_{Z}(k_{1},\tau_{r}) < 1$ depends on the value of $\xi_{r}$.  As long as  $\Omega^{(st)}_{Z}(k_{1},\tau_{r}) < 1$ we will also have that $\Omega^{(st)}_{Z}(k,\tau_{r}) <1$ for all the $k < k_{1}$. For instance for $k= k_{r} = a_{r} H_{r}$ we will have that 
\begin{equation}
\frac{\Omega^{(st)}(k_{r}, \tau_{r})}{\Omega^{(st)}(k_{1}, \tau_{r})} = \xi_{r}^{(\beta-1) ( 4 - 2 \gamma - 2 \delta)},
\label{YYY7}
\end{equation}
which is always smaller than $1$ provided 
\begin{equation}
\overline{\xi}_{r} \leq  \xi_{r} < 1,\qquad \qquad (\gamma + \delta) <2, \qquad 
\qquad \delta \ll \gamma, \qquad \delta \ll \beta<1/2.
\label{YYY8}
\end{equation}
It is finally useful to recall that $k/(a_{1} H_{1})$ depends on the comoving scale but also on the 
cosmological parameters:
\begin{eqnarray}
\frac{k}{a_{1}\, H_{1}} = 10^{-23.05}\,\, \biggl(\frac{k}{\mathrm{Mpc}^{-1}}\biggr)\, \biggl(\frac{r_{T}}{0.01}\biggr)^{-1/4}\,\,\biggl(\frac{h_{0}^2 \Omega_{R0}}{4.15\times 10^{-5}}\biggr)^{-1/4} \,\,\biggl(\frac{{\mathcal A}^{(s)}}{2.41\times10^{-9}}\biggr)^{-1/4} \,\, \xi_{r}^{1/2-\beta}
\label{YYY8a}
\end{eqnarray}
showing that for the typical scales relevant for the magnetogenesis problem $k/(a_{1} H_{1}) < {\mathcal O}(10^{-23})$ 
since $\beta< 1/2$ and $\xi_{r}\ll1 $;  this happens for the same reason 
already mentioned in connection with Eq. (\ref{UUU9N}) where we argued that the 
existence of a stiff phase affects the determination of the maximal number of $e$-folds 
presently accessible to large-scale observations \cite{QUINT12,QUINT13}. 

\subsubsection{Constraints from the radiation stage}
The physical possibilities described in the present and in the previous 
subsections are somehow mutually exclusive: either the gauge coupling freezes in the stiff phase 
(i.e. $\delta \ll \gamma$ and $\delta\ll \beta$) or it freezes during the radiation epoch (implying
$\zeta \ll \delta$ and $\zeta\ll \beta$). Furthermore if $\sigma$ flattens out for $\tau> \tau_{r}$  we will also have that $(\gamma + \delta ) <2$, with $\delta>0$; in other words in this case $\delta$ will not have to be  
much smaller than $1$ as it happens instead in Eq. (\ref{YYY8}). Since $\gamma <2$ (because of the inflationary limit) we will have that 
$\gamma< 2 - \delta$. With these caveats, the  expression of $\Omega^{(rad)}_{Z}(k,\tau)$ 
follows from Eqs. (\ref{UUU10a})--(\ref{UUU11a}) and (\ref{UUU13a})--(\ref{UUU14a}): 
\begin{equation}
\Omega^{(rad)}_{Z}(k,\tau) = \frac{2 \,H_{1}^4\,a_{1}^4}{3 H^2 \, a^4\, M_{P}^2} \overline{C}(\tau_1,\tau_{r}, \gamma,\delta, \zeta)
\biggl(\frac{k}{H_{r} \, a_{r}}\biggr)^{4 - 2 \gamma - 2\zeta}\biggl(\frac{a_{r} \, H_{r}}{ a_{1}\, H_{1}}\biggr)^{4 - 2 \gamma- 2\delta}  {\mathcal V}(k, \tau, \tau_{r}, \zeta), 
\label{YYY9}
\end{equation}
where, for the sake of conciseness, we introduced 
\begin{equation}
{\mathcal V}(k, \tau, \tau_{r},\zeta)= Q_{E}^2(k\tau, k\tau_{r}, \zeta) + Q_{B}^2(k\tau, k\tau_{r}, \zeta).
\label{YYY9a}
\end{equation}
Since in the limit $\zeta\ll \gamma$ (and for fixed values of the other arguments) 
we have that ${\mathcal V}(k, \tau, \tau_{r},\zeta) \to 1$,  when deriving the relevant bounds we shall 
omit ${\mathcal V}(k, \tau, \tau_{r},\zeta)$ which will be instead taken into account in the numerical discussion.  
The other useful strategy is to introduce explicitly $\xi_{r}$ which is always much smaller than $1$;
 in this case we can use the limit (\ref{UUU12a}) and Eq. (\ref{YYY9}) becomes, after 
some algebra:
\begin{equation}
\Omega^{(rad)}_{Z}(k,\tau) = \frac{2}{3} \biggl(\frac{H_{1}}{M_{P}}\biggr)^2 \, \biggl(\frac{a_{1}}{a_{r}}\biggr)^{4 - 2/\beta} \overline{C}( \gamma,\delta, \zeta)
\biggl(\frac{k}{H_{r} \, a_{r}}\biggr)^{4 - 2 \gamma - 2\zeta} \xi_{r}^{(1-\beta)(4 - 2\gamma - 2\delta)}.
\label{YYY10}
\end{equation}
Since the above expression applies for $k \leq k_{r}$ and $\tau \geq \tau_{r}$ 
the most constraining bound will arise for $k = {\mathcal O}(a_{r} H_{r})$; in this case Eq. (\ref{YYY10})
becomes:
\begin{equation}
\Omega^{(rad)}_{Z}(k_{r},\tau) = \frac{\pi}{24} \, r_{T} \, {\mathcal A}^{(s)}\, \xi_{r}^{4\beta - 2} \overline{C}( \gamma,\delta, \zeta) \xi_{r}^{(1-\beta)(4 - 2\gamma - 2\delta)}.
\label{YYY11}
\end{equation}
It follows that $\Omega^{(rad)}_{Z}(k_{r},\tau) < 1$ 
provided $\overline{\xi}_{r} \leq  \xi_{r} < 1$. In fact we will recover the same condition of Eq. (\ref{YYY8}) with one difference: from Eq. (\ref{YYY11}) we had $ (\gamma + \zeta) <2$; now this condition is equivalent, in practice, to 
$\gamma <2$ since $\zeta \ll 1$. 
\begin{figure}[!ht]
\centering
\includegraphics[height=8cm]{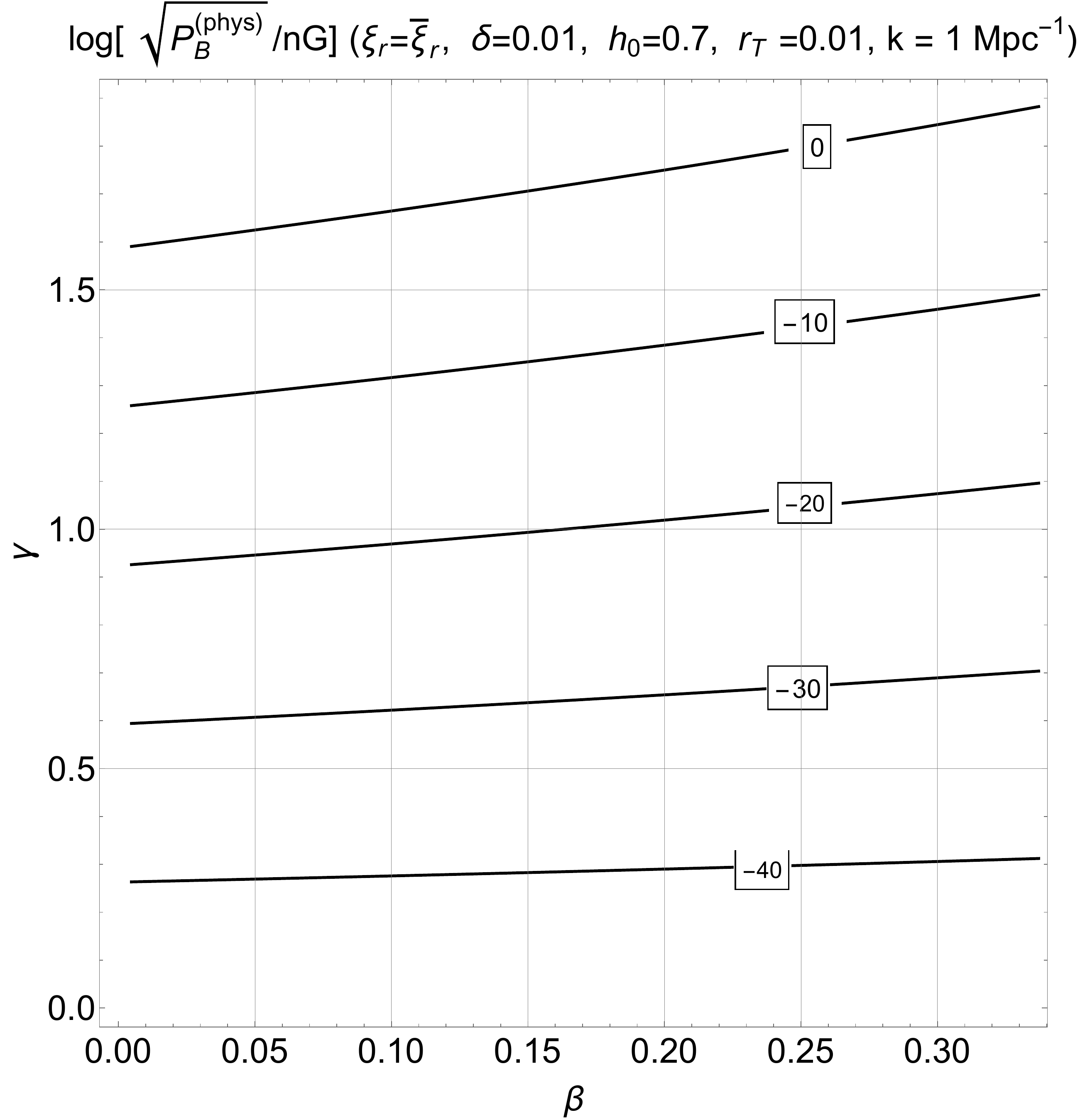}
\includegraphics[height=8cm]{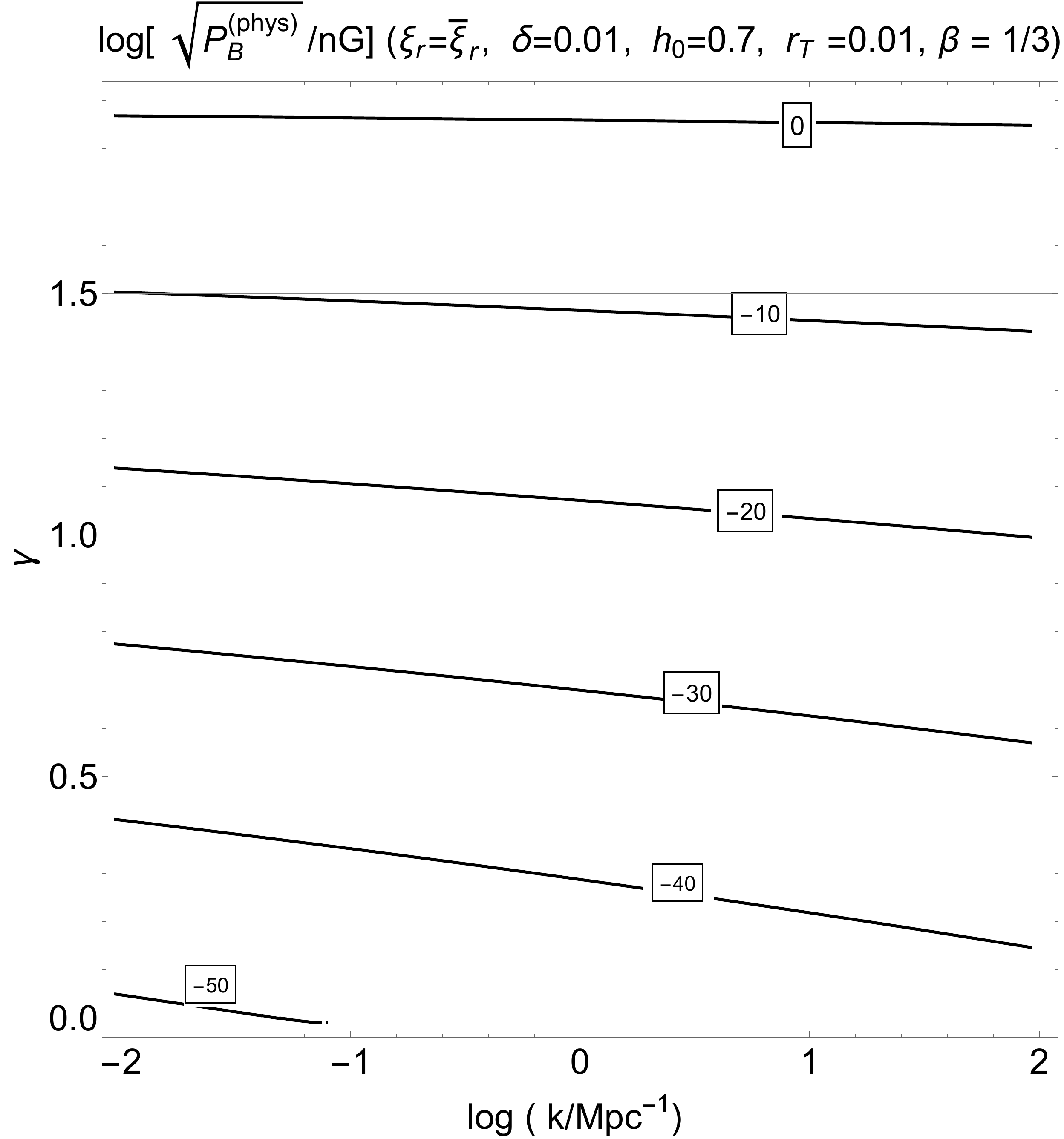}
\caption[a]{We illustrate the physical power spectrum by fixing $\xi_{r}$ to its 
minimal value compatible with the backreaction constraints. 
The common logarithms of the square root of the physical 
power spectra (in nG units) are illustrated on the various curves appearing in both plots.}
\label{FIG1}      
\end{figure}
\subsection{The magnetogenesis requirements} 
The non-screened vector modes of the hypercharge
field project on the electromagnetic fields through the cosine of the Weinberg angle so that the hypermagnetic  power spectra must be multiplied  by $\cos^2{\theta_{W}}$ as soon as they reenter the Hubble radius.  Defining with $\tau_{k}=1/k$ the reentry time of a generic wavelength, the ratio between $\tau_{k}$ and the time of matter-radiation equality $\tau_{eq}$ is given by  $(\tau_{k}/\tau_{eq}) = \sqrt{2} (H_{0}/k) \Omega_{M0}/\sqrt{\Omega_{R0}}$; in more explicit terms we have: 
\begin{equation}
\frac{\tau_{k}}{\tau_{eq}} = 1.06 \times 10^{-2} \biggl(\frac{h_{0}^2 \Omega_{M0}}{0.1386} \biggr) 
\biggl(\frac{h_{0}^2 \Omega_{R 0}}{4.15\times 10^{-5}}\biggr)^{-1/2} \, \biggl( \frac{k}{\mathrm{Mpc}^{-1}}\biggr)^{-1}.
\label{WWW1}
\end{equation}
Equation (\ref{WWW1}) implies that the wavenumbers relevant to the magnetogenesis considerations [i.e. $k = {\mathcal O}(\mathrm{Mpc}^{-1})$] reentered prior to matter radiation equality.  The evolution of the mode functions in the presence of the conductivity 
has been discussed in the past and we just remind the main points. For $\tau > \tau_{k}$ the conductivity dominates so that the evolution of the magnetic and of the electric mode functions is modified as:
\begin{equation}
g_{k}^{\prime} = - k^2 f_{k} - \sigma_{em} \, g_{k}, \qquad f_{k}^{\prime} = g_{k}.
\label{condadd1}
\end{equation}
In Eq. (\ref{condadd1}) $\sigma_{em}$ denotes the electromagnetic conductivity 
since the modes of the field reenter {\em after} symmetry breaking.
According to Eq. (\ref{condadd1}) the electric fields are suppressed by the finite value of the conductivity, the magnetic fields are not dissipated for typical scales smaller than the magnetic diffusivity scale. The evolution described by  Eq. (\ref{condadd1}) can be systematically solved as an expansion in $(k/\sigma_{em})$ by setting initial conditions 
at $\tau =\tau_{k}$. Let us assume, for instance, that the gauge coupling freezes during 
the radiation stage:
\begin{equation}
f_{k}(\tau) = B_{g\,f}(k, \tau_{1}, \tau_{k}) \frac{\overline{g}_{k}}{k} e^{- \frac{k^2}{k_{\sigma}^2}}, 
\qquad\qquad 
g_{k}(\tau) = \biggl(\frac{k}{\sigma_{em}}\biggr) B_{g\,g}(k, \tau_{1}, \tau_{k}) \overline{g}_{k} e^{- \frac{k^2}{k_{\sigma}^2}},
\label{condadd2}
\end{equation}
where  the magnetic diffusivity scale $k_{\sigma}$ has been defined as
 $k_{\sigma}^{-2}  = \int_{\tau_{k}}^{\tau} \, d z /\sigma_{em}(z)$. While the estimate of $k_{\sigma}$ 
can be made accurate by computing the transport coefficients of the plasma in different regimes (see e.g. \cite{cond0}), 
for the present purposes this is not necessary since the ratio $(k/k_{\sigma})^2$  is so small, for the 
phenomenologically interesting scales, that the negative exponentials  in Eq. (\ref{condadd2}) 
evaluate to $1$. In fact by taking $\tau = \tau_{\mathrm{eq}}$ we have that $k_{\sigma}$ turns out to be:
\begin{equation}
\biggl(\frac{k}{k_{\sigma}}\biggr)^2 = \frac{4.75 \times 10^{-26}}{ \sqrt{2 \, h_{0}^2 \Omega_{M0} (z_{\mathrm{eq}}+1)}} \, \biggl(\frac{k}{\mathrm{Mpc}^{-1}} \biggr)^2,
\label{condadd3}
\end{equation}
where $\Omega_{M0}$ is the present critical fraction in matter and $z_{\mathrm{eq}} + 1 = a_{0}/a_{\mathrm{eq}}\simeq {\mathcal O}(3200)$ is the redshift of matter-radiation equality. 
The conductivity has been discussed in the past in various papers. None of them are specifically concerned with the specific scenario discussed here. However, as a reference, 
the approach used here reproduces the one of Refs. \cite{PSC7,cond1}. These 
analyses have been later revisited in \cite{cond1,cond2}. In both papers it is suggested 
that an extra phase could be patched after inflation. This perspective has been 
originally discussed in Ref. \cite{PSC10}.

According to the standard lore the observed large-scale fields in galaxies (and to some extent in clusters)  should have been much smaller before the gravitational collapse of the protogalaxy. 
Compressional amplification enhances the initial values of the magnetic seeds 
by $4$ or even $5$  (see e.g. \cite{rev3}): when the protogalactic matter 
collapsed by gravitational instability over a typical scale ${\mathcal O}(\mathrm{Mpc})$
the mean matter density before collapse was of the order of $\rho_{crit}$. 
Conversely right after the collapse the mean matter density 
became, approximately, six orders of magnitude larger than the critical density.
Since the physical size of the patch decreases from $1$ Mpc to 
$30$ kpc the magnetic field increases, because of flux conservation, 
of a factor $(\rho_{a}/\rho_{b})^{2/3} \sim 10^{4}$ 
where $\rho_{a}$ and $\rho_{b}$ are, respectively the energy densities 
right after and right before gravitational collapse. 

\begin{figure}[!ht]
\centering
\includegraphics[height=8cm]{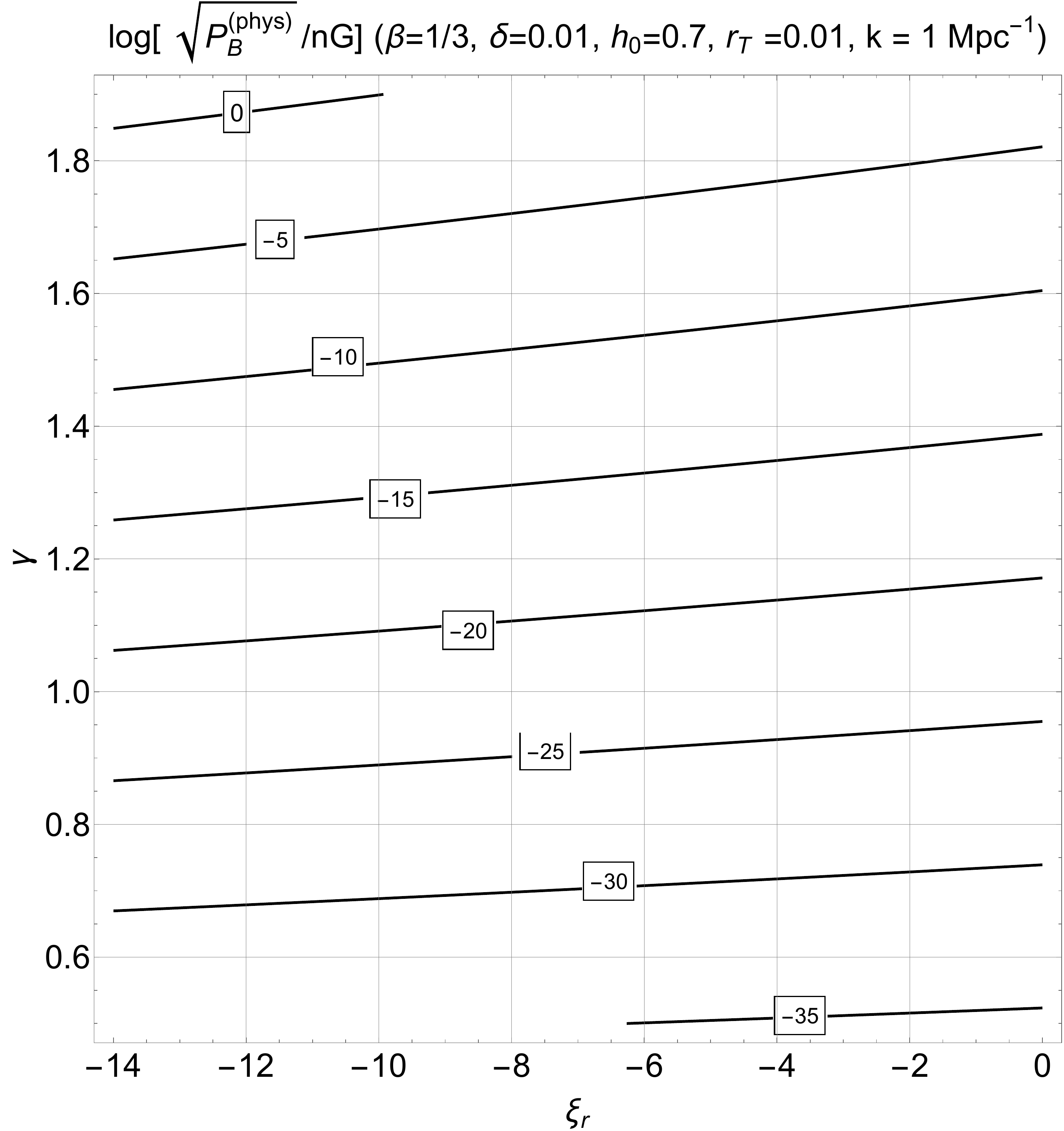}
\includegraphics[height=8cm]{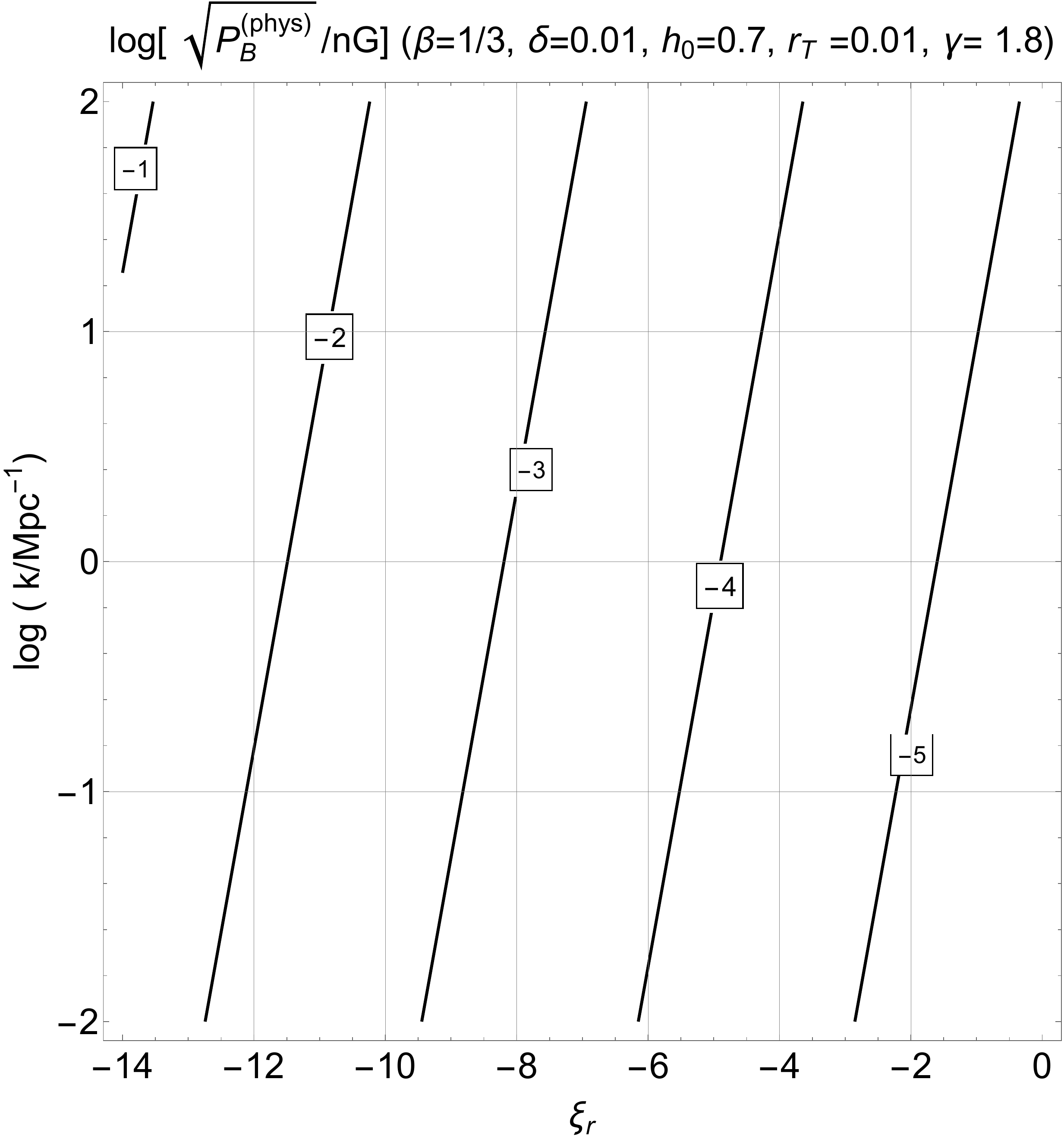}
\caption[a]{The values of the physical power spectra have been illustrated when $\beta$ is fixed to its fiducial theoretical value (i.e. $\beta =1/3$) while $\xi_{r}$ is free to vary in the range 
$\xi_{r} \geq \overline{\xi}_{r}$.}
\label{FIG2}      
\end{figure}
Most of the work in the context of the dynamo 
theory focuses on reproducing the correct features of the 
magnetic field of our galaxy (see, e.g. \cite{b1,b2,b3}). The dynamo term may  be responsible for the origin of the magnetic field of the galaxy eve if, in what follows, 
we shall not assume the existence of a strong galactic dynamo action but rather 
establish a range of initial conditions that might be complemented by the dynamo activity.
In short the idea is the following; the typical rotation period of the galaxy is 
${\mathcal O}(10^{8})$ yrs and comparing this figure with the typical age of the galaxy, ${\cal O}(10^{10} {\rm yrs})$, it can be comcluded that the galaxy performed about $30$ rotations since the time of the protogalactic collapse. The achievable amplification produced by the 
dynamo instability can be at most of $10^{13}$, i.e. about $30$ $e$-folds. Thus, if 
the present value of the galactic magnetic field is ${\mathcal O}(\mu\mathrm{G})$, its value 
right after the gravitational collapse of the protogalaxy might have 
been as small as ${\mathcal O}(10^{-10})$ nG over a typical scale of $30$--$100$ kpc.
\begin{figure}[!ht]
\centering
\includegraphics[height=8cm]{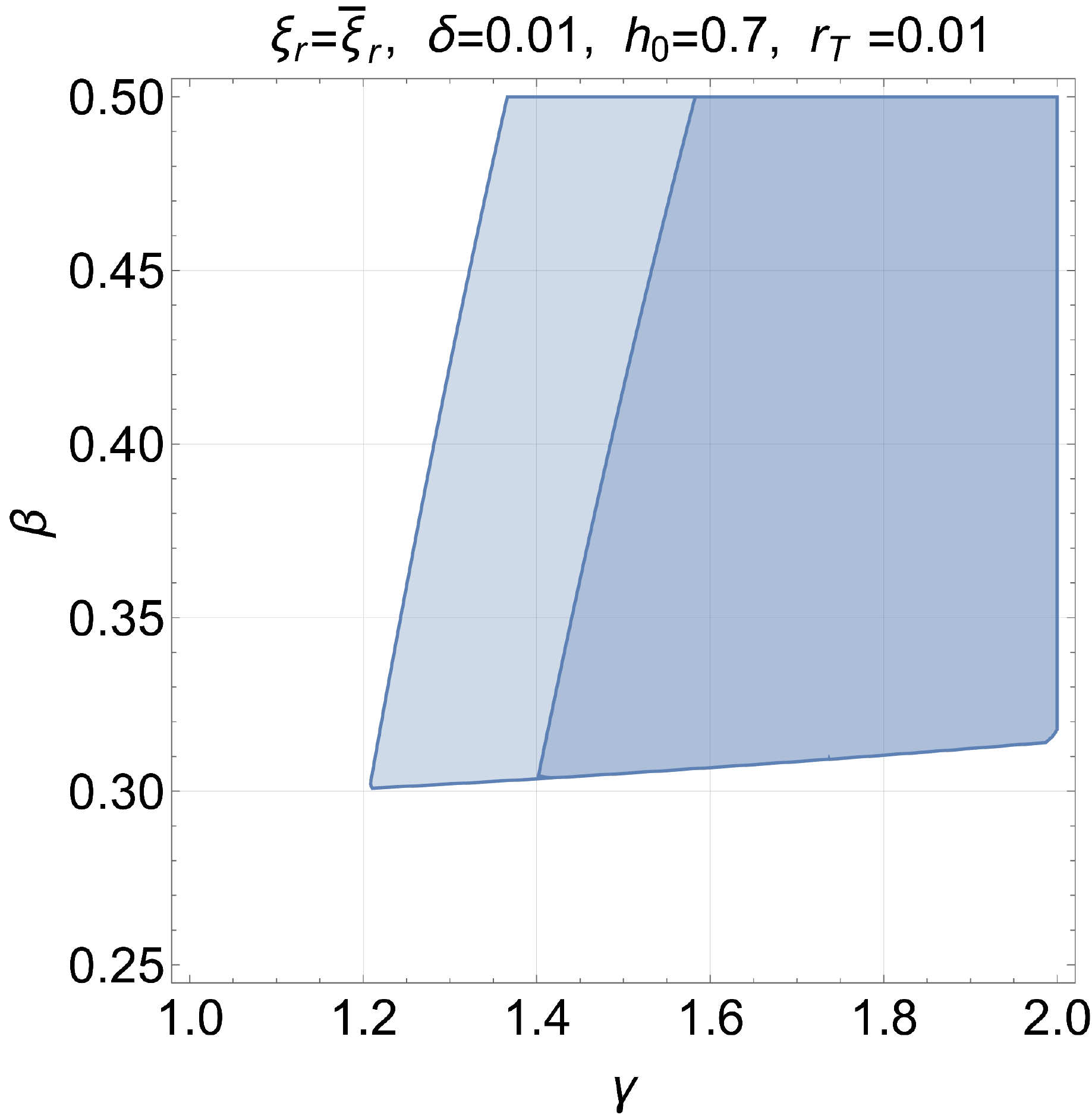}
\includegraphics[height=8cm]{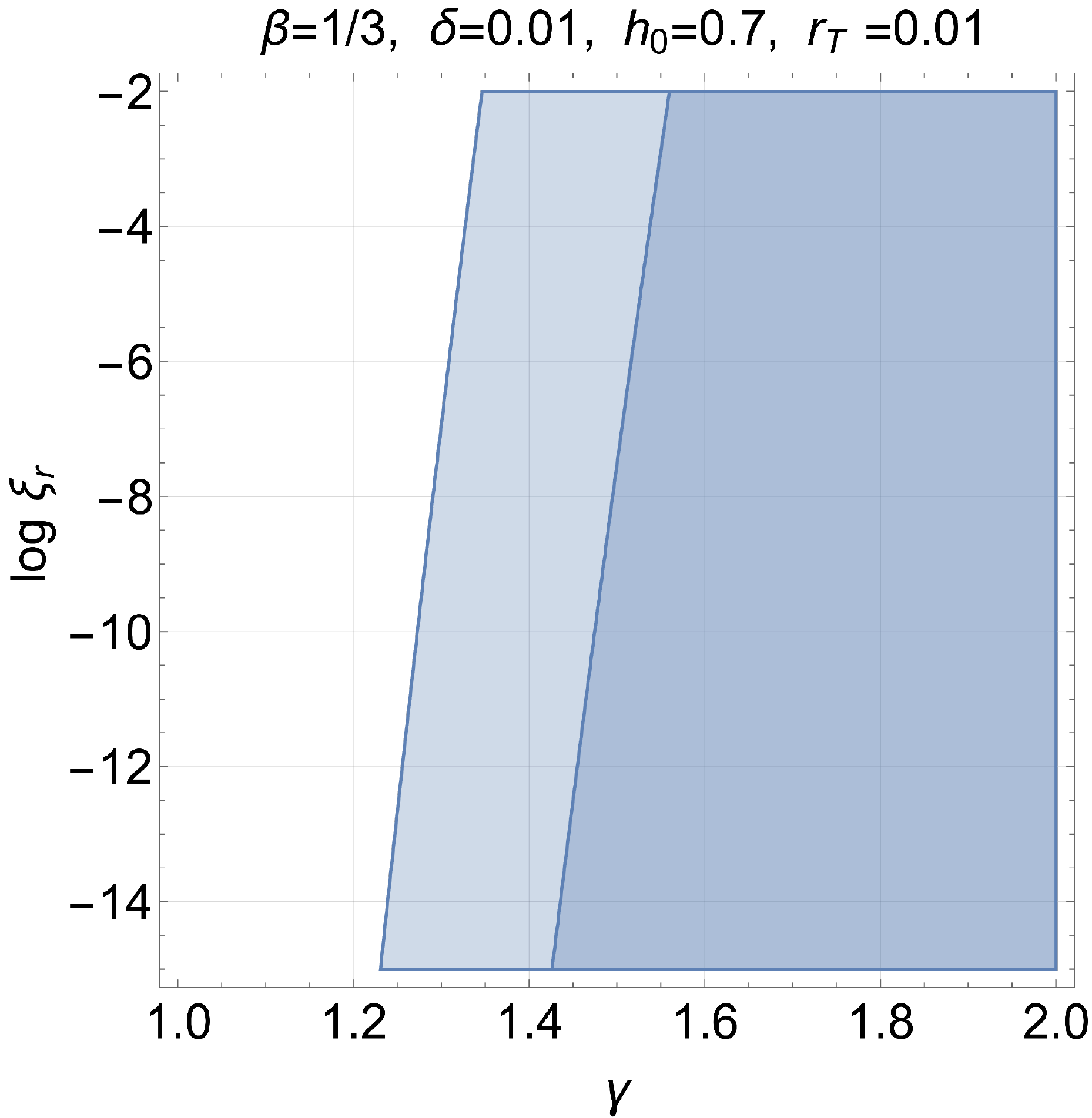}
\caption[a]{We illustrate the allowed regions of the parameter space where the critical energy density bounds and the magnetogenesis requirements are simultaneously satisfied.}
\label{FIG3}      
\end{figure}
The compressional amplification and the dynamo action are typically combined 
together so that the magnetogenesis requirements 
 roughly demand that the magnetic fields at the time of the gravitational collapse of the protogalaxy should be approximately larger than a (minimal) power spectrum  between ${\mathcal O}(10^{-32})\,\mathrm{nG}^2$ and ${\mathcal O}(10^{-22})\, \mathrm{nG}^2$:
\begin{equation}
\log{\biggl(\frac{\sqrt{P^{(phys)}_{B}}}{\mathrm{nG}} \biggr)} \geq - \omega, \qquad 11 < \omega < 16.
\label{mg3}
\end{equation}
The least demanding requirement  of Eq. (\ref{mg3}) (i.e. $\sqrt{P^{(phys)}_{B}} \geq 10^{-16}\,\mathrm{nG}$) follows by assuming 
that, after compressional amplification, every rotation of the galaxy increases the initial magnetic field of one $e$-fold. According to some this requirement is not completely since it takes more than one $e$-fold  to increase the value of the magnetic field by one order of magnitude  and this is the rationale for the most demanding condition of Eq. (\ref{mg3}), i.e. $\sqrt{P^{(phys)}_{B}} \geq 10^{-11}\,\mathrm{nG}$.

\subsubsection{Freezing during the stiff stage}
If the freezing occurs during the stiff phase we can safely ignore 
the dependence of the gauge coupling for $\tau > \tau_{r}$. 
Furthermore during the stiff stage we will have that $0<\delta \ll \gamma$. This 
means that the magnetogenesis constraints will depend essentially on four 
distinct parameters: $\xi_{r}$ and $\beta$ ( parametrizing the evolution during the stiff phase), 
$\gamma$ (i.e. the rate of evolution of the gauge coupling during the inflationary stage), 
and the comoving wavenumber $k$. For typical values of ${\mathcal N} = {\mathcal O}(100)$ 
and for $r_{T} < 0.07$ \cite{RT1,RT2,RT2a} we have that $\overline{\xi}_{r} = {\mathcal O}(10^{-14})$. 
We can therefore fix $ \xi_{r} = \overline{\xi}_{r}$ and consider the variation of $\beta$, $\gamma$ and $k$. This analysis is illustrated in Fig. \ref{FIG1}. In both plots of Fig. \ref{FIG1} we report the common logarithm of $\sqrt{P_{B}^{(phys)}(k, \tau_{k})}$ in units of nG where 
$\tau_{k}$ defines the reentry of the scales relevant for magnetogenesis.  
The various contours in both plots corresponds to the curves where 
the magnetic power spectrum takes the same values. In the left plot of Fig. \ref{FIG1} the $(\beta, \, \gamma)$ plane is illustrated while in the plot at the 
right we consider instead the $(\gamma, \, k)$ plane. We see that for $1< \gamma < 2$, $\beta = {\mathcal O}(1/3)$ and $k = {\mathcal O}(\mathrm{Mpc}^{-1})$ the magnetogenesis requirements of Eq. (\ref{mg3}) are approximately satisfied. 
\begin{figure}[!ht]
\centering
\includegraphics[height=8cm]{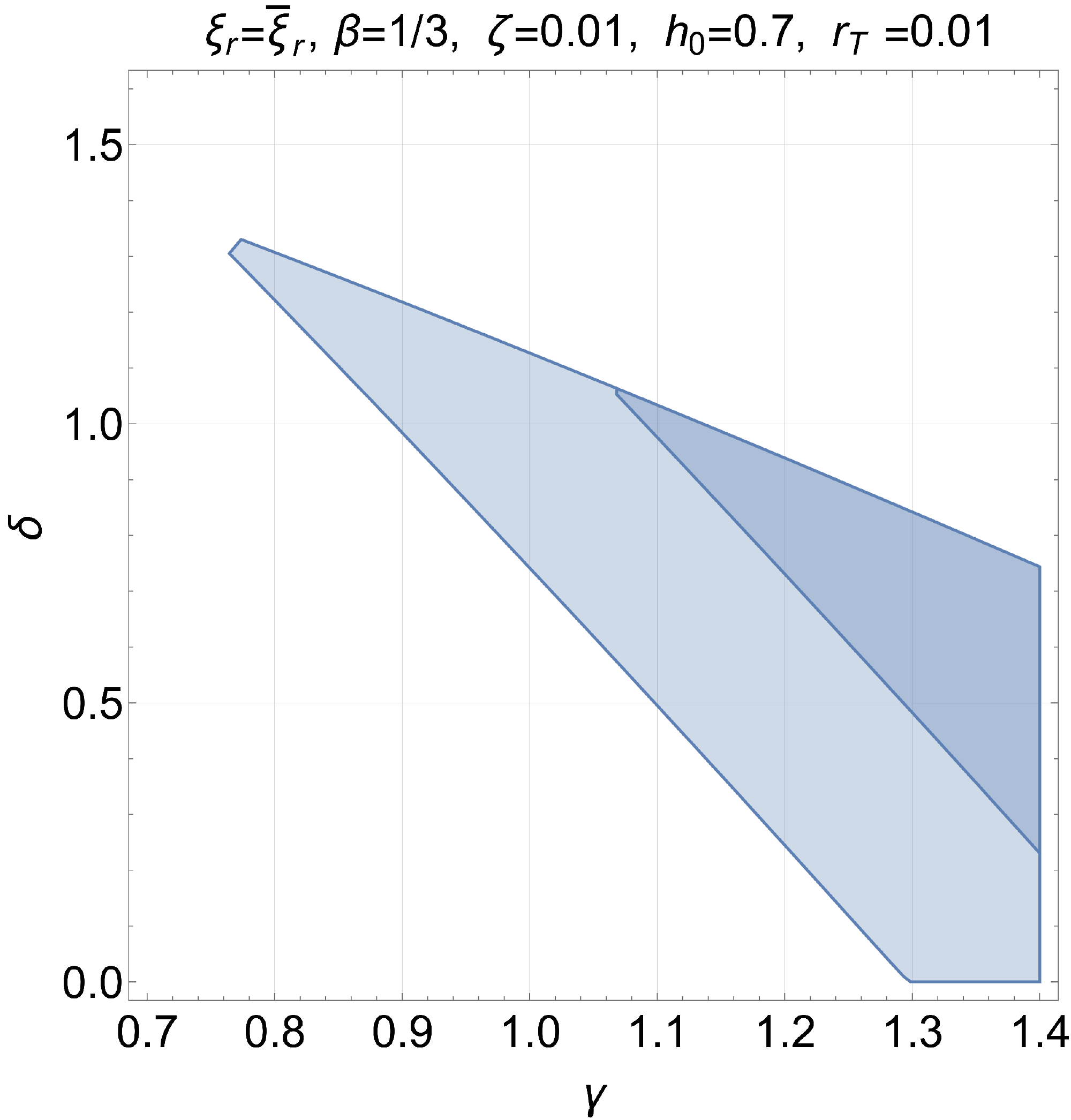}
\includegraphics[height=8cm]{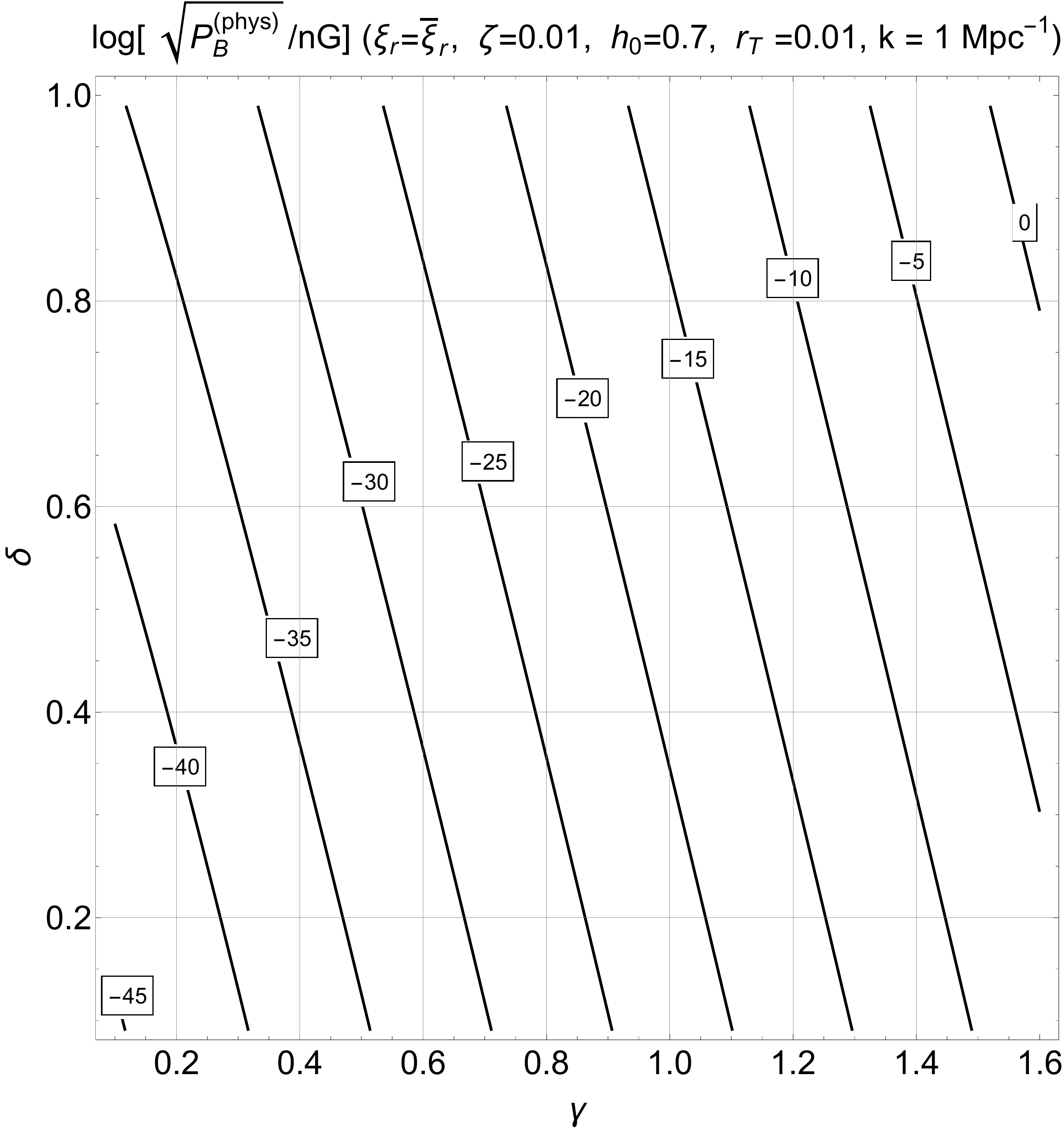}
\caption[a]{We illustrate the region where the critical energy density bound and the magnetogenesis 
constraints are satisfied. In this case all the parameters are fixed except $\gamma$ and $\delta$.}
\label{FIG4}      
\end{figure}
Since the requirements of Eq. (\ref{mg3}) are satisfied for $\beta \leq {\mathcal O}(1/3)$ and $1 <\gamma <2$,
it is interesting to fix exactly $\beta \to 1/3$ and consider $\xi_{r} \geq \overline{\xi}_{r}$. This complementary slice of the parameter space is illustrated in Fig. \ref{FIG2}.
Larger values of $\xi_{r}$ (i.e. shorter stiff phases) 
practically do not change the actual values of the power spectra if the scale is fixed. Conversely 
for fixed $\gamma$ (see the plot at the right in Fig. \ref{FIG2}) the values 
of the power spectrum decreases sharply when\footnote{Here we simply mean that for fixed $k$ and $\gamma$ an increase in $\xi_{r}$ does not change 
     the value of the power spectrum as it is clear from the left plot of Fig. \ref{FIG2} 
     where the contours 
     are approximately constant with $\xi_{r}$. When $\gamma$ and $\beta$ are fixed 
     an increase of $\xi_{r}$ entails a decrease of the power spectrum (see the right plot of Fig. \ref{FIG2}).} $ \xi \gg \overline{\xi}_{r}$.
Having analyzed the orders of magnitude of the problem, in Fig. \ref{FIG3} we concentrate on the interplay between the requirements of Eq. (\ref{mg3}) and the constraints stemming 
from the spectral energy density deduced in Eqs. (\ref{YYY4a})--(\ref{YYY4}) and in Eqs. (\ref{YYY7})--(\ref{YYY8}). In the left
plot  we consider the $(\beta,\, \gamma)$ plane. The regions with different shadings 
correspond to the classes of limits mentioned in Eq. (\ref{mg3}): the darker 
area illustrate the most constraining requirement while the lighter 
shading accounts for the less constraining conditions appearing in Eq. (\ref{mg3}). 
With the same notations the plane $(\xi_{r}, \gamma)$ is illustrated in the right plot. In both plots of Fig. \ref{FIG3} we took into account all the relevant backreaction constraints during the stiff phase.
In summary, when the gauge coupling 
freezes during the stiff epoch the magnetogenesis constraints are satisfied provided the following 
conditions are met:
\begin{equation}
1/3 \leq \beta < 1/2, \qquad \xi\ge \overline{\xi}_{r}, \qquad 1.2\leq \gamma < 2, \qquad \delta \ll \gamma.
\label{FINAL1}
\end{equation}
We finally conclude by noting that the variation of $r_{T}$ has clearly some 
impact on the magnetogenesis requirements so that we consistently 
selected for all the plots $r_{T} = 0.01$ \cite{RT1,RT2,RT2a}. Smaller values of $r_{T}$ 
do not change the conclusions discussed here. The recent analyses of Ref. \cite{RT2a} 
seem to imply further reductions of $r_{T}$ but even an excursion of $1$ order 
of magnitude (i.e. $r_{T} \to 0.001$) remains overall compatible with the general trends
observed here.

\subsubsection{Freezing during the radiation stage}
\begin{figure}[!ht]
\centering
\includegraphics[height=8cm]{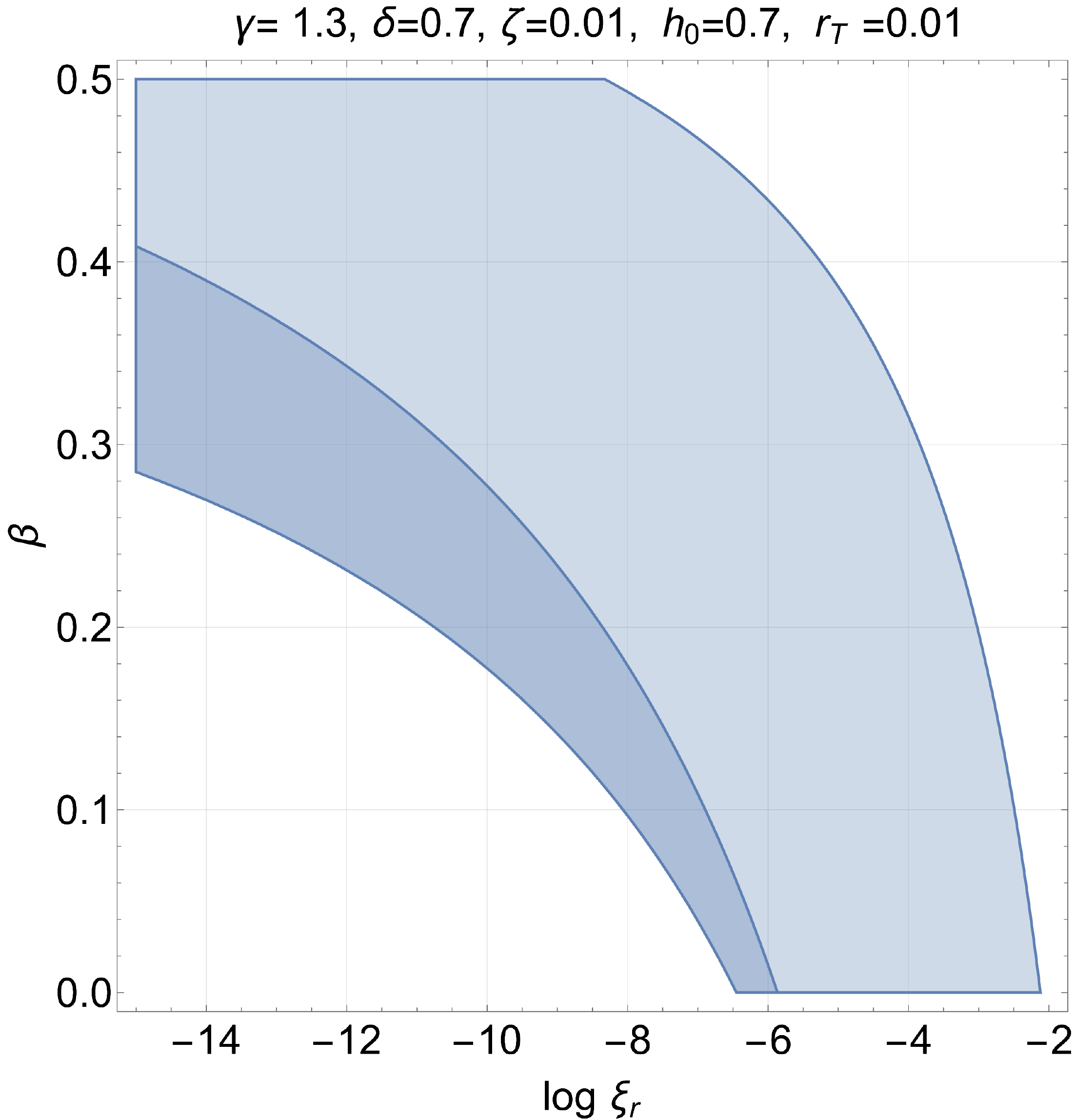}
\includegraphics[height=8cm]{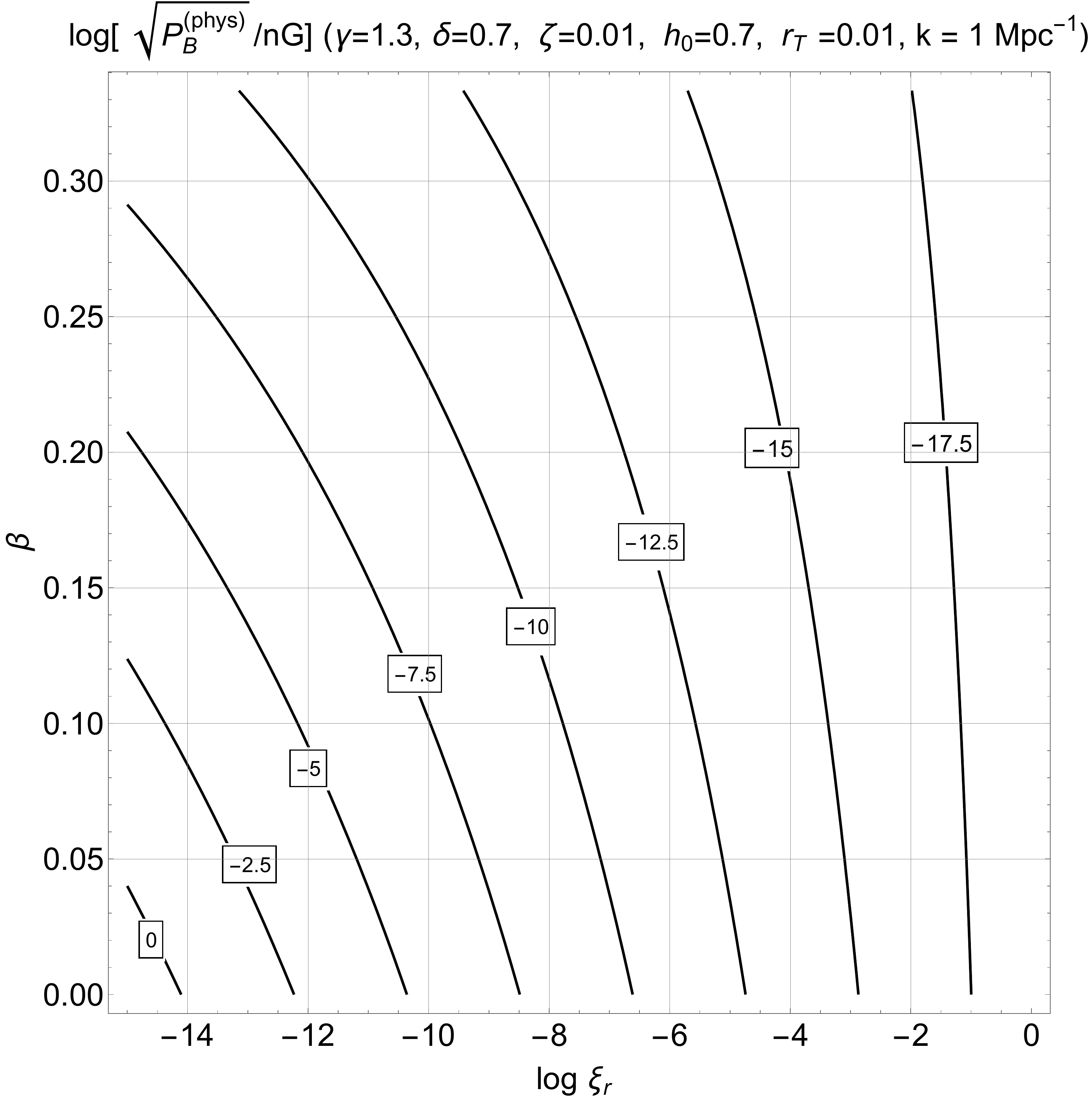}
\caption[a]{We illustrate the region where the critical energy density bound and the magnetogenesis constraints are satisfied. In this case all the parameters are fixed except $\beta$ and $\xi_{r}$.}
\label{FIG5}      
\end{figure}
If the freezing of the gauge coupling takes place during the radiation stage
we must focus on the case $\zeta \ll \delta$ while $\delta$ can be comparable with $\gamma$.
This means that the backreaction constraints of Eqs. (\ref{YYY10})--(\ref{YYY11}) and the limits of Eq. (\ref{mg3}) will not only depend on 
 $\xi_{r}$, $\beta$ and $\gamma$ but also on $\delta$.
This case is illustrated in Figs. \ref{FIG4} and \ref{FIG5}. 
The notations are exactly the ones already established in the 
previous figures. Both in Figs. \ref{FIG4} and \ref{FIG5} the plots at the left 
illustrate the allowed region of the parameter space where all the constraints and 
the magnetogenesis requirements are satisfied. In the plots at the right (both in Figs. \ref{FIG4} and \ref{FIG5}) we instead illustrate the same section of the parameter space by reporting the corresponding values of the power spectra in nG units.
In the left plot of Fig. \ref{FIG4} the values of $\xi_{r}$ and $\beta$ have been fixed to their fiducial values. In the $(\gamma, \,\delta)$ plane the region where the magnetogenesis constraints are satisfied coincides, broadly speaking, with the region where $\gamma = {\mathcal O}(\delta)$. 

In Fig. \ref{FIG5} we finally illustrated the situation where all parameters 
are fixed except $\beta$ and $\xi_{r}$. In the left plot we present the allowed 
region of the parameter space. In the plot at the right we instead present the 
values of the power spectra.
We can the conclude that in the case when the gauge coupling freezes during the radiation stage the fiducial region of the parameters is given by 
\begin{equation}
\zeta \ll \gamma = {\mathcal O}(\delta), \qquad 1/3\leq \beta < 1/2, \qquad \xi\geq \overline{\xi}_{r}.
\end{equation}
Again also in this case we could choose different values of $r_{T}$ and $h_{0}$. However 
the differences will be irrelevant for the present purposes. In spite of the fact 
that $\beta$ and $\xi_{r}$ have been considered as free parameters, it is interesting 
to remark that the allowed region of the parameter space pins down 
an area that roughly coincides with the theoretical expectation: a fairly long stiff 
phase dominated by kinetic energy of the inflaton/quintessence field. 

\newpage
\renewcommand{\theequation}{6.\arabic{equation}}
\setcounter{equation}{0}
\section{Concluding considerations}
\label{sec6}
The addition of higher-order curvature corrections is quite common in various classes of inflationary models and in the effective theory of inflation. While the addition of nonlinear contributions to the gravitational action is customarily discussed within the metric approach, in this paper 
we investigated the possibility of large-scale magnetogenesis in a
 quintessential inflationary scenario based on the Palatini formulation.

In the perspective developed here the addition of a nonlinear term does not affect the scalar modes of the geometry during the inflationary phase but the tensor-to-scalar ratio is suppressed. Among the various models  that could be analyzed within the Palatini formulation there are the dual potentials where the inflaton and the quintessence fields are identified. Provided the late-time potential has a quintessential form, it is natural to have a long stiff stage following the standard inflationary phase. Both 
aspects of this scenario have a  positive impact for the magnetogenesis considerations.
 The suppression of $r_{T}$ and the presence of long stiff phase affect
 the maximal number of $e$-folds accessible to large-scale observations 
 and also increase the energy density of the magnetic fields which remains nonetheless bounded by the 
 backreaction constraints. The objective has been to constrain the rate of 
evolution of the gauge coupling by imposing all the relevant limits 
together with the magnetogenesis requirements.

After examining in detail the evolution during the inflationary stage and in the post-inflationary phase, we computed the physical gauge spectra which turn out to be essential for the phenomenological purposes. We argued that the gauge coupling may freeze either during the stiff phase or in the radiation epoch but, in both cases, the magnetogenesis requirements are met.  The values of the magnetic power spectra may reach $0.01\mathrm{nG}^2$ roughly corresponding to values of the physical fields  ${\mathcal O}(0.1)$ nG over typical length scales between a fraction of the Mpc and $100$ Mpc prior to the gravitational collapse of the protogalaxy. The magnetogenesis requirements are therefore satisfied together with all the backreaction constraints both during and after inflation. The idea discussed here can also be applied to the scenarios characterized by different forms of the dual potential and by complementary parametrizations of the nonlinear action. While the overall logic and the general results 
are expected to be similar the details of the models might lead to some modifications.

\section*{Acknowledgments} 
It is a pleasure to thank T. Basaglia, A. Gentil-Beccot, S. Rohr and J. Vigen of the CERN Scientific Information Service for their kind help throughout various stages of this investigation.
\newpage

\begin{appendix}

\renewcommand{\theequation}{A.\arabic{equation}}
\setcounter{equation}{0}
\section{Explicit form of the matrix elements}
\label{APPA}
The explicit expressions of the matrix elements appearing in Eq. (\ref{VVV3}) are:
\begin{eqnarray}
A_{f\, f}(k,\tau, \tau_{1}) &=& \frac{\pi}{2} \sqrt{q_{1} x_{1}} \sqrt{ k y} \biggl[ Y_{\mu -1}( q_{1} x_{1}) J_{\mu}(k y) - J_{\mu-1}(q_{1} x_{1}) Y_{\mu}(k y) \biggr],
\nonumber\\
A_{f\, g}(k, \tau, \tau_{1}) &=& \frac{\pi}{2} \sqrt{q_{1} x_{1}} \sqrt{ k y} \biggl[ J_{\mu}( q_{1} x_{1}) Y_{\mu}(k y) - Y_{\mu}(q_{1} x_{1}) J_{\mu}(k y) \biggr],
\nonumber\\
A_{g\, f}(k, \tau, \tau_{1}) &=& \frac{\pi}{2} \sqrt{q_{1} x_{1}} \sqrt{ k y} \biggl[ Y_{\mu - 1}( q_{1} x_{1}) J_{\mu -1}(k y) - J_{\mu-1}(q_{1} x_{1}) Y_{\mu-1}(k y) \biggr],
\nonumber\\
A_{g\, g}(k, \tau, \tau_{1}) &=& \frac{\pi}{2} \sqrt{q_{1} x_{1}} \sqrt{ k y} \biggl[ J_{\mu}( q_{1} x_{1}) Y_{\mu-1}(k y) - Y_{\mu}(q_{1} x_{1}) J_{\mu-1}(k y) \biggr].
\label{VVV4}
\end{eqnarray}
In Eqs. (\ref{VVV4}) $J_{\alpha}(z)$ and $Y_{\alpha}(z)$ are the standard Bessel functions \cite{abr1,abr2}.
For the sake of conciseness the various arguments  appearing in Eq. (\ref{VVV4}) have been defined in the following manner:
\begin{equation}
x_{1} = k \tau_{1}, \qquad y(\tau) = \tau + (q_{1} +1)\tau_{1}, \qquad\qquad q_{1} = \delta/\gamma,\qquad\qquad \mu = \delta +1/2.
\label{VVV5}
\end{equation}
Equation (\ref{VVV5}) shows that, within the present notations,  $y(-\tau_{1}) = q_{1} \tau_{1}$ which also 
implies (by definition of $x_{1}$) that  $k y(-\tau_{1}) = q_{1} \, k\, \tau_{1} = q_{1} x_{1}$. As expected from the continuity of the mode functions, we also have that 
\begin{equation}
A_{f\,g}(k, - \tau_{1}, \tau_{1}) =A_{g\,f}(k, - \tau_{1}, \tau_{1})=0,\qquad 
 A_{f\,f}(k, - \tau_{1}, \tau_{1}) =A_{g\,g}(k, - \tau_{1}, \tau_{1})=1. 
 \label{VVV6a}
 \end{equation}
 While the first two results of Eq. (\ref{VVV6a}) follow from the definitions of Eq. (\ref{VVV5}), the Wronskian of the Bessel functions \cite{abr1,abr2} is required to deduce the second pair of relations. The 
matrix appearing Eq. (\ref{VVV3}) is unitary since: 
\begin{equation}
A_{f\, f}(k, \tau, \tau_{1}) A_{g\, g}(k, \tau, \tau_{1}) - A_{f\, g}(k, \tau, \tau_{1}) A_{g\, f}(k, \tau, \tau_{1}) =1,
\label{VVV6}
\end{equation}
as it  can also be verified from the explicit matrix elements of Eq. (\ref{VVV4}) and from  standard recurrence relations involving the Bessel functions and their Wronskians. 
The explicit expressions of the matrix elements appearing in Eq. (\ref{UUU1}) will be instead given by:
 \begin{eqnarray}
B_{f\, f}(k, \tau, \tau_{1}, \tau_{r}) &=& \frac{\pi}{2} \sqrt{q_{2} x_{r}} \sqrt{ k z} \biggl[ Y_{\nu -1}( q_{2} x_{r}) J_{\nu}(k z) - J_{\nu-1}(q_{2} x_{r}) Y_{\nu}(k z) \biggr],
\nonumber\\
B_{f\, g}(k, \tau, \tau_{1}, \tau_{r}) &=& \frac{\pi}{2} \sqrt{q_{2} x_{r}} \sqrt{ k z} \biggl[ J_{\nu}( q_{2} x_{r}) Y_{\nu}(k z) - Y_{\nu}(q_{2} x_{r}) J_{\nu}(k z) \biggr],
\nonumber\\
B_{g\, f}(k, \tau, \tau_{1}, \tau_{r}) &=& \frac{\pi}{2} \sqrt{q_{2} x_{r}} \sqrt{ k z} \biggl[ Y_{\nu - 1}( q_{2} x_{r}) J_{\nu -1}(k z) - J_{\nu-1}(q_{2} x_{r}) Y_{\nu-1}(k z) \biggr],
\nonumber\\
B_{g\, g}(k, \tau, \tau_{1}, \tau_{r}) &=& \frac{\pi}{2} \sqrt{q_{2} x_{r}} \sqrt{ k z} \biggl[ J_{\nu}( q_{2} x_{r}) Y_{\nu-1}(k z) - Y_{\nu}(q_{2} x_{r}) J_{\nu-1}(k z) \biggr].
\label{UUU2a}
\end{eqnarray}
As in the case of Eq. (\ref{VVV5}), the various arguments appearing in Eq. (\ref{UUU2a}) have been 
defined as:
\begin{equation}
x_{r} = k \tau_{r}, \qquad z(\tau) = \tau + (q_{2} - 1)\tau_{r}, \qquad q_{2} = \biggl[\frac{\zeta}{\delta}+ \zeta\biggl(\frac{\tau_{1}}{\tau_{r}}\biggr) \biggl(\frac{\delta + \gamma}{\gamma\delta}\biggr)\biggr],\qquad \nu= \zeta +1/2.
\label{UUU3a}
\end{equation}
It can be immediately checked that, as in the case of Eq. (\ref{VVV6}) the coefficients 
of Eq. (\ref{UUU2a}) obey:
\begin{equation}
B_{f\, f}(k, \tau, \tau_{1}, \tau_{r}) B_{g\, g}(k, \tau, \tau_{1},\tau_{r}) - B_{f\, g}(k, \tau, \tau_{1}, \tau_{r}) B_{g\, f}(k, \tau, \tau_{1}, \tau_{r}) =1.
\label{UUU3ab}
\end{equation}
Furthermore full analogy with Eq. (\ref{VVV6a}) we will also have that 
\begin{equation}
B_{f\,g}(k, \tau_{r}, \tau_{1}, \tau_{r}) = B_{g\,f}(k, \tau_{r},  \tau_{1}, \tau_{r})=0,
 \qquad B_{f\,f}(k, \tau_{r}, \tau_{1}, \tau_{r}) = B_{g\,g}(k, \tau_{r}, \tau_{1}, \tau_{r})=1,
 \label{UUU3abc}
 \end{equation}
as it can be explicitly seen by using Eq. (\ref{UUU3a}) and the Wronskians of the 
Bessel functions \cite{abr1,abr2}.

\end{appendix}


\newpage

\begin{thebibliography}{99}

\itemsep -5pt

\bibitem{Hone1} N. D. Birrel and P. C. W. Davies, {\it Quantum Fields in Curved Space} (Cambridge University Press, Cambridge, England, 1984).

\bibitem{Hone2} L. Parker and D. Toms, {\it Quantum Field Theory in Curved Spacetime: Quantized Fields and Gravity} (Cambridge University Press, Cam- bridge, England, 2009).

\bibitem{Htwo1}  D. G. Boulware and S. Deser, Phys. Rev. Lett. {\bf 55}, 2656 (1985);  Phys. Lett. B {\bf 175}, 409 (1986).

\bibitem{Htwo2} R. R. Metsaev and A. A. Tseytlin, Phys. Lett. B {\bf 191}, 115 (1987); Nucl. Phys. B {\bf 293}, 385 (1987).

\bibitem{Htwo3}  C. G. Callan, E. J. Martinec, M. J. Perry, and D. Friedan, Nucl. Phys. B {\bf 262}, 593 (1985); A. Sen, Phys. Rev. Lett. {\bf 55}, 1846 (1985).

\bibitem{Htwo4}  J. Madore, Phys. Lett. A {\bf 110}, 289 (1985); Phys. Lett. A {\bf 111}, 283 (1985).

\bibitem{Htwo5} B.~Zwiebach, Phys. Lett. B {\bf 156}, 315-317 (1985).

\bibitem{Htwo6}  S. Weinberg, Phys. Rev. D {\bf 77}, 123541 (2008). 

\bibitem{Htwo7} E. Elizalde, A. Jacksenaev, S. D. Odintsov, and I. L. Shapiro, Phys. Lett. B {\bf 328}, 297 (1994).

\bibitem{Htwo8} K.~S.~Stelle, Gen. Rel. Grav. {\bf 9}, 353 (1978).

\bibitem{Htwo9a} G. Stephenson, Nuovo Cim.\ {\bf 9}, 263 (1958).

\bibitem{Htwo9b}  P.~W.~Higgs, Nuovo Cim.\  {\bf 11},  816 (1959).

\bibitem{PAL1} I.~Antoniadis, A.~Karam, A.~Lykkas and K.~Tamvakis, JCAP {\bf 11}, 028 (2018).

\bibitem{PAL2} I.~Antoniadis, A.~Karam, A.~Lykkas, T.~Pappas and K.~Tamvakis, JCAP {\bf 03}, 005 (2019).

\bibitem{PAL3} V.~M.~Enckell, K.~Enqvist, S.~Rasanen and L.~P.~Wahlman, JCAP {\bf 02}, 022 (2019).

\bibitem{PAL4} M.~Giovannini, Class. Quant. Grav. {\bf 36},  235017 (2019).

\bibitem{PAL4a} I. D. Gialamas and A. Lahanas,  Phys. Rev. D {\bf 101} 084007 (2020).

\bibitem{PAL5} I.~Antoniadis, A.~Lykkas and K.~Tamvakis, JCAP {\bf 04},  033 (2020).

\bibitem{PAL6} I.~D.~Gialamas, A.~Karam and A.~Racioppi, JCAP {\bf 11}, 014 (2020).

\bibitem{PAL7} I.~Antoniadis, A.~Guillen and K.~Tamvakis, JHEP {\bf 08}, 018 (2021).

\bibitem{PAL8} I.~D.~Gialamas, A.~Karam, T.~D.~Pappas and V.~C.~Spanos, Phys. Rev. D {\bf 104},  023521 (2021).

\bibitem{PAL9}  A. A. Starobinsky, Phys. Lett. B {\bf 91}, 99 (1980).

\bibitem{RT1}  Y.~Akrami {\it et al.} [Planck Collaboration], Astron. Astrophys. {\bf 641}, A10 (2020).

\bibitem{RT2}  N.~Aghanim \textit{et al.} [Planck Collaboration], Astron. Astrophys. {\bf 641}, A6 (2020).

\bibitem{RT2a} P.~A.~R.~Ade {\it et al.} [BICEP/Keck], Phys. Rev. Lett. {\bf 127}, 151301 (2021).

\bibitem{QUINT1} P. J. E. Peebles and A. Vilenkin, Phys.Rev. D {\bf 59}, 063505 (1999).

\bibitem{QUINT2} P.~J.~E.~Peebles, Phys. Rev. D {\bf 62}, 023502 (2000).

\bibitem{QUINT3} P. Peebles, B. Ratra, Cosmology with a time variable cosmological constant, Astrophys. J. {\bf 325}, L17 (1988).

\bibitem{QUINT4} R. R. Caldwell, R. Dave, and P. J. Steinhardt, Phys. Rev. Lett. {\bf 80}, 1582 (1998).

\bibitem{RT3} G.~Hinshaw {\it et al.} [WMAP Collaboration],   Astrophys.\ J.\ Suppl.  {\bf 208}, 19 (2013).

\bibitem{RT4} C. L. Bennett, {\it et.al.}  [WMAP Collaboration],  Astrophys.\ J.\ Suppl.  {\bf 208}, 20B (2013).

\bibitem{QUINT5}  J.~Haro, W.~Yang and S.~Pan, JCAP {\bf 1901},  023 (2019).

\bibitem{QUINT6} M. Giovannini, Phys. Rev. D {\bf 58}, 083504 (1998). 

\bibitem{QUINT7} Ya. Zeldovich, Sov. Phys. Usp. {\bf 6}, 475 (1964) [Usp. Fiz. Nauk. {\bf 80}, 357 (1963)].

\bibitem{QUINT8} A. D. Sakharov, Sov. Phys. JETP {\bf 22}, 241 (1966) [Zh. Eksp. Teor. Fiz. {\bf 49}, 345 (1965)].

\bibitem{QUINT9} L.P. Grishchuk, Annals N. Y. Acad. Sci. {\bf 302}, 439 (1977).

\bibitem{QUINT10} L. H. Ford, Phys. Rev. D {\bf 35}, 2955  (1987).

\bibitem{QUINT11} B. Spokoiny, Phys. Lett. B {\bf 315}, 40 (1993).

\bibitem{QUINT12} A.R. Liddle, S.M. Leach, Phys. Rev. D {\bf 68}, 103503 (2008).

\bibitem{QUINT13} M.~Giovannini, Phys. Rev. D \textbf{67}, 123512 (2003).

\bibitem{rev1}  K. Enqvist, Int.\ J.\ Mod.\ Phys.\  D  {\bf 7}, 331 (1998).

\bibitem{rev2} M.~Giovannini,  Int.\ J.\ Mod.\ Phys.\  D {\bf 13}, 391 (2004).

\bibitem{rev3} J.~D.~Barrow, R.~Maartens and C.~G.~Tsagas,  Phys.\ Rept.\  {\bf 449}, 131 (2007).

\bibitem{PSC1}  S. Carroll, G. Field and R. Jackiw, Phys. Rev. D {\bf 41},  1231 (1990).

\bibitem{PSC2}  W. D. Garretson, G. Field and S. Carroll,  Phys. Rev. D {\bf 46}, 5346 (1992).

\bibitem{PSC3} G. Field and S. Carroll Phys. Rev. D, {\bf 62}, 103008 (2000).

\bibitem{PSC4} B. Ratra,  Astrophys.\, J.\, Lett.  {\bf 391}, L1 (1992).

\bibitem{PSC5} M.~Gasperini, M.~Giovannini, and G.~Veneziano, Phys. Rev. Lett. {\bf 75}, 3796 (1995).

\bibitem{PSC7} M. Giovannini, Phys.\ Rev.\ D {\bf 61}, 063004 (2000).

\bibitem{PSC9}   M.~Giovannini and M.~E.~Shaposhnikov,  Phys.\ Rev.\ D {\bf 57}, 2186 (1998).

\bibitem{PSC10} M. Giovannini, Phys.\ Rev.\ D {\bf 61}, 063502 (2000).

\bibitem{PSC11} M.~Giovannini, Phys. Rev. D {\bf 92}, 121301 (2015).

\bibitem{abr1}  A. Erdelyi, W. Magnus, F. Oberhettinger, and F. R. Tricomi {\em Higher Trascendental Functions} (Mc Graw-Hill, New York, 1953).

\bibitem{abr2} M. Abramowitz and I. A. Stegun, {\em Handbook of Mathematical Functions} (Dover, New York, 1972).

\bibitem{NINEaa} S. Deser and C. Teitelboim, Phys. Rev. D 13, 1592 (1976).

\bibitem{TENaa}  S. Deser, J. Phys. A {\bf 15}, 1053 (1982).
  
\bibitem{fordcoul} L. H. Ford, Phys. Rev. D {\bf 31}, 704 (1985).

\bibitem{mginc1} M.~Giovannini, Phys. Rev. D {\bf 85}, 101301 (2012).

\bibitem{mginc2} M.~Giovannini, Phys. Rev. D {\bf 86}, 103009 (2012).

\bibitem{cosmow} S. Weinberg, {\it Cosmology} (Oxford University Press, Oxford 2008).

\bibitem{cond0} M.~Giovannini and N.~Q.~Lan, Phys. Rev. D \textbf{80}, 027302 (2009).

\bibitem{cond1} M. Giovannini, Phys.\ Lett.\  B {\bf 659}, 661 (2008).

\bibitem{cond2} T.~Fujita and R.~Durrer, JCAP {\bf 09}, 008 (2019).

\bibitem{cond3} T.~Kobayashi and M.~S.~Sloth, Phys. Rev. D {\bf 100}, 023524 (2019).

\bibitem{b1}  H. Alfv\'en and C.-G. F\"althammer, {\it Cosmical Electrodynamics}, 2nd edn., (Clarendon press, Oxford, 1963).

\bibitem{b2} E. N. Parker, {\it Cosmical Magnetic Fields} (Clarendon Press, Oxford, 1979).

\bibitem{b3} Ya. B. Zeldovich, A. A. Ruzmaikin, D.D. Sokoloff  {\it Magnetic Fields in Astrophysics} (Gordon  Breach Science, New York, 1983).

\end{thebibliography}
\end{document}